\pdfoutput=1
\documentclass[11pt,arxiv]{suny-thesis}
\usepackage[utf8x]{inputenc} 
\usepackage{amsmath}
\usepackage{amssymb}
\usepackage{mathtools}
\usepackage{cancel}
\usepackage{textcomp}
\usepackage[numbers,sort&compress]{natbib} 
\usepackage{soul}
\usepackage{mathrsfs}
\usepackage{enumitem}
\usepackage{setspace}
\usepackage{esint}
\SetMathAlphabet{\mathbf}{normal}{OML}{mdput}{b}{n} 
\DeclareMathAlphabet{\mathpzc}{OT1}{pzc}{m}{it}
\renewcommand{\dotsm}{\!\cdot\!\!\cdot\!\!\cdot\!}
\renewcommand{\dotsb}{\cdot\!\cdot\!\cdot}
\newcommand{\pp}{\textit{pp.}}
\newcommand{\latin}[1]{\textit{#1}\!}
\DeclareMathOperator{\Amp}{\mathcal{A}}
\DeclareMathOperator{\Prob}{\operatorname{Pr}}
\DeclareMathOperator{\Ident}{\mathcal{I}}
\newcommand{{\Schroedinger}}{Schr{\"o}dinger}
\renewcommand{\Vec}[1]{\boldsymbol{\mathbf{#1}}}
\newcommand{\abs}[1]{\left|{#1}\right|}
\newcommand{\GradVar}[2]{\Vec{\nabla}\!_{#2}#1}
\newcommand{\Grad}[1]{\GradVar{#1}{}}
\newcommand{\Curl}{\Grad{\!\times\!}}
\newcommand{\Z}{\mathbb{Z}}
\newcommand{\R}{\mathbb{R}}
\newcommand{\Lagrangian}{\mathcal{L}}
\newcommand{\Hamiltonian}{\mathcal{H}}
\newcommand{\Action}{\mathcal{S}}
\newcommand{\Kernel}{\mathcal{K}}
\newcommand{\VecPot}{\Vec{A}}
\newcommand{\VecPotExc}{\VecPot_{\textsl{exc}}}
\newcommand{\ScaPot}{\phi}
\newcommand{\Flux}{\Phi}
\newcommand{\FluxExc}{\Phi_{\textsl{exc}}}
\newcommand{\D}[1]{\text{d}#1}
\newcommand{\DotD}[1]{\!\cdot\!\D{#1}}
\newcommand{\DD}[2]{\text{d}^{#1}#2}
\newcommand{\PathD}[1]{\mathpzc{D}\{#1\}}
\newcommand{\Partial}[2]{\frac{\partial#1}{\partial#2}}

\newcommand{\Perms}[1]{\mathbb{S}_{#1}}
\newcommand{\Mesh}{\mathcal{C}}
\newcommand{\Weight}{\chi}
\newcommand{\Rep}{D}
\newcommand{\bra}[1]{\left<#1\right|}
\newcommand{\ket}[1]{\left|#1\right>}
\newcommand{\braket}[2]{\left<#1\middle|#2\right>}
\newcommand{\Eqref}[1]{Eq.~\eqref{#1}}
\newcommand{\Figref}[1]{Figure~\ref{#1}}
\DeclareMathOperator{\Source}{\mathpzc{s}}
\DeclareMathOperator{\Target}{\mathpzc{t}}
\newcommand{\Grp}[1]{\mathrm{#1}}
\DeclareMathOperator{\SpOr}{\Grp{SO}}
\newcommand{\SecRef}[1]{Section~\ref{#1}}
\newcommand{\ChapRef}[1]{Chapter~\ref{#1}}
\newcommand{\Hfunc}{\mathit{H}}
\newcommand{\SymmOp}{\leftrightarrow}
\newcommand{\ThetaHat}{\Vec{\hat{\theta}}}
\newcommand{\RhoHat}{\Vec{\hat{\rho}}}
\allowdisplaybreaks
\numberwithin{equation}{chapter}
\author{Klil H. Neori}
\title{Identical Particles in Quantum Mechanics: Operational and Topological Considerations}
\college{College of Arts \& Sciences}
\department{Physics}
\date{2015} 
\begin{document}

\frontmatter

\maketitle
\renewcommand*{\thefootnote}{\fnsymbol{footnote}}
\begin{abstract}
\addcontentsline{toc}{chapter}{Abstract}
This dissertation reports our investigation into the existence of anyons, which interpolate between bosons and fermions, in light of the Symmetrization Postulate, which states that only the two extremes exist. The Symmetrization Postulate can be understood as asserting that there are only two consistent ways of combining the behavior of distinguishable particles to obtain the behavior of identical ones. We showed that anyonic behavior then arises because of the way in which the probability amplitudes of distinguishable particles in two dimensions are affected by the topology of the space. These can then be combined in one of the ways arising from the Symmetrization Postulate, to form identical anyons. Therefore, anyons do not invalidate the Symmetrization Postulate, but are entirely compatible with it. In order to show that anyons can arise without particle identity, we investigated how distinguishable particles can gain particular properties from the topology of their configuration space. We managed to do so, and in the process, discovered a novel approach for quantizing multiply-connected spaces, in a way that is more suited to anyons than the standard approach.
\end{abstract}
\begin{acknowledgments}
\addcontentsline{toc}{chapter}{Acknowledgments}
A doctorate is much like an apprenticeship, and must be lead by a good master. Philip Goyal encouaged me to apply to UAlbany, and has been an inexhaustible source of ideas and critiques, even before I finally committed to pursue my dissertation with him. I owe most of the rigor of this work to his keen eye for detail, and all of my novelty to the creative tension which we enjoyed through the years.

I would never have met Philip, nor come to UAlbany, had Avshalom C. Elitzur and Daniel Rohrlich not helped me get back into Physics. Since coming to Albany, I have spent many hours discussing ideas with Ariel Caticha and Oleg Lunin in person, and with Larry Schulman and John Skilling via email. Marco Varisco provided feedback for the groupoid work at a crucial moment, and Ronnie Brown and Peter A. Horv{\'a}thy contacted us unsolicited to provide further advice and context for it, which was very encouraging.

There is a phrase in Hebrew: without bread, there is no Torah. The Physics Department provided me with teaching assistantships for most of my tenure, and Philip provided me with research assistantships, both with his startup funds and through a grant from the John Templeton Foundation\footnote{``Quantifying Relations as a Foundation for Physics'', Grant ID 24435.}. The GSA provided a stipend for my work as Lead Senator. Travel and conference support came from the Department, from the John Templeton Foundation, and from the Jaynes Foundation at Boise State University. Finally, I was honored to have been chosen to receive the 2012 Mr. \& Mrs. Masood Alam Comprehensive Exam Award and the 2015 Akira Inomata Excellence in Research Award.

Good staff is vital for any academic endeavor. Paul Labate, David Liguori, and the indispensable Leslie Saint~Vil kept the money and information flowing, the office and class-space allocated, and the technology functioning. Librarians and their assistants provided electronic and physical copies of essential source materials, vital for foundations work.

I wish to thank my fellow graduate students, both at the Physics Department and through the GSA. 
I could count on any of them for moral support, comfort, and stress-relief. I would not be where I am without them.

Finally, my generous and long-suffering family. My parents, who have supported me emotionally and financially from birth, and encouraged me relentlessly on this journey, and my siblings, who persist in their affection despite my rarely punctuated absence in the last five years.
\end{acknowledgments}
\chapter*{Preface}\label{ch:preface}
\addcontentsline{toc}{chapter}{Preface}
The following is a PhD dissertation, reporting research carried out by Klil H. Neori under the supervision of Philip Goyal, at the University at Albany, SUNY, from from the summer of 2011 to the fall of 2015. The research topic was identical particles in quantum mechanics. A short overview of the entire project was presented as~``From Operational Identical Particles to Distinguishable Anyons and Back Again'' at~PASCAL 2015\footnote{Physics All-Student Conference at Albany 2015, March 7, 2015, University at Albany, SUNY, Albany, New York.}.

In~\ChapRef{ch:dissintro} we provide an introduction to the conceptual and topological background needed for placing our main results in context. First we present the more traditional approach to identical particles, where quantization precedes symmetrization. The restriction to bosons and fermions was eventually recognized to require a Symmetrization Postulate. \SecRef{sec:identpart}, in particular, refers to an operational proof of this postulate published in~\citet{Goyal2015}. Preliminary results focusing on the non-interacting, two-particle case were presented as~``On the Origin of the Quantum Rules for Identical Particles'' at~{QTRF-6}\footnote{Quantum Theory: Reconsideration of Foundations - 6, June 11-14, 2012, Linnaeus University, V{\"a}xj{\"o}, Sweden.} and~MaxEnt 2012\footnote{32$^{\text{nd}}$ International Workshop on Bayesian Inference and Maximum Entropy Methods in Science and Engineering, July 15-20, 2012, MPP, Garching near Munich, Germany.}, and published in the proceedings of the latter as~\citet{NeoriGoyal2013}. As explained in~Ref.~\cite{Goyal2015}, this result concerns the combination of distinguishable-particle transition amplitudes into those of identical particles, which is important for the application to anyons.

The reduced configuration space approach to identical particles attempts to derive the Symmetrization Postulate by considering particles to be identical before quantization, and then using topological tools to restrict their behavior, rather than identity being imposed as a symmetry after quantization~\cite{Souriau1967,LaidlawDeWitt1971,LeinaasMyrheim1977}. This approach leads to additional possibilities, called anyons, in two dimensions. Using the operational understanding of the Symmetrization Postulate, we show that anyonic behavior is solely due to topology, rather than particle identicality. An overview of these results was presented as ``Anyons in the Operational Formalism'' at~MaxEnt 2014\footnote{34$^{\text{th}}$ International Workshop on Bayesian Inference and Maximum Entropy Methods in Science and Engineering September 21-26, 2014, Clos Luc{\'e}, Amboise, France.}, published in its proceedings as~\citet{NeoriGoyal2015a}. The derivation of quantization in multiply-connected spaces using the fundamental groupoid, taken up in~\ChapRef{ch:groupoidresults}, was submitted for publication as~\citet{NeoriGoyal2015b}. This is followed by an explicit formation of distinguishable anyons, and their combination into identical anyons, in~\ChapRef{ch:distidentanyons}. \ChapRef{ch:histover} closes the novel contributions of this dissertation with a discussion of the historical and philosophical implications of the work. The latter two chapters comprise the contents of~\citet{NeoriGoyal2015c}, which is in draft form.

\newpage
\tableofcontents
\listoffigures
\mainmatter
%
\renewcommand*{\thefootnote}{\arabic{footnote}}
\setcounter{footnote}{0}
\chapter{Introduction}\label{ch:dissintro}
Elementary particles are usually assumed to belong to one of two classes:~\emph{bosons}, which are limited to symmetric multiple-particle wave-functions; and~\emph{fermions}, which are limited to anti-symmetric multiple-particle wave-functions.

Initial works in the new quantum mechanics did not venture beyond fermions and bosons~\cite{Dirac1926,Heisenberg1926}. However, it was soon noted that there is more to be said about the subject.~\citet{Wigner1927a,Wigner1927b} was the first to explore other possibilities. He treated particle exchanges as generators of the symmetric group~$\Perms{N}$ of all permutations over~$N$ particles, and used group representation theory to perform a more general analysis. He found that, indeed, for two particles only bosons and fermions are possible, but even three particles~\cite{Wigner1927a} make the situation more complicated: while bosons and fermions emerge as one-dimensional representations of~$\Perms{N}$, there are two more two-dimensional representations, which might correspond to a different type of particle.~\citet{Pauli1980} provided a similar development, and considered the fact that these other options do not come about to be a mystery yet to be resolved. \citet{Dirac1930} similarly used group representation theory to treat identical particles, but never actually developed even proto-paraparticles. While he acknowledged that other possibilities are mathematically possible, he noted that only bosons and fermions were so far needed to explain natural phenomena.

For a few decades, only a limited number of \textit{ad hoc} attempts to push beyond these two possibilities were published. \citet{Gentile1940} posited identical particle types allowing for an arbitrary number~$d$ in each single-particle state. However, this formulation depends on the choice of basis, so it was not pursued further. \citet{Green1953} made a similar extension in quantum field theory, but this, again, was somewhat \latin{ad hoc} and did not immediately lead to further developments.

The issue was only seriously and rigorously taken up in the 1960s. At first, several authors attempted to prove that there are only two types of identical particles, using the basic rules of quantum mechanics combined with the assumption that the particles are indistinguishable~\cite{Jauch1960,JauchMisra1961,Galindoetal1962,Pandres1962}. However, \citet{MessiahGreenberg1964} responded by showing that this was not a direct consequence of particle indistinguishability. Instead, limiting the situation to bosons and fermions requires an additional assumption, which they termed \emph{the Symmetrization Postulate}. In the context of non-relativistic quantum mechanics, they proposed to call particles which, as anticipated by~Refs.~\cite{Wigner1927a,Wigner1927b,Pauli1980}, live in multi-dimensional representations of the symmetric group, \emph{paraparticles}. They then showed how difficult it would be to test for this kind of behavior with most common experiments, while suggesting new ones which would nevertheless allow deviations to be detected. The impetus for this investigation was the introduction by~\citet{Greenberg1964} of the idea that quarks are paraparticles: specifically, that they live a two-dimensional irreducible representation of~$\Perms{3}$, which helps explain how they are able to share a ground state in nucleons, even though they otherwise share properties with fermions. \citet{Greenberg2015} later argued that this was an important step on the way to the ultimate resolution of this conundrum, namely, the introduction of color as an additional internal degree of freedom for quarks~\cite{GreenbergZwanziger1966}. This direction continued to be explored and extended into a field-theoretical generalization called~\emph{quons}~\cite{Greenberg1991}.

In addition to these extensions, attempts to prove the Symmetrization Postulate, or at least to break it down to less blatant subsidiary assumptions, continued. One notable example is~\citet{Tikochinsky1988}. This work proved the Symmetrization Postulate for non-interacting particles by working from the Feynman rules for transition amplitudes~\cite{Feynman1948}. We were able to improve upon the rigor of that result by starting from the operational approach to reconstructing the Feynman rules developed in~\citet{GoyalKnuthSkilling2010}. We published preliminary results, for two non-interacting particles, in~\citet{NeoriGoyal2013}. A generalization to~$N$ interacting particles can be found in~\citet{Goyal2015}.

An important alternative attempt at proving the Symmetrization Postulate, which takes an entirely different tack, creates a challenge for the operational approach. A common feature of the non-relativistic approaches described so far is that they start from particles which are labeled, and whose exchange is then treated as a physical process, akin to rotation or translation. Particle identity is then imposed, just like spherical symmetry or space homogeneity.~\citet{Dirac1930} is particularly explicit about this point. However, as was most clearly expressed by~\citet{Mirman1973}, if particles are identical, then there is no physical meaning to their labels, nor to the exchange thereof. Consequently, there is no reason to assume the existence of unitary symmetry operators corresponding to particle label permutations. One way to resolve this is by treating exchanges as the end results of paths undertaken by particles. That is, the particles start out in a certain state, move through their configuration space, and then end up in an end-state which differs from their initial state only by a permutation of the particles. Particle identity can then be expressed by taking this classical configuration space and identifying points which only differ by such a permutation. The resulting, \emph{reduced} configuration space, can now be quantized. The hope is that then the Symmetrization Postulate can be proven as resulting from quantization over this space. This \emph{reduced configuration space approach} was popularized by~\citet{LeinaasMyrheim1977}, although it was preceded by~\citet{Souriau1967} and~\citet{LaidlawDeWitt1971}.

In the reduced configuration space, exchange paths correspond to certain types of paths starting and ending at the same point, that is, to loops. The degree of freedom which allows for bosons and fermions now comes from the fact that there are several types of paths, called homotopy classes. Each class consists of all paths which can be continuously deformed to one another without changing the end points. If we restrict ourselves to loops, paths starting and ending at the same point, then we have a group structure called the~\emph{fundamental group}, which is an algebraic topological property of the space. The representations of this group are what corresponds to statistics in the label-based formalism. Unlike there, the possible statistics depend on the dimension of the space in which the particles dwell. 

All three of the original works~\cite{Souriau1967,LaidlawDeWitt1971,LeinaasMyrheim1977} came to the same conclusions about identical particles in three dimensions. As long as particle incidence points are removed, the fundamental group is simply the symmetric group, meaning that there are two one-dimensional representations, corresponding to bosons and fermions. However, Ref.~\cite{LeinaasMyrheim1977} found that in two dimensions or less, there is a whole family of particle types interpolating between bosons and fermions. \citet{Wilczek1982b} coined the term~\emph{anyons} to describe this new range of possibilities in two dimensions. When anyons are exchanged, their wave-function accrues a certain complex amplitude, which depends on the particular loop of exchange, and is of the form~$e^{ik\varphi}$, where~$e^{i\varphi}$ is the generator of the statistics of the anyons in question, in analogy to the~$\pm1$ accrued by exchanged bosons and fermions, respectively\footnote{\citet{Girardeau1965} did anticipate the importance of topology as early as~\citeyear{Girardeau1965}, but only treated connectivity issues. He focused on the contrast between one- and three-dimensional spaces, skipping over two.}.

On the surface, this could be considered as simply a failure to prove the Symmetrization Postulate. However, anyonic behavior is not just theoretical: the Fractional Quantum Hall Effect, which occurs in thin layers in condensed matter~\cite{TsuiStormerGossard1982,Laughlin1983a,ArovasSchriefferWilczek1984,Stormer1999,Laughlin1999}, results in excitations which have anyonic statistics~\cite{CaminoZhouGoldman2005,KimLawlerVishveshwaraFradkin2005}. Therefore, any attempt to deal with identical particles must provide for this kind of behavior in two dimensions. This, however, does not mean that anyonic behavior necessarily requires identical particles. As was noted by \citet{GoldinMenikoffSharp1985}, even distinguishable particles may have topological exchange amplitudes, as it is the multiple-connectedness of the configuration space that creates them, and that happens in two dimensions as soon as the coincidence points are removed. It is very common to generate the topological amplitudes through the use of non-local interaction~\cite{Lerda1992,Morandi1992,Zee1995}.

Nevertheless, the references to~\cite{LeinaasMyrheim1977} have imbued anyons with identicality, and connected them to the Symmetrization Postulate. An overwhelming portion of the literature about anyons, such as~Refs.~\cite{Wilczek1982b,SudarshanImboShahImbo1988,Schulman1993,Wu1984,Stern2008}, takes reducing the configuration space as their starting point (Ref.~\cite{Wilczek1982b} resorts to it in implying complications in generalization for more than two particles, even though the author does not need this for two particles).

Now that we have provided a general motivational overview,  we will dive into individual issues that build up the understanding necessary for appreciating the novel results reported herein~(\ChapRef{ch:litrev}). We will explore the strangeness of quantum mechanics~(\SecRef{sec:strangeqm}), and the reconstruction efforts it motivated~(\SecRef{sec:recon}). We will then study the relation between the~\Schroedinger{} and Feynman formalisms~(\SecRef{sec:sch2feyn}), followed by the operational reconstruction which uses the latter~(\SecRef{sec:opapproach}), and its application to the issue of identical particles~(\SecRef{sec:identpart}).

Following that we will discuss the reduced configuration space approach to identical particles~(\SecRef{sec:redconf}), which lead to anyons, followed by an explanation of the most striking evidence for their existence, the Fractional Quantum Hall Effect~(\SecRef{sec:fqheanyons}). Moving through the Ahoronov-Bohm Effect~(\SecRef{sec:abeffect}), we will explain homotopy, an important topological tool, and will specifically discuss a relatively unusual tool, the fundamental groupoid~(\SecRef{sec:homotopy}).

This background chapter concluded, there will come our novel results. First, we explore the possibilities of quantization in multiply-connected spaces through the representations of the fundamental groupoid~(\ChapRef{ch:groupoidresults}). This can be used to choose specific topological amplitudes for paths, not just for loops as with the fundamental group. We use this to provide consistent amplitudes for distinguishable particle exchanges, which, in the special case of the punctured plane, gives us distinguishable anyons. These can be combined using the operational results mentioned in~\SecRef{sec:identpart} to create identical anyons~(\ChapRef{ch:distidentanyons}). This cleanly breaks up the behavior of identical anyons into the topological level, which is what grants them the ability to interpolate between bosons and fermions, and the operational level, which gives rise to the bosons and fermions due to identicality.

Finally, we will explore the persistence of the view that the reduced configuration space is vital for anyons and their exploration, providing some philosophical background having to do with the idea of identity~(\ChapRef{ch:histover}). We will find that the attempt to avoid the use of labels as unmeasurable in themselves simply leads to other kinds of unmeasurable superstructure, whether it is in covering spaces or fiber bundles. This approach offers no escape from auxiliary degrees of freedom. 
\chapter{Detailed Background on Identical Particles}\label{ch:litrev}
\section{The Strangeness of Quantum Mechanics}\label{sec:strangeqm}
Quantum mechanics has been a counter-intuitive theory since its inception. The link between its abstract mathematical formalism and experienced reality is through measurements, for which it only provides a statistical distribution~\cite{Born1926}, yet it does not include an underlying deterministic theory. It satisfies complementarity~\cite{Bohr1937}, so that attempts to ascribe qualities to systems outside of measurements lead to paradoxes, like that of~\citet{EPR1935}. Identical quantum-mechanical particles behave strangely, either satisfying the Pauli Exclusion Principle~\cite{Pauli1925} or Bose-Einstein statistics~\cite{Bose1924,Einstein1924}.

Nevertheless, it has been extremely successful at explaining phenomena which were not amenable to classical physical models, and was able to generate many successful predictions, which then lead to additional phenomena worth investigating. As a result, most of the intellectual energy dedicated to quantum mechanics has been focused upon solving problems with it, rather than better understanding its foundations.

For those who did wish for quantum mechanics to be more intuitive, there was a sense that in the background to quantum mechanics there might lie a theory that is classical, albeit statistical and very complicated. From the Bohm-de Broglie pilot wave theory~\cite{Bohm1952a,Bohm1952b} to Nelson's stochastic mechanics~\cite{Nelson1966}, it seemed like there must be a way to dig in and find an underlying deterministic theory. However, results such as the theorems of~\citet{Bell1964} and~\citet{KochenSpecker1968} showed that, even if such a theory were to be found, it would be highly non-local, as well as context-dependent. Therefore, it would be nothing like classical models which we previously found conducive to our understanding.
\section{Reconstructing Quantum Mechanics}\label{sec:recon}
These issues have lead researchers to investigate other routes of rebuilding or reconstructing quantum mechanics, in ways which bring into focus the physical intuitions required to understand it. Hardy's 5 axioms~\cite{Hardy2001} are an attempt to use continuous unitary transformations as the highlight of quantum behavior, while working within the realm of states and operators. Caticha's Entropic Dynamics~\cite{Caticha2011} tries to go as far as possible with ideas stemming from information processing, before adding physical assumptions, resulting in the wave-function formalism through a Hamilton-Jacobi formulation.

Our work follows upon the operational approach introduced in~\citet{GoyalKnuthSkilling2010} to reconstruct specifically the Feynman rules for quantum mechanics~\cite{Feynman1948}. Before we delve into it, let us provide a brief overview of how the Feynman rules relate to the more familiar Dirac-\Schroedinger{} formulation.
\section{\Schroedinger{} \textit{vs.} Feynman Rules in Quantum Mechanics}\label{sec:sch2feyn}
The usual entity of importance in quantum mechanics is the wave-function, or, more generally, the wave-vector, which is a description of the physical state of a system. We start with the state in which a system is prepared. In Dirac notation, we have:
\begin{equation}
\ket{\psi_0}\text{;}
\end{equation}
the wave-function for a particle is written as~$\psi_0(x)=\braket{x}{\psi_0}$. The prepared system then evolves with time, with an optional interaction with an outside field, which is expressed by transforming this wave-vector or wave-function through the application of a unitary transformation generated by a Hamiltonian~$\Hamiltonian$, which, for simplicity, we take to be time-independent:
\begin{equation}
\ket{\psi(t)}=e^{-i\Hamiltonian t/\hbar}\ket{\psi_0}\text{;}
\end{equation}
the current wave-function is then~$\psi(x,t)=\braket{x}{\psi(t)}$. Finally, the system is measured, using a complete orthonormal set of wave-functions~$\ket{e}$ representing the possible classical outcomes. The probability for each outcome~$e$ is then the absolute value squared of the component of the resulting wave-function on the member of the unitary basis reflecting that outcome (Born's Rule):
\begin{equation}
\Prob(e)=\abs{\braket{e}{\psi(t)}}^2\text{.}\label{eq:measbraket}
\end{equation}

The Feynman approach focuses on the relation between preparation and measurement. We can rewrite the inner product between the final wave-function and a measurement state as a matrix element of a propagator indexed by the initial state and the final state:
\begin{equation}
\Amp(e)=\braket{e}{\psi(t)}=\bra{e}e^{-i\Hamiltonian t/\hbar}\ket{\psi_0}
\end{equation}
so that:
\begin{equation}
\Prob(e)=\abs{\Amp(e)}^2\text{.}
\tag{\ref{eq:measbraket}'}\label{eq:measamp}
\end{equation}
Full calculation comes through the Feynman path-integral approach, in which one would write:
\begin{align}
\psi(x, t)= \braket{x}{\psi(t)}=\bra{x}e^{-i\Hamiltonian t/\hbar}\ket{\psi_0}&=\int\D{x'}\bra{x}e^{-i\Hamiltonian t/\hbar}\ket{x'}\braket{x'}{\psi_0}\notag\\
&=\int\D{x'}\Kernel(x, t\,; x')\psi_0(x')\text{,}
\end{align}
and the propagator~$\Kernel(x,t\,; x')$ becomes the interface with physical theory. Feynman's rules then relate to how these propagators act when additional measurements are interposed between the preparation and the final measurement. 

Acting through transition amplitudes or propagators makes many quantum mechanical phenomena easier to express. For example, if we have a transition amplitude~$\Amp_U$ for a particle passing from a point~$x$ on the left of a barrier to a point~$y$ to the right of it when only one hole is open, and another transition amplitude~$\Amp_D$ for passing from one to the other when only another hole is open, then the amplitude for this transitions when both holes are open is~$\Amp_U+\Amp_D$. This is generally called the~\emph{sum rule} for transition amplitudes. The fact that it is the amplitudes rather than the probabilities which add is what causes an interference pattern to appear on a screen to the right. On the other hand, if a measurement were to decide whether a particle was going through one hole or the other, then the probabilities would add, as would be expected for classical particles. 

The other Feynman rules are the~\emph{product rule}, which states that the amplitudes of two processes done in succession multiply~(which can easily be seen from the propagator expression, since successive developments in time simply multiply by~$e^{-iHt_n/\hbar}$);~\emph{reciprocity}, which states that the amplitude of a set of measurements or a path taken in reverse order (within a short enough timeframe, so that~$e^{-i\Hamiltonian t/\hbar}\to\Ident$) is the complex conjugate; and, finally, Born's rule for calculating the probabilities from the amplitudes, which states that the transition probability is given by the modulus-squared of the corresponding amplitude.
\section{The Operational Approach to Quantum Mechanics}\label{sec:opapproach}
The aim of the operational approach in~Ref.~\cite{GoyalKnuthSkilling2010} is to derive the Feynman rules for these transition amplitudes from more fundamental principles. Focusing on amplitudes rather than states or measurement operators has the advantage of inter-relating physically measurable results, namely measurement outcomes. The idea is that sufficiently many physical assumptions may result in some constraints on how these amplitudes can be combined, and then we will end up with Feynman's rules, from which the rest of Quantum Mechanics can be derived. 

This seems to have first been attempted by~\citet{Tikochinsky1988a,Tikochinsky1988b}, and independently by~\citet{Caticha1998}. The idea was simple: start from sequences of measurement outcomes as your basic objects, and assume that their probability is derived from complex transition amplitudes. Then use the ways in which these sequences can be combined to find symmetries in the functions relating the respective amplitudes, which lead to functional equations. Solutions to these equations lead to the sum, product, and complex conjugate rules. Three main distinctions between~Refs.~\cite{Tikochinsky1988a,Tikochinsky1988b} and~Ref.~\cite{Caticha1998} were that the latter, unlike the former, had a more careful definition of ``and'' and ``'or'', did not assume determinism, and did not use reciprocity, which corresponds to negation from~\citet{Cox1946}, and which they would rather avoid. The third distinction is acknowledged in the retrospective~\citet{TikochinskyGull2000}.

Nevertheless, both of these derivations assume complex numbers, and moreover, that the functions in their functional equations are analytic, so that it makes sense to differentiate them by their complex arguments. In both cases, analyticity does a lot of the heavy lifting for them, and there is a risk that the importance of other physical or logical assumptions is neglected as a result. Indeed,~Ref.~\cite{Tikochinsky1988a} itself notes that, once analyticity has been used, one of the algebraic symmetries is automatically satisfied. Analyticity also does most of the work in~\citet{Tikochinsky1988}.

\citet{GoyalKnuthSkilling2010} is notable for starting with weaker hypotheses: instead of assuming analyticity, or even complex numbers, they posit that two real numbers are fundamentally needed to calculate the probability of a process, an assumption which they argue is a formalization of complementarity~\cite{Bohr1937}. That, combined with several other assumptions that are physically or logically motivated, leads to the Feynman rules, without assuming analyticity.
\section{Behavior of Identical Particles in Quantum Mechanics}\label{sec:identpart}
Once the basics of quantum mechanics have been derived by a certain framework, the best way to bolster its credibility is to apply it to additional problems in quantum mechanics, and to see whether it can shed any light on them. Indeed,~\citet{Tikochinsky1988} did so for the issue of identical particles in quantum mechanics.

Current elementary particles fall into two types: \emph{bosons} and \emph{fermions}. Bosons can be pushed into the same state without limit. Photons are bosons, which is why electromagnetic waves are so easily treated classically: the overabundance of photons in the same state masks the quantum nature of light. In wave-function terms, this property is expressed in their overall wave-function being symmetric; for example:
\begin{align}
\Psi(x_2,x_1,x_3)&=+\Psi(x_1,x_2,x_3)\text{.}\label{eq:labelexbosons}
\intertext{Fermions, meanwhile, refuse to be in the same state. Electrons are fermions, which is why atoms are built up by progressively filling up shells of electrons. This property is expressed by their wave-function being anti-symmetric:}
\Psi(x_2,x_1,x_3)&=-\Psi(x_1,x_2,x_3)\text{.}\label{eq:labelexfermions}
\end{align}

The~\emph{Symmetrization Postulate}, as noted in \ChapRef{ch:dissintro}, states that these are the only types possible. Ref.~\cite{Tikochinsky1988} proved that this is the case for non-interacting particles, starting with the Feynman rules Tikochinsky previously derived in~Refs.~\cite{Tikochinsky1988a,Tikochinsky1988b}, and using the same framework. Here, too, analyticity played a key role, as did several implicit assumptions. 

The research reported in~\citet{NeoriGoyal2013} and~\citet{Goyal2015} was able to overcome the dependence on analyticity. Additionally, one of the implicit assumptions in~Ref.~\cite{Tikochinsky1988}, namely, that the amplitude for an identical particle transition had a functional dependence on the single-particle amplitudes, was made explicit in the special case of two, non-interacting particles in~Ref.~\cite{NeoriGoyal2013}. In~Ref.~\cite{Goyal2015}, this was generalized to a form which allows the incorporation of interactions:  the amplitude of the identical-particle transition was now assumed to have a functional dependence on the amplitudes for the corresponding transitions of the distinguishable particle system. This leads to a different view of what the Symmetrization Postulate means: instead of limiting to two types of particle behaviors, it limits to two types of combinations of distinguishable-particle behaviors.
\section{The Reduced Configuration Space}\label{sec:redconf}
Now comes a significant jump in subject and tone from the preceding, which focused on the standard, textbook approach to dealing with identical particles in non-relativistic quantum mechanics: one starts by assuming that one knows how to quantize a system of distinguishable particles; one then takes the elements of this quantum theory, whether these were wave-functions or transition amplitudes, and imposes particle identity as a form of additional symmetry, which leads to constraints as to the types of statistics possible. This is the most common way of addressing identical particles.

However, as we noted in~\ChapRef{ch:dissintro}, three foundational papers,~\citet{Souriau1967},~\citet{LaidlawDeWitt1971} and~\citet{LeinaasMyrheim1977}, took a different approach. They assumed that the classical particles were themselves identical, and then derived the resulting properties of identical quantum particles from the nuances of the quantization of an already fully symmetrized system. Ref.~\cite{LeinaasMyrheim1977} argued most strongly for this point, responding to the questioning of the very idea of particle labels rased in~\citet{Mirman1973}, and citing the Gibbs paradox~\cite{Gibbs1960} as additional evidence. These papers initiated the the~\emph{reduced configuration space approach}. This was presaged by a discussion at the 5th Solvay Conference in 1927, where Einstein and Lorentz noted a problem with using the full configuration space for~$N$ particles, and Pauli responded with a solution involving field theory~(see~\citet[\pp~181--183,442--443]{BacciagaluppiValentini2009}). Nevertheless, our discussion will continue to restrict itself to the non-relativistic regime, with a constant number of particles.

Since the developments in this approach rely heavily on the properties of the space of the classical particles, they had to make that an explicit parameter of their analysis, choosing:~$\R^n$, for~$n\ge1$. All three papers reproduce the symmetrization postulate for identical particles in~$\R^3$, considered the most physical single-particle configuration space. In fact, the Symmetrization Postulate is recovered for any~$n\ge3$. However, the situation is different in two and one dimensions. This is merely referred to in passing in~Refs.~\cite{Souriau1967,LaidlawDeWitt1971}, but is developed at length in~Ref.~\cite{LeinaasMyrheim1977}. Their treatment shows that in those cases, there is a continuous interpolation between fermions and bosons. The two-dimensional case in particular leads to particles dubbed~\emph{anyons}. They capture the properties of excitations in two-dimensional quantum systems, and their physical manifestation is found in the Fractional Quantum Hall Effect~\cite{TsuiStormerGossard1982,Laughlin1983a,ArovasSchriefferWilczek1984,Stormer1999,Laughlin1999}.

As already presaged in~\citet{Schulman1971}, where it was noted that the fundamental group for a twice-punctured plane is non-abelian, further theoretical development of this approach included the incorporation of parastatistics, in this case particles existing in multiple-dimensional representations of the braid group. Another direction was the extension from~$\R^n$ to circles, tori and spheres. These, and the use of more advanced mathematical tools, are summarized in~\citet{ImboShah-ImboMahajanSudarshanca.1990}. 

In the next section, we will examine the clearest expression of anyons in Nature, the fractional quantum Hall effect for low-temperature two-dimensional systems.
\section{The Fractional Quantum Hall Effect and Anyons}\label{sec:fqheanyons}
The Fractional Quantum Hall Effect furnishes the most reliable evidence for the existence of anyons in Nature. We will build up to them, starting from the Classical Hall Effect, for which the only form of quantization needed is the fact that charges are discrete, passing through the Integer Quantum Hall Effect, in which electrons are quantum-mechanical, but non-interacting, and finally to the Fractional Quantum Hall Effect, in which electron-electron interaction is fundamental, and over which anyonic excitations can form.

\begin{figure}[ht!]
\centering
\includegraphics{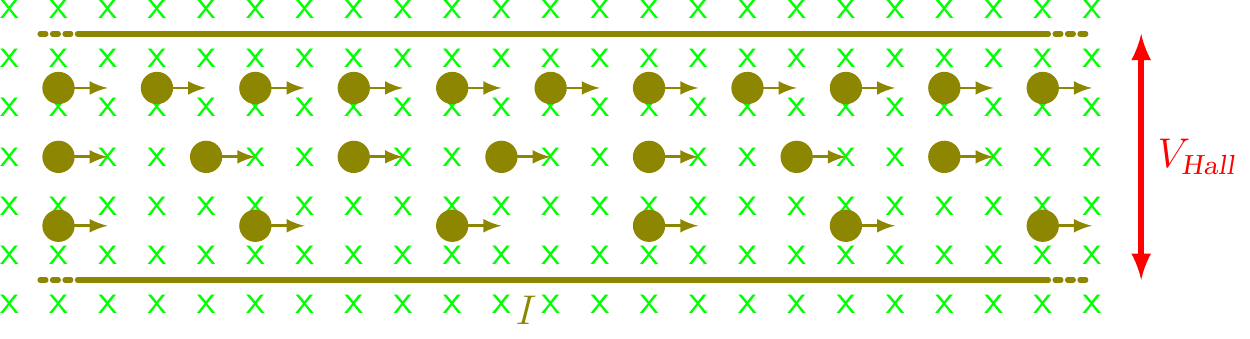}
\caption[Hall effect]{Negative charge carriers passing horizontally through a conductor subject to a magnetic field pointing towards the page create a vertical Hall voltage.}\label{fig:fqhe}
\end{figure}

The Hall effect~\cite{Hall1879} is the existence of a transverse voltage drop over an approximately flat conductor in the direction perpendicular to a magnetic field passing through it. After charges realign so that the system is in a steady state, charges are expected to move along the wire, so that the forces in the transverse direction, being the Lorentz force from the magnetic field in one direction, and the electrostatic force from the inhomogeneous charge distribution in the other, will cancel out. This results in a~\emph{Hall voltage}:
\begin{equation}
V_{\textsl{Hall}}=\frac{IB}{qnh}\text{,}
\end{equation}
where~$I$ is the current, $B$ is the magnetic field component perpendicular to the plane of the wire,~$q$ is the unit charge of the current carriers,~$n$ is the density of the carriers in the material, and~$h$ is the wire's height.~(See~\Figref{fig:fqhe})

This can be rewritten as the~\emph{Hall coefficient}:
\begin{equation}
R_{\textsl{Hall}}\triangleq\frac{V_{\textsl{Hall}} h}{IB} = \frac{1}{nq}\text{,}
\end{equation}
so that macroscopic measurements in the middle expression is a measure of an important property of the material on the right: namely, the charge density of the carriers, and very importantly, their sign.

In usual conductors, such as copper, the sign is negative, as clearly current is carried by electrons, so~$q=-e$. One indication that the situation in semiconductors is peculiar, and requires the use of the quantum theory of solids, is the existence of~\emph{positive} Hall coefficients, meaning that they act as if they have positively-charged current carriers, even though the positive ions do not move around. The solution is that~\emph{holes}, or absences of electrons in almost-filled energy bands, act like positive carriers.

Nevertheless, this is insufficient for understanding the \emph{Quantum} Hall Effects. As their name would suggest, they depend on the explicit quantum behavior of electrons in solids under a magnetic field. Let us start with the Integer Quantum Hall Effect.

In the classical case, we expect the Hall conductance~$\sigma_{\textsl{Hall}}$ to depend linearly on the reciprocal of the external magnetic field~$1/B$:
\begin{equation}
\sigma_{\textsl{Hall}}=\frac{I}{V_{\textsl{Hall}}}=\frac{qnh}{B}\text{;}
\end{equation}
however, at low temperatures, and under sufficiently strong magnetic fields, the situation becomes quite different. Let us rewrite the expression in terms of the London unit of flux ($\Phi^{q}_0\triangleq 2\pi\hbar c/q$):
\begin{equation}
\sigma_{\textsl{Hall}}=\frac{qnh}{B}=\frac{q(N/V)h}{B}=\frac{qN}{BA}=\frac{qN}{\Flux}=\frac{qN}{k\frac{2\pi\hbar c}{q}}=\frac{N}{k}\cdot \frac{q^2}{2\pi\hbar c}\text{,}
\end{equation}
where~$A$ is the area of the conductor in the direction perpendicular to the magnetic field.

The ratio~$\nu\triangleq N/k$, loosely the number of electrons per flux quantum, is called the~\emph{filling factor}. When this is an integer, it means that a Landau level is full. Now, if the flat wire were completely ordered and pure, then we would not expect there to be any difference between this situation and a classical one, as is argued in~\citet{Laughlin1999}, since we should be able to substitute any current with a relativistic boost of a resting wire, so quantum phenomena would not enter into it. However, if there are imperfections in the sample, these cause there to be localized states, which create gaps between the Landau levels. This leads to plateaus, as there are regions in which an increase in the magnetic field does not lead to additional empty states in the next Landau level. These plateaus are the manifestation of the integer quantum Hall effect measured by~\citet{KlitzingDordaPepper1980}. Notice that the discussion so far has treated electrons as non-interacting fermions (hence the importance of filled or unfilled Landau levels). Under these circumstances, the integer effect is the most we can comprehend. In order to explain the fractional effect, we will need to incorporate collective behavior.

Following the exposition in~\citet{Stormer1999}, for a non-integer~$\nu=N/k$, we can think of the system in the two-dimensional conductor as consisting of~$N$ electrons per~$k$ vortices generated by flux quanta; furthermore, the vortices represent a many-particle configuration in which there is a gap in the electron cloud. If we take Coulomb interactions into account, it is energetically favorable for electrons to coincide with these vortices, and the more vortices surrounding each electron, the farther away are the rest of the electrons, so the lower the energy. The lowest-energy configuration will have the vortices spread evenly between the electrons. For example, for a filling factor of~$\nu=\frac{1}{3}$, the case measured in~\citet{TsuiStormerGossard1982} and theoretically explained by~\citet{Laughlin1983a,Laughlin1983b}, the favored state has three vortices surrounding each electron. The resulting electrons with fluxes are called~\emph{composite particles}.

The flux quanta generating these vortices have important consequences for statistics. Each flux quantum changes the sign of an exchange. So an electron (a fermion) with an odd number of flux quanta becomes a boson, while if it has an even number of flux quanta, it remains a fermion. In the~$\nu=\frac{1}{3}$ case, the number of flux quanta is odd, so we get composite bosons. That means that they can condensate, and we get a very stable multi-particle ground state. Like in the integer case, there is a plateau, hence the fractional quantum Hall effect. Note that for~$\nu=\frac{1}{2}$, each electron would be coupled with two flux quanta at the ground state; this would make the composite particle a fermion, and these fill up the Landau levels until the Fermi energy. The situation is similar to that at zero field, so the system behaves as then: without a plateau.

So far we have only discussed composite particles, which are fermions with flux quanta transforming them either into bosons or back into fermions. The objects which have fractional statistics are quasiparticles which are \emph{excitations} of the condensate. We again discuss the~$\nu=\frac{1}{3}$ case. Suppose we change the magnetic field a little. Then we may increase it, adding a flux quantum, equivalent to a deficit of one-third of an electron, or a quasiparticle of charge~$+\frac{1}{3}e$. Alternately, if the magnetic field is slightly less than necessary for this filling factor, then a few electrons will be missing their third flux, so there will be surpluses of thirds of electrons, or quasiparticles of charge~$-\frac{1}{3}$. As shown in~\citet{Halperin1984,Halperin1984e} and~\citet{ArovasSchriefferWilczek1984}, these particles have an anyonic phase of~$e^{i\pi/3}$ (and generally, for an excitation over filling factor~$\nu$, the anyonic phase is~$e^{i\pi\nu}$). It is also there that the connection with~Ref.~\cite{Wilczek1982b} and, in the latter, with~Ref.~\cite{LeinaasMyrheim1977}, is made. 

Since the fractional quasiparticles are excitations of collective states of a material, they cannot simply be extracted, interfered with one another, and then detected in convenient bubble chambers in order to establish their statistics. Instead, interference must happen within the same two-dimensional material, at low temperature, and the magnetic field needed to establish the condensate over which they arise. It took more than two decades, but this was eventually achieved by two groups:~\citet{CaminoZhouGoldman2005} and~\citet{KimLawlerVishveshwaraFradkin2005}. The former provides particular credence to promoting distinguishable anyons, as the evidence is provided through the interference of one~$\nu=\frac{1}{3}$ particle around a tube full of~$\nu=\frac{2}{5}$ particles. This is seen as evidence of anyon ``statistics'' even though these particles are distinguishable. This is another example of the fact that, at least in the experimental literature, ``statistics'' is used interchangeably with a topological interaction between the particles, rather than some property hinging on particle identity, or the reduction of the configuration space. We will pursue anyonic behavior for distinguishable particles through strictly topological tools in~\SecRef{sec:distanyons}.

But is this evidence of anyons, or just of excitations which act anyonically? After all, these are not fundamental particles, but \emph{quasi}particles over a complicated ground-state. Their existence depends upon interactions between more fundamental particles, namely electrons and ions in a solid, as well as an external magnetic field. However, even in Quantum Field Theory, which is the theory where many of what we call fundamental particles are expressed, there is an underlying assumption that it is an effective theory of excitations over something more fundamental. That is the basis for renormalization: the theory does not express behavior all the way to arbitrarily high energies or arbitrarily short distances, so it is valid to simply stop integration at a certain maximal energy, in order to avoid the ultra-violate divergences. In that sense, the fractional statistics excitations over the Fractional Quantum Hall Effect ground state are explained by an anyonic effective theory, and are measurable, so they are evidence of such particles existing, and therefore any foundational approach to quantum mechanics would have to be able to take that into account.

In order to facilitate our ability to do so, we will introduce a simpler example of the importance of topology in quantum mechanics, the Aharonov-Bohm effect, followed by an introduction to relevant mathematical tools necessary for the rest of this Part, namely path homotopy and the fundamental groupoid.
\section{The Aharonov-Bohm Effect}\label{sec:abeffect}
We now step back from the 1980s to the late 1950s. One of the first hints that global aspects of the configuration space create curious situations in quantum mechanics came from the~Aharonov-Bohm effect~\cite{AharonovBohm1959}. Unlike their lesser-known predecessor~\citet{EhrenbergSiday1949}, covering similar ground, Ref.~\cite{AharonovBohm1959} cited topological concerns explicitly. The issue in question was a peculiar property of electromagnetism in the quantum formalism. Recall that in classical electromagnetism, it is the fields~$\Vec{E}$ and~$\Vec{H}$ which are fundamental and appear in the equations of motion:
\begin{equation}
m\ddot{\Vec{r}}=e\Vec{E}+\frac{e}{c}\left(\dot{\Vec{r}}\times\Vec{B}\right)\text{,}
\end{equation}
while the potentials~$\ScaPot$ and~$\VecPot$, related to the former through:
\begin{equation}
\Vec{E}=-\Grad{\ScaPot}\text{ and }\Vec{A}=\Curl{\VecPot}\text{,}
\end{equation} 
are relegated to formal convenience when expressing such systems in the Lagrange or Hamilton formalisms. In quantum mechanics, however, it is in fact the Hamiltonian which appears in the main dynamical equation, whether that of~\Schroedinger{} or of~Heisenberg, and similarly it is the Lagrangian which appears in the Feynman path integral approach. Both of these feature the potentials directly.

Now, quantum mechanics is gauge invariant; in the~\Schroedinger{} formalism, this is expressed by saying that if~$\Psi$ is a solution of:
\begin{equation}
i\hbar\Partial{}{t}\Psi(\Vec{r})=\left[\frac{1}{2m}\left[-i\hbar\Grad{}-\frac{q}{c}\VecPot(\Vec{x},t)\right]^2+q\ScaPot(\Vec{x},t)\right]\Psi(\Vec{r})\text{,}\label{eq:schrnrem}
\end{equation}
then~$\Psi'=e^{\frac{iq}{\hbar c}\Lambda}\Psi$ is a solution of:
\begin{equation}
i\hbar\Partial{}{t}\Psi'(\Vec{r})=\left[\frac{1}{2m}\left[-i\hbar\Grad{}-\frac{q}{c}\VecPot'(\Vec{x},t)\right]^2+q\ScaPot'(\Vec{x},t)\right]\Psi'(\Vec{r})\text{,}\tag{\ref{eq:schrnrem}$'$}\label{eq:schrnremgt}
\end{equation}
with:
\begin{equation}
\VecPot'=\VecPot+\Grad{\Lambda}\text{ and }\ScaPot'=\ScaPot-\frac{1}{c}\Partial{}{t}\Lambda\text{,}
\end{equation}
for arbitrary~$\Lambda(\Vec{x},t)$. Since the only effect on the wave-function is multiplying by a phase, the potentials themselves are not significant directly; however, in some cases one is forced to have a non-zero potential in a region where no field operates; for example, around an isolated solenoid containing a magnetic field. Does this field then have a physical effect, even though the electron is excluded from the region in which the field itself acts?

The surprising answer is that it does, in fact, have an effect: the Aharonov-Bohm effect. The interference pattern created by an electron scattering around an isolated solenoid depends in a measurable way on the flux through this same solenoid. Let us show how this comes about.

\begin{figure}[ht!]
\centering
\includegraphics{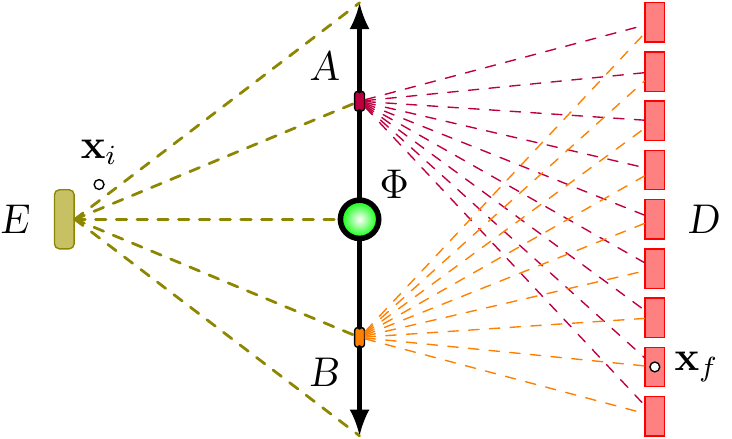}
\caption[Aharonov-Bohm setup]{\textsl{E} is an electron emitter, \textsl{D} is an array of electron detectors on the right, \textsl{S} is a screen with two slits, \textsl{A} and~\textsl{B}, and a solenoid which is isolated from the electrons, and through which there passes a magnetic flux $\Phi$ controlled via the current. $\Vec{x}_i$ is an arbitrary point near the emitter, and~$\Vec{x}_f$ is the location of a detector.}\label{fig:abeschr}
\end{figure}

We start with the~\Schroedinger{} formalism. Let us use a setup as in~\Figref{fig:abeschr}. We have an electron emitter~\textsl{E} on the left, and an array~\textsl{D} of electron detectors on the right. In between, we have a screen~\textsl{S}, presumed to extend arbitrarily high and low, with two slits, \textsl{A} and~\textsl{B}, between which is embedded a solenoid which is isolated from the electrons, and through which passes a magnetic flux, controlled by adjusting the current running through it. Finally, we single out some point~$\Vec{x}_i$ on the left side of the screen, and a representative point~$\Vec{x}_f$ at one of the detectors.

The wave-function value for a point~$\Vec{x}_f$ on the screen for the packet created when~\textsl{B} is closed and there is no flux is~$\Psi_{\textsl{A}}^{(0)}(\Vec{x}_{\textsl{f}})$, while when~\textsl{A} is closed  we have~$\Psi_{\textsl{B}}^{(0)}(\Vec{x}_{\textsl{f}})$. Due to the superposition principle, when both slits are open, the total wave-function value will be:
\begin{equation}
\Psi_{\Sigma}^{(0)}(\Vec{x}_{\textsl{f}})=\Psi_{\textsl{A}}^{(0)}(\Vec{x}_{\textsl{f}})+\Psi_{\textsl{B}}^{(0)}(\Vec{x}_{\textsl{f}}).
\end{equation}

Now we perform the same experiment with a current running through the solenoid, creating a total flux~$\FluxExc$, while there is still no field outside of the excluded area. With~\textsc{B} blocked, there are no longer any holes in the region allowed for the electrons; furthermore, there is no magnetic field in this region, so that the vector potential satisfies:
\begin{equation}
\Curl{\VecPot}=0\text{;}
\end{equation}
this means that we can apply Stokes' theorem to show that the line integral of the vector potential between two points only depends on the end-points, or, equivalently, that the vector potential is a conservative field, and is generated by a scalar function:
\begin{equation}
\VecPot=\Grad{\Lambda_{\textsl{A}}}\text{,}
\end{equation} 
where we choose:
\begin{equation}
\Lambda_{\textsl{A}}(\Vec{x})\triangleq\int_{\Vec{x}_{\textsl{i}}\xrightarrow[\textsl{A}]{}\Vec{x}}\!\!\!\!\!\!\!\!\!\!\!\!\VecPotExc(\Vec{x}')\DotD{\Vec{x}'}\text{.}
\end{equation}

\begin{figure}[ht!]
\centering
\includegraphics{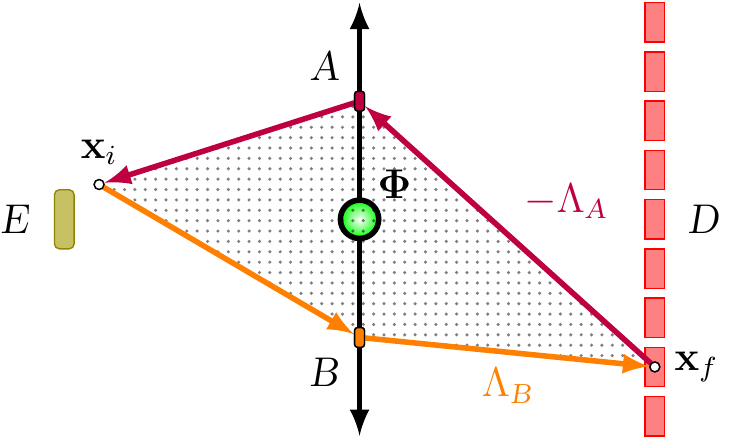}
\caption[Aharonov-Bohm using~\Schroedinger{} approach]{Subtracting $\Lambda_{\textsl{A}}$ from~$\Lambda_{\textsl{B}}$ results in a closed line integral around the flux, which leads to a surface integral including the flux via Stokes' theorem.}\label{fig:abeschrints}
\end{figure}

We can then use a gauge transformation to express the effect on the wave-function at the detector through multiplication by an appropriate phase depending on this generating function:
\begin{equation}
\Psi_{\textsl{A}}^{(\FluxExc)}(\Vec{x}_{\textsl{f}})=e^{\frac{iq}{\hbar c}\Lambda_{\textsl{A}}(\Vec{x}_{\textsl{f}})}\Psi_{\textsl{A}}^{(0)}(\Vec{x}_{\textsl{f}})\text{;}
\end{equation}
similarly:
\begin{equation}
\Psi_{\textsl{B}}^{(\FluxExc)}(\Vec{x}_{\textsl{f}})=e^{\frac{iq}{\hbar c}\Lambda_{\textsl{B}}(\Vec{x}_{\textsl{f}})}\Psi_{\textsl{B}}^{(0)}(\Vec{x}_{\textsl{f}}),
\end{equation}
where:
\begin{equation}
\Lambda_{\textsl{B}}(\Vec{x})\triangleq\int_{\Vec{x}_{\textsl{i}}\xrightarrow[\textsl{B}]{}\Vec{x}}\!\!\!\!\!\!\!\!\!\!\!\!\VecPotExc(\Vec{x'})\DotD{\Vec{x'}}
\end{equation}
Using simple superposition, we get for the total wave-function:
\begin{equation}
\begin{aligned}
\Psi_T^{(\FluxExc)}(\Vec{x}_{\textsl{f}})&=e^{\frac{iq}{\hbar c}\Lambda_{\textsl{A}}(\Vec{x}_{\textsl{f}})}\Psi_{\textsl{A}}^{(0)}(\Vec{x}_{\textsl{f}})+e^{\frac{iq}{\hbar c}\Lambda_{\textsl{B}}(\Vec{x}_{\textsl{f}})}\Psi_{\textsl{B}}^{(0)}(\Vec{x}_{\textsl{f}})=\\
&=e^{\frac{iq}{\hbar c}\Lambda_{\textsl{A}}(\Vec{x}_{\textsl{f}})}\left[\Psi_{\textsl{A}}^{(0)}(\Vec{x}_{\textsl{f}})+e^{\frac{iq}{\hbar c}[\Lambda_{\textsl{B}}(\Vec{x}_{\textsl{f}})-\Lambda_{\textsl{A}}(\Vec{x}_{\textsl{f}})]}\Psi_{\textsl{B}}^{(0)}(\Vec{x}_{\textsl{f}})\right].
\end{aligned}
\end{equation}
Now (see~\Figref{fig:abeschrints}):
\begin{equation}
\Lambda_{\textsl{B}}(\Vec{x}_{\textsl{f}})-\Lambda_{\textsl{A}}(\Vec{x}_{\textsl{f}})=\varointctrclockwise_{\Vec{x}_{\textsl{i}}\leftrightharpoons\Vec{x}_{\textsl{f}}}\!\!\!\!\!\!\!\!\!\!\!\VecPotExc(\Vec{x})\DotD{\Vec{x}}=
\iint_{\mathcal{S}}\Curl{\VecPotExc(\Vec{x}})\DotD{\Vec{a}}=
\iint_{\mathcal{S}}\Vec{B}\DotD{\Vec{a}}=\FluxExc,\label{eq:primitiveabfluxint}
\end{equation}
so:
\begin{equation}
\Psi_T^{\FluxExc}(\Vec{x}_{\textsl{f}})=e^{\frac{iq}{\hbar c}\Lambda_{\textsl{A}}(\Vec{x}_{\textsl{f}})}\left[\Psi_{\textsl{A}}^{(0)}(\Vec{x}_{\textsl{f}})+e^{2\pi i\left(\FluxExc/\Flux_0^q\right)}\Psi_{\textsl{B}}^{(0)}(\Vec{x}_{\textsl{f}})\right]\text{;}\label{eq:abinterferencesch}
\end{equation}
we see that the magnetic flux in the solenoid, which the electron does not encounter directly, nevertheless can be probed by the electron through the interference pattern on the detector array. This effect has the period $\Phi^{q}_0\triangleq\frac{2\pi\hbar c}{q}$, called London's unit of flux, which we mentioned in~\SecRef{sec:fqheanyons}.

Let us now show how the path integral formalism, which we will be using extensively later, is applied to this case. Recall that there, the focus is on calculating propagators:
\begin{equation}
\Psi_f(\Vec{x}_f, t_f)=\int \Kernel(\Vec{x}_{\textsl{f}}, t_{\textsl{f}}\,;\Vec{x}_{\textsl{i}}, t_{\textsl{i}}) \Psi_f(\Vec{x}_i, t_i)\DD{3}{x},
\end{equation}
that is,~$\Kernel$ is the transition amplitude between one point to the other. This is calculated through the path integral:
\begin{equation}
\Kernel(\Vec{x}_{\textsl{f}}, t_{\textsl{f}}\,;\Vec{x}_{\textsl{i}}, t_{\textsl{i}}) = \int\exp\left(\frac{i\Action[\Vec{x}(t)]}{\hbar}\right)\PathD{\Vec{x}(t)},\label{eq:intprop}
\end{equation}
where $\PathD{\bullet}$ is used instead of $\D{\bullet}$ in the integral in order to emphasize the sum over paths. The action functional $\Action$ is defined as:
\begin{equation}
\Action[\Vec{x}(t)] = \int\limits_{t_{\textsl{i}}}^{t_{\textsl{f}}}\!\!\Lagrangian\left(\Vec{x}(t), \dot{\Vec{x}}(t),t\right)\D{t}.
\end{equation}

For a charged particle in a general electromagnetic field affecting it directly, denoted by the scalar potential $\ScaPot_0$ and the vector potential $\VecPot_0$, along with an additional time-constant magnetic field excluded from the space of the particle, denoted by $\VecPotExc$:
\begin{equation}
\Lagrangian(\Vec{x},\dot{\Vec{x}},t)= \frac{m}{2}\dot{\Vec{x}}^2-q\ScaPot_0(\Vec{x},t)+\frac{q}{c}\left[\VecPot_0(\Vec{x},t)+\VecPotExc(\Vec{x})\right]\cdot\dot{\Vec{x}},
\end{equation}

Since the vector potential for the excluded flux is time-independent, if we integrate the Lagrangian into an action, we get an additive term depending only on the path in space, rather than on the path in space-time:
\begin{equation}
\begin{aligned}
\Action[\Vec{x}(t)] &= \int\limits_{t_{\textsl{i}}}^{t_{\textsl{f}}}\left(\frac{m}{2}\dot{\Vec{x}}^2-q\ScaPot_0(\Vec{x},t)+
\frac{q}{c}\VecPot_0(\Vec{x},t)\cdot\dot{\Vec{x}}+\frac{q}{c}\VecPotExc(\Vec{x})\cdot\dot{\Vec{x}}\right)\D{t} =  \\
&=\int\limits_{t_{\textsl{i}}}^{t_{\textsl{f}}}\left(\frac{m}{2}\dot{\Vec{x}}^2-q\ScaPot_0(\Vec{x},t)
+ \frac{q}{c}\VecPot_0(\Vec{x},t)\cdot\dot{\Vec{x}}\right)\D{t}
+\frac{q}{c}\int\limits_{\Vec{x}_{\textsl{i}}}^{\Vec{x}_{\textsl{f}}}\VecPotExc(\Vec{x})\DotD{\Vec{x}}.
\end{aligned}
\end{equation}

If we now define the first term, which only depends on the particle's behavior and the fields acting upon it, as $\Action_0[\Vec{x}(t)]$, and see the integral in the second term as a generalization of the~$\Lambda$ functions from earlier into a functional~$\Lambda[\Vec{x}(t)]$, and insert these into the exponent in~\Eqref{eq:intprop}, we get the following suspiciously familiar result:
\begin{equation}
\Kernel(\Vec{x}_{\textsl{f}}, t_{\textsl{f}}\,;\Vec{x}_{\textsl{i}}, t_{\textsl{i}}) = \int\exp\left(\frac{i}{\hbar}\Action_0[\Vec{x}(t)]\right)\cdot\exp\left(\frac{iq}{\hbar c}\Lambda[\Vec{x}(t)]\right)\PathD{\Vec{x}(t)};\label{eq:abpropabstract}
\end{equation}

\begin{figure}[ht!]
\centering
\includegraphics{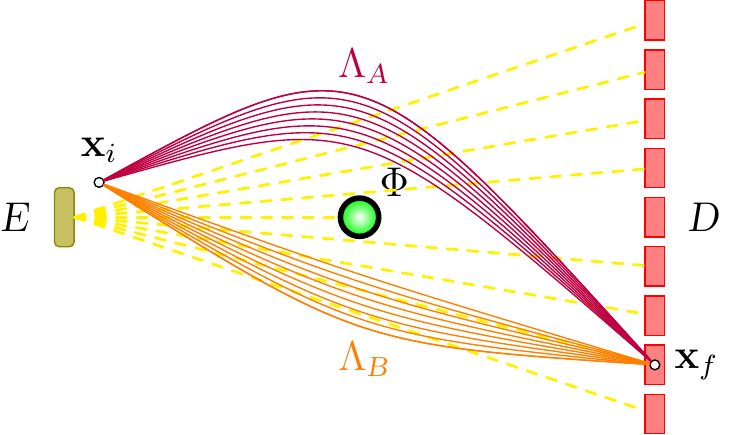}
\caption[Aharonov-Bohm using path-integral approach]{Electrons emanate from the emitter~\textsl{E} to the detectors~$\textsl{D}$, while being excluded from a solenoid containing a flux~$\Phi$. Two distinct families of paths are illustrated: the line integral for those above is~$\Lambda_{\textsl{A}}$, for those below~$\Lambda_{\textsl{B}}$.}\label{fig:abefeyn}
\end{figure}

Now, in~\Figref{fig:abefeyn}, we return to the setup above, this time dispensing with the screen and slits.

Suppose we take a path~$\Vec{x}_{\textsl{A}}(t)$ which passes above the solenoid, without winding around it, like one of those included in~$\Lambda_{\textsl{A}}$ in~\Figref{fig:abefeyn}, and compare it to another path,~$\Vec{x}'_{\textsl{A}}(t)$, which can be smoothly deformed into it, while still having the same end-points. Then if we look at the surface~$\mathcal{S}_{\textsl{A}}$ between them, it does not contain any net flux, meaning that, using a process similar to~\Eqref{eq:primitiveabfluxint}:
\begin{equation}
\Lambda[\Vec{x}'_{\textsl{A}}(t)]-\Lambda[\Vec{x}_{\textsl{A}}(t)]=\varointctrclockwise_{\partial\mathcal{S}_{\textsl{A}}}\!\VecPotExc(\Vec{x})\DotD{\Vec{x}}=
\iint_{\mathcal{S}_{\textsl{A}}}\Vec{B}\DotD{\Vec{a}}=\Flux(\mathcal{S}_{\textsl{A}})=0\text{.}
\end{equation} 
Therefore, we can ascribe a constant value, which we will call~$\Lambda_{\textsl{A}}$, to all paths which can be smoothly deformed into one that goes from above. Similarly, we can do the same for all paths that pass below without winding, overall, that is, which can be smoothly deformed into one of those marked by~$\Lambda_{\textsl{B}}$ in~\Figref{fig:abefeyn}, again, while retaining the end-points~$\Vec{x}_i$ and~$\Vec{x}_f$.

However, if we compare a path from below to a path from above, we will get, as in~\Eqref{eq:primitiveabfluxint}, that:
\begin{equation}
\Lambda[\Vec{x}_{\textsl{B}}(t)]-\Lambda[\Vec{x}_{\textsl{A}}(t)]=\FluxExc\text{.}
\end{equation}

In fact, there is an infinity of classes of paths. Each class is defined by the winding number it has in relation to the paths with~$\Lambda_{\textsl{A}}$, or number of times that it is going around the flux in the center, equivalent to the times it counts the excluded flux in its net flux. If we define~$\Kernel_0^{\textsl{A}}(\Vec{x}_{\textsl{f}}, t_{\textsl{f}}\,;\Vec{x}_{\textsl{i}}, t_{\textsl{i}})$, $\Kernel_0^{\textsl{B}}(\Vec{x}_{\textsl{f}}, t_{\textsl{f}}\,;\Vec{x}_{\textsl{i}}, t_{\textsl{i}})$, \latin{etc.}, to be the path integral:
\begin{equation}
\int\exp\left(\frac{i}{\hbar}\Action_0[\Vec{x}(t)]\right)\PathD{\Vec{x}(t)}
\end{equation}
limited to paths of the class~$\textsl{A}$, $\textsl{B}$, \latin{etc.}, respectively, then we can break up~\Eqref{eq:abpropabstract} into a sum by these classes:
\begin{equation}
\Kernel(\Vec{x}_{\textsl{f}}, t_{\textsl{f}}\,;\Vec{x}_{\textsl{i}}, t_{\textsl{i}})= e^{\frac{iq}{\hbar c}\Lambda_{\textsl{A}}} \Kernel_0^{\textsl{A}}(\Vec{x}_{\textsl{f}}, t_{\textsl{f}}\,;\Vec{x}_{\textsl{i}}, t_{\textsl{i}})+
e^{\frac{iq}{\hbar c}\Lambda_{\textsl{B}}}  \Kernel_0^{\textsl{B}}(\Vec{x}_{\textsl{f}}, t_{\textsl{f}}\,;\Vec{x}_{\textsl{i}}, t_{\textsl{i}})+\dotsb\text{.}
\end{equation}
If we remove the explicit dependence on start and end points for brevity, we get a result similar to that in~\Eqref{eq:abinterferencesch}:
\begin{equation}
\Kernel=e^{\frac{iq}{\hbar c}\Lambda_{\textsl{A}}} \left[\Kernel_0^{\textsl{A}}+e^{\frac{iq}{\hbar c}(\Lambda_{\textsl{B}}-\Lambda_{\textsl{A}})} \Kernel_0^{\textsl{B}})+\dotsb\right]=e^{\frac{iq}{\hbar c}\Lambda_{\textsl{A}}} \left[\Kernel_0^{\textsl{A}}+e^{2\pi i\left(\FluxExc/\Flux_0^q\right)}\Kernel_0^{\textsl{B}})+\dotsb\right]\text{.}
\end{equation}
the magnitude of the probability amplitude therefore oscillates with period~$\Phi^{q}_0$, as we saw from the~\Schroedinger{} picture. 

An advantage of the path integral approach is that it is easier to get to the crux of the matter. For let us return to the reason that the vector potential~$\VecPot$ cannot be identically zero outside the solenoid. The field is zero, and~$\Vec{B}=\Curl{\Vec{A}}$, so a null vector potential would be consistent with this requirement. Furthermore, Stokes' theorem, relating the line integral of a vector field around a region with the surface integral of the curl of this field through that region would suggest that a zero potential will be valid wherever this theorem would apply, as the line integral should be zero anyway. However, when we take a path around the solenoid, regardless of whether the electron passes through it, the electromagnetic field does exist within it, so that a use of Stokes' theorem results in the line integral equaling the total flux through the solenoid, times the number of turns of the path. Furthermore, this total flux is a parameter that affects the interference pattern of an electron scattered through this system. In effect, the electron is probing a portion of the physical electromagnetic field with which it does not interact directly, simply because this field goes through a ``hole'' in this region, so that without it the region is no longer~ \emph{simply-connected}, a term we shall explain in the next section.

Now, in itself this is merely another piece of evidence for the non-locality of quantum mechanics. Moreover, the closed line integral itself, depending as it does on field values, is already a gauge-invariant quantity in the classical theory as well, so that is not an issue. However, it does raise the question of what happens when the configuration space of a system is \emph{really} multiply-connected. What if we find a space in which there are real, intrinsic holes, where absolutely nothing, neither fields nor matter, exists? Then there seems to be an ambiguity in the definition of the problem. If we take the physical fields and extend them to the hole, then depending on what flux this creates in the now ``filled-out'' region, we could have different constraints on the behavior outside. Initial conditions and fields in the configuration space are no longer enough information; we must add a new piece of information, in this case, the line integral of the vector potential around each ``hole''. But is that always the case? What if the space cannot be described as a plane with holes in it? For example, the topological group of rotations in three dimensions,~$\SpOr(3)$, is doubly-connected; that is, if you take a path once around an axis, Stokes' theorem does not apply, but compounding this path with itself leads to one in which this can again be done. What is needed is a language which generalizes the notion of ``holes'', and then an approach to quantum mechanics which will allow us to quantize systems with such generalized spaces. That language is \emph{homotopy theory}, and these spaces are multiply-connected configuration spaces.
\section{Homotopy and the Fundamental Groupoid}\label{sec:homotopy}
In the previous section, there was a property of paths in the configuration space which made it possible to conclude that the line integral of the potential around the them was zero. It seems to have had to do with the fact that the surface that they captured between them was completely within the space, and we knew we could apply Stokes' theorem to show that the properties of the curl inside that surface are what matters. Another way of putting it is that these paths could be gradually deformed one into the other, while retaining the same endpoints. This is not the case for the paths on opposite sides of the solenoid. This is a special case of a more general connection between paths which is called homotopy~(see~Refs.~\cite{Spanier1981,SingerThorpe1967}). Here we will give a short review of the subject, while focusing on an uncommonly used algebraic structure, the fundamental groupoid, to be defined below~(see~Ref.~\cite{Higgins1971}).

\begin{figure}[ht!]
\centering
\includegraphics{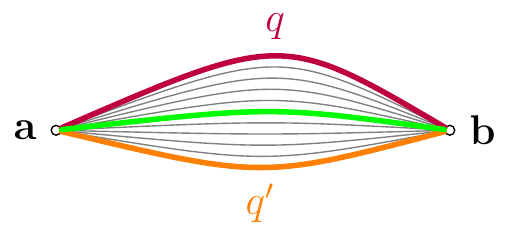}
\caption[Homotopic paths]{Homotopic paths from $\Vec{a}$ to $\Vec{b}$: $[q]=[q']$.}\label{fig:homotab}
\end{figure}

We start with a space~$X$ which has some topology which we will keep implicit. Let us single out two points in that space,~$\Vec{a}$ and~$\Vec{b}$, which may be identical. A~\emph{path from}~$\Vec{a}$ \emph{to}~$\Vec{b}$ is a smooth function from the unit interval to the space:
\begin{equation}
q:[0,1]\to X
\end{equation}
such that~$q(0)=\Vec{a}$ and~$q(1)=\Vec{b}$. We then write:
\begin{equation}
\Source(q)\triangleq q(0),\quad\Target(q)\triangleq q(1),\label{eq:srctgtpathdef}
\end{equation}
the former being the path's~\emph{source}, \emph{from} which it is coming, the latter its~\emph{target}, \emph{to} which it is going. We say that two paths,~$q$ and~$q'$, with the same source and target are~\emph{homotopic} to each other if they can be smoothly deformed one into the other, while retaining the same end-points. Mathematically speaking, this means that there is a smooth function:
\begin{equation}
F:[0,1]\times[0,1]\to X
\end{equation}
such that~$F(t,0)\equiv\Vec{a}$,~$F(t,1)\equiv\Vec{b}$,~$F(0,t)\equiv q(t)$, and~$F(1,t)\equiv q'(t)$. 
This is an equivalence relation, and we can talk about the~\emph{homotopy classes} of paths between~$a$ and~$b$; that of the path~$q$ is~$[q]$, so this property can be written as:
\begin{equation}
[q]=[q']
 \end{equation}
that is, the homotopy classes of~$q$ and~$q'$ are equal~(see~\Figref{fig:homotab}).

\begin{figure}[ht!]
\centering
\includegraphics{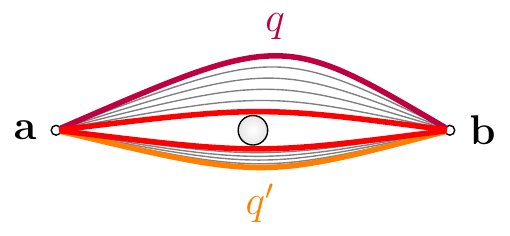}
\caption[Non-homotopic paths]{Non-homotopic paths from $\Vec{a}$ to $\Vec{b}$: $[q]\neq[q']$.}\label{fig:nhomot}
\end{figure}

When the two paths are not homotopic, we instead write:
\begin{equation}
[q]\neq[q']\text{,}
 \end{equation}
that is, the homotopy classes of the two paths are different~(see \Figref{fig:nhomot}).

For a homotopy class~$[q]$, we extend the definitions for~$\Source$ and~$\Target$ in~\Eqref{eq:srctgtpathdef} to:
\begin{equation}
\Source([q])\triangleq\Source(q),\quad\Target([q])\triangleq\Target(q);
\end{equation}
these are well-defined because paths in the same homotopy class share their source and target.

For convenience, if~$\Vec{x}$ is any point in X, then~$[\Vec{x}]$ is the homotopy class of the constant path at~$\Vec{x}$.

\begin{figure}[ht!]
\centering
\includegraphics{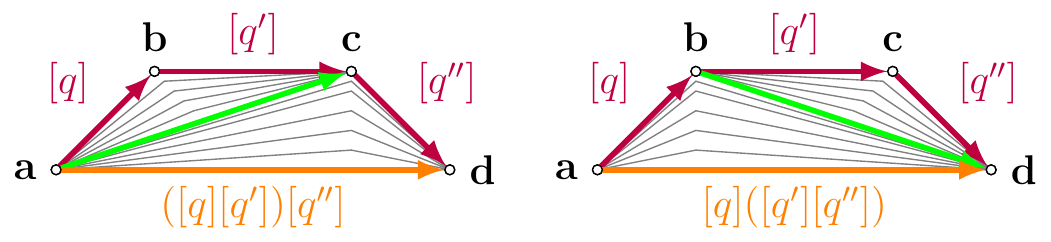}
\caption[Associativity of concatenation]{Concatenation. Illustrating associativity:~$([q][q'])[q'']=[q]([q'][q''])$.}\label{fig:homotconcat}
\end{figure}

Given two paths,~$q$ and~$p$, such that~$\Target(q)=\Source(p)$, their \emph{concatenation}~$qp$ is defined as:
\begin{equation}
(qp)(t)=\begin{cases}q(2t),&0\le t\le \frac{1}{2}\\
p(2t-1),&\frac{1}{2}<t\le1
\end{cases}
\end{equation}
with~$[qp]=[q][p]$; this operation can be shown to be associative~(see~\Figref{fig:homotconcat}).
The homotopy classes of a space~$X$ with concatenation form an algebraic structure, called its~\emph{fundamental groupoid}, and written as~$\Pi(X,X)$. Let us go on a short detour to explain this structure, as it is not commonly known.

The shortest definition of a groupoid is that it is a category where every morphism is invertible. In our case, the objects of the category would be the points of the space~$X$, and the morphisms between two points are the homotopy classes of paths between them.

In more group-theoretic terms, a groupoid is a collection~$\{\alpha\}$ on which a partial operation is defined, usually designated by a product, as are a \emph{source} function~$\Source(\bullet)$ and a \emph{target} function~$\Target(\bullet)$ projecting to another space~$X$, such that:
\begin{enumerate}
\item $\alpha\beta$ is defined if and only if~$\Source(\beta)=\Target(\alpha)$.
\item The operation is \emph{associative}: whenever~$(\alpha\beta)\gamma$ is defined, then so is~$\alpha(\beta\gamma)$, and they are equal.
\item For each~$\alpha$ there is a \emph{reverse}~$\alpha^{-1}$, such that~$\alpha\alpha^{-1}=\Ident_{\Source(\alpha)}$ and~$\alpha^{-1}\alpha=\Ident_{\Target(\alpha)}$, where~$\Ident_{\Vec{x}}$ is an element called~\emph{the identity at}~$\Vec{x}$, so that~$\Source(\Ident_{\Vec{x}})=\Target(\Ident_{\Vec{x}})=\Vec{x}$, and~$\Ident_{\Source(\beta)}\beta=\beta\Ident_{\Target(\beta)}=\beta$.

Normally we would call this the~\emph{inverse}, but inversion is also a symmetry of some very important spaces, one example we of which we shall see in~\SecRef{sec:groupoidproof}. The term \emph{reverse} has the advantage of being suitable for a discussion of paths.
\end{enumerate}
For a more thorough discussion, see~\citet{Higgins1971}. A proof of the necessity of~$\Source$ and~$\Target$ is in the \hyperref[ch:grpdappdx]{Appendix}.

\begin{figure}[ht!]
\centering
\includegraphics{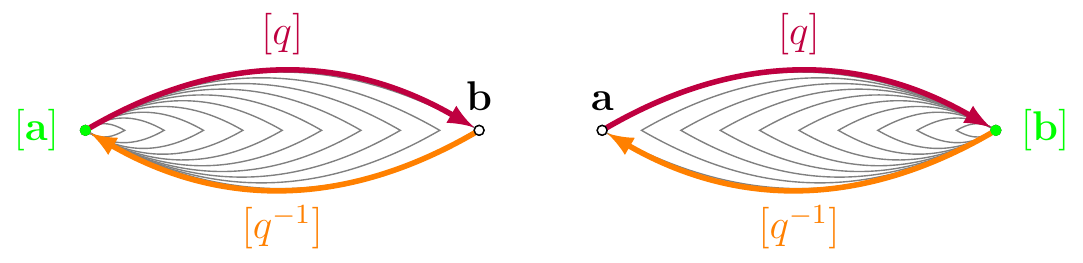}
\caption[Path reversal]{Reverse. We see that~$[\Vec{a}]=[q][q^{-1}]$ and~$[\Vec{b}]=[q^{-1}][q]$.}\label{fig:homotreverse}
\end{figure}

We have already shown that the operation is well-defined and associative. To complete the list of requirements for a groupoid, we note that aside from concatenation, we also have \emph{reversal}: for each path~$q$ we may define a reverse,~$q^{-1}$, as follows:
\begin{equation}
(q^{-1})(t)=q(1-t)\text{,}
\end{equation}
and the two satisfy:
\begin{equation}
[qq^{-1}]=[\Source(q)]\quad;\quad[q^{-1}q]=[\Target(q)]\text{,}
\end{equation}
so we may define~$[q]^{-1}\triangleq[q^{-1}]$, in the sense that~$[\Vec{x}]$ acts as~$\Ident_{\Vec{x}}$~(see~\Figref{fig:homotreverse}).

\begin{figure}[ht!]
\centering
\includegraphics{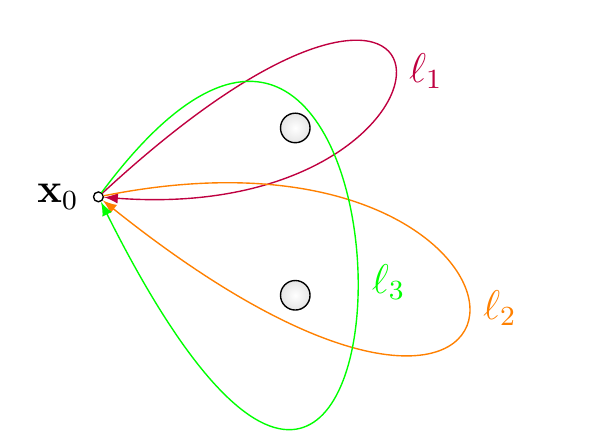}
\caption[Loop concatenation]{Loops through~$\Vec{x}_0$. $[\ell_3]=[\ell_1][\ell_2]$}\label{fig:loops}
\end{figure}

Let us now choose a point~$\Vec{x}_0$ in~$X$. The fundamental \emph{group} based at that point is:
\begin{equation}
\Pi(X,\Vec{x}_0)=\left\{[q]\middle| \Source(q)=\Target(q)=\Vec{x}_0\right\}\text{;}
\end{equation}
that is, the classes of all the paths which start and finish at that point. They will also be called loops~(see~\Figref{fig:loops}).
Generally,~for a subset~$A\subseteq X$, $\Pi(X,A)$ is the subgroupoid of all~$[q]$ where~$\Source([q]),\Target([q])\in A$. Another special case used in~Ref.~\cite{LaidlawDeWitt1971} is~$\Pi(X,\Vec{a},\Vec{b})$, the collection of homotopy classes~$[q]$ with~$\Source([q])=\Vec{a}$ and~$\Target([q])=\Vec{b}$.

Fundamental groups at different points are not entirely independent, as long as there is a path connecting them. This is always the case when the space is~\emph{path-connected}, which is a postulate common to~Ref.~\cite{LaidlawDeWitt1971} and to their successors. Indeed, let~$\Vec{x}_1$ and~$\Vec{x}_2$ be arbitrary points in~$X$. Let~$l$ be a loop based at~$\Vec{x}_1$, that is,~$\Source(l)=\Target(l)=\Vec{x}_1$, and let~$q$ be a path such that~$\Source(q)=\Vec{x}_1$ and~$\Target(q)=\Vec{x}_2$. Then~$q^{-1 }l q$ is a loop based at~$\Vec{x}_2$, and~$[q]^{-1}[l][q]$ is an element of the fundamental group there, so~$[q]$ induces a~\emph{homomorphism} between the fundamental groups, since if~$l$ and~$k$ are loops based at~$\Vec{x}_1$, then:
\begin{equation}
[q^{-1}lkq]=[q^{-1}][l][k][q]=[q^{-1}][l][q][q^{-1}][k][q]=[q^{-1}lq][q^{-1}kq]\text{;}
\end{equation}
similarly, the reverse homotopy class~$[q]^{-1}$ induces a homomorphism in the opposite direction, and:
\begin{equation}
([q]^{-1})^{-1}([q^{-1}][l][q])[q]^{-1}=[qq^{-1}lqq^{-1}]=[l];
\end{equation}
therefore, the homotopy class~$[q]$ induces an \emph{isomorphism} between the two fundamental groups. For any two homotopy classes,~$[q]$ and~$[p]$, such that~$\Source([q])=\Source([p])=\Vec{x}_1$ and~$\Target([q])=\Target([p])=\Vec{x}_2$, the respective isomorphisms induced by them are related by:
\begin{equation}
[p^{-1} l p]=[p^{-1}q q^{-1} l q q^{-1} p]=[p^{-1}q][q^{-1}l q][p^{-1}q]^{-1}\text{,}\label{eq:inneriso}
\end{equation}
which is an inner isomorphism, that is, of the form~$\alpha\to\beta\alpha\beta^{-1}$, since~$[p^{-1}q]\in\Pi(X,\Vec{x}_2)$. We will see in~\ChapRef{ch:groupoidresults} that this leads to a simplification for abelian groups and for one-dimensional representations.

Finally, we can explain what we mean by simply-connected, multiply-connected, \latin{etc.} Always speaking of connected, path-connected spaces, a~\emph{simply-connected} space is one whose fundamental group is trivial, that is, every closed loop in it can be contracted to a point, while a~\emph{multiply-connected} space is one whose fundamental group is non-trivial. So the two-dimensional plane without~``holes'' is simply-connected. A space with a hole, such as the space to which electrons are restricted in the Aharonov-Bohm setup, is multiply-connected. In fact, it is~\emph{infinitely-connected}, in the sense used in~\latin{e.g.}~Ref.~\cite{Schulman1968}: its fundamental group is isomorphic to an infinite group, in this case the infinite cyclic group~$\Z$ of integers.

Note that the Feynman path integral approach is very well-suited for exploring these topological properties of spaces, as paths, rather than points, are its building blocks.
\chapter{Quantization in Multiply-Connected Spaces}\label{ch:groupoidresults}
\section{Motivation and Relation to Homotopy}\label{sec:lmdwexmotivation}
Before we present the result of this chapter, let us provide a review of some work with which it contrasts, and from which it extracts some vital tools. 

The path integral approach to multiply-connected spaces was originally introduced in~\citet{Schulman1967}, and further disseminated through~\citet{Schulman1968}. Those works developed a justification for discrete spin under the path integral formalism. The configuration space which is quantized for that purpose, that of~$\SpOr(3)$ taken as a topological group, is not simply-connected, and that is where the discussion of path integrals in multiply-connected spaces becomes necessary. The standard definition of the path-integral becomes ambiguous under these circumstances: in a singly-connected space, all paths between two given points are in the same homotopy class, so they are smoothly deformable into each other, while retaining the same endpoints, and thus their contributions also go smoothly into each other. Therefore, aside from an overall phase ambiguity, the sum over paths making up the transition amplitude is defined as:
\begin{equation}
\Kernel({\Vec{b}}, t_{\Vec{b}}\,; {\Vec{a}}, t_{\Vec{a}})=\int\PathD{\Vec{x}(t)}\exp\left(i\Action\{\Vec{x}(t)\}/\hbar\right)\label{eq:pathintdef}\text{,}
\end{equation}
where~$\Vec{x}(t)$ runs over all smooth paths between~$\Vec{a}$ and~$\Vec{b}$.

If there is more than one homotopy class, then the situation is different: paths in one class can no longer be smoothly deformed into those in another, so that neither do the phases, and we in fact have to split the integral into a sum over integrals of these components:
\begin{equation}
\Kernel({\Vec{b}}, t_{\Vec{b}}\,; {\Vec{a}}, t_{\Vec{a}})=\smashoperator[r]{\sum\limits_{[q]\in\Pi(X,\Vec{a},\Vec{b})}}\widetilde{\Weight}([\Vec{x}(t)])\widetilde{\Kernel}^{[q]}({\Vec{b}}, t_{\Vec{b}}\,; {\Vec{a}}, t_{\Vec{a}})\text{,}\label{eq:amphtpcls}
\end{equation}
with the \emph{partial amplitudes} being defined as the path integral limited to the homotopy classes:
\begin{equation}
\widetilde{\Kernel}^{[q]}({\Vec{b}}, t_{\Vec{b}}\,; {\Vec{a}}, t_{\Vec{a}})\triangleq\int\PathD{\Vec{x}(t): [\Vec{x}(t)]=[q]}\exp\left(i\Action\{\Vec{x}(t)\}/\hbar\right)\text{,}\label{eq:partialintdef}
\end{equation}
and the~\emph{weights}~$\widetilde{\Weight}([q])$ yet to be determined. Note that, unlike in~\SecRef{sec:homotopy}, the parameterizations of the paths matter, which we indicate by writing them as~$\Vec{x}(t)$. The homotopy class of the path~$[\Vec{x}(t)]$ is then that of the path reparamaterized to the unit interval:
\begin{equation}
q(t)=\Vec{x}\left((t_{\Vec{b}}-t_{\Vec{a}})t+t_{\Vec{a}}\right)\text{;}
\end{equation}
we will use this notation in~\SecRef{sec:groupoidproof}, as well.

\begin{figure}[ht!]
\centering
\includegraphics{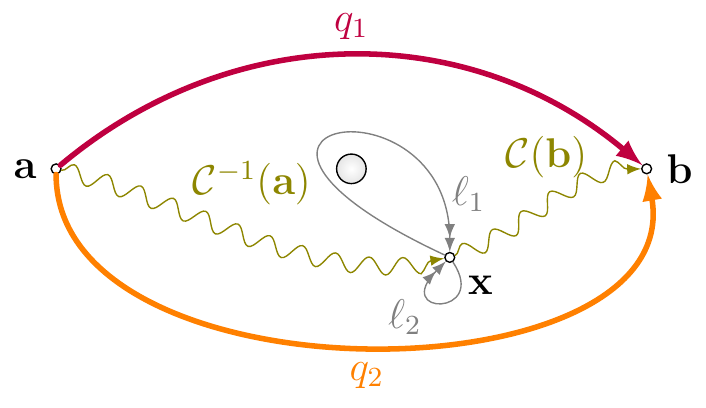}
\caption[Mesh transforms loops into paths]{Loops become paths through a mesh: $[q_i]=f_{\Vec{ab}}([\ell_i])=[\Mesh^{-1}(\Vec{a})][\ell_i][\Mesh(\Vec{b})]$.}\label{fig:mesh}
\end{figure}

In our adapted notation,~\Eqref{eq:amphtpcls} is the result in~Ref.~\cite{Schulman1968} from which~Refs.~\cite{LaidlawDeWitt1971,Laidlaw1971} start their analysis. However, they do not parameterize the partial amplitudes with homotopy classes directly, but instead through members of the fundamental group at some arbitrary point~$\Vec{x}_0$; hence the use of the tilde. Now, we have seen in~\SecRef{sec:homotopy} that the fundamental groups at different points for a path-connected space are related; this paper goes a step further, capturing \emph{all} homotopy classes, that is, members of the fundamental groupoid, by just the fundamental group at a single point,~$\Vec{x}_0$. This is done using a~\emph{mesh}, a choice of path from~$\Vec{x}_0$ to each point~$\Vec{a}$: 
\begin{align}
&\Mesh:X\to\left\{q\middle|\,q:[0,1]\to X\right\}\label{eq:meshdef}\\
&\Mesh(\Vec{a})(0)=\Vec{x}, \Mesh(\Vec{a})(1)=\Vec{a}\text{.}\notag
\end{align}
This can be done because $X$ is path-connected. Once a mesh has been chosen, a natural one-to-one correspondence between elements of the fundamental group and those of the fundamental groupoid, given the end-points, is given by:
\begin{align}
&f_{\Vec{ab}}:\Pi(X,\Vec{x})\to \Pi(X, \Vec{a}, \Vec{b})\\
&f_{\Vec{ab}}(\alpha)=[\Mesh^{-1}(\Vec{a})]\alpha[\Mesh(\Vec{b})]\text{,}
\end{align}
where $\Mesh^{-1}(\Vec{a})$ is the reverse of the path~$\Mesh(\Vec{a})$ (See~\Figref{fig:mesh}).

This works for any $\Vec{a}$ and $\Vec{b}$, allowing a consistent definition of the partial amplitudes, as well as their coefficients in the expression for the total amplitude, indexed by the fundamental group:
\begin{align}
\Kernel^{\alpha}(\Vec{b}, t_{\Vec{b}}\,; \Vec{a}, t_{\Vec{a}})&\triangleq\widetilde{\Kernel}^{f_{\Vec{ab}}(\alpha)}(\Vec{b}, t_{\Vec{b}}\,; \Vec{a}, t_{\Vec{a}})\\
\Weight(\alpha)&\triangleq\widetilde{\Weight}(f_{\Vec{ab}}(\alpha))
\intertext{providing them with an amended version of~\Eqref{eq:amphtpcls}:}
\Kernel({\Vec{b}}, t_{\Vec{b}}\,; {\Vec{a}}, t_{\Vec{a}})&=\smashoperator[r]{\sum\limits_{\alpha\in\Pi(X,\Vec{x}_0)}}\Weight(\alpha)\Kernel^{\alpha}({\Vec{b}}, t_{\Vec{b}}\,; {\Vec{a}}, t_{\Vec{a}})\text{,}\label{eq:amphtfgrp}
\end{align}
as their new starting point. For them the partial amplitudes~$\Kernel^{\alpha}$ are black boxes: they do not use the formulas~\Eqref{eq:pathintdef} or~\Eqref{eq:partialintdef}, so  a significant portion of the paper is dedicated to investigating their properties as a stepping-stone to the more relevant results constraining~$\Weight(\alpha)$. We will see in~\SecRef{sec:groupoidproof} that a direct investigation of the integrand of~\Eqref{eq:pathintdef} will save us much of this trouble.

Returning to their method, we note that $f$ respects the groupoid operation:
\begin{equation}
f_{\Vec{ab}}(\alpha)f_{\Vec{bc}}(\beta)=f_{\Vec{ac}}(\alpha\beta)\text{;}
\end{equation}
we find it helpful to see it as a parametrized local inverse to a homomorphism from the fundamental groupoid to the fundamental group at~$\Vec{x}_0$:
\begin{align}
&g:\Pi(X,X)\to\Pi(X,\Vec{x}_0)\\
&g([q])\triangleq[\Mesh(\Source([q]))][q][\Mesh^{-1}(\Target([q]))]\text{,}
\end{align}
so~$g([q][q'])=g([q])g([q'])$ and:
\begin{equation}
g(f_{\Vec{ab}}(\alpha))=\alpha\quad;\quad f_{\Source([q])\!\Target([q])}(g([q]))=[q]\text{.}
\end{equation}

The main result of the work in~Refs.~\cite{LaidlawDeWitt1971,Laidlaw1971} is a proof that the weights~$\Weight(\alpha)$ are a fundamental group representation. They then show that the choice of mesh does not change the physical predictions, but rather only the transition amplitudes by an overall phase, so that it is the choice of group representation which embodies the additional, physically meaningful topological degree of freedom.

However, this result has a limitation, which we have had to overcome, in order to apply these methods to the problem of distinguishable anyons. These anyons would be required to have a consistent topological phase for every simple, counter-clockwise exchange. Since such an exchange would not be a closed loop, methods which only provide topological phases  representing members of the fundamental group will not be suitable.

To see this explicitly, let~$\Rep(\bullet)$ be the fundamental group representation, and let there be a mesh~$\Mesh$ and base-point~$\Vec{x}_0$ chosen as well, so that~$g$ is well defined. Then:
\begin{equation}
\widetilde{\Weight}([q])=\Rep(g([q]))\text{,}\label{eq:ldwgroupoidrep}
\end{equation}
that is, a fundamental groupoid representation has implicitly been chosen as well.

\begin{figure}[ht!]
\centering
\includegraphics{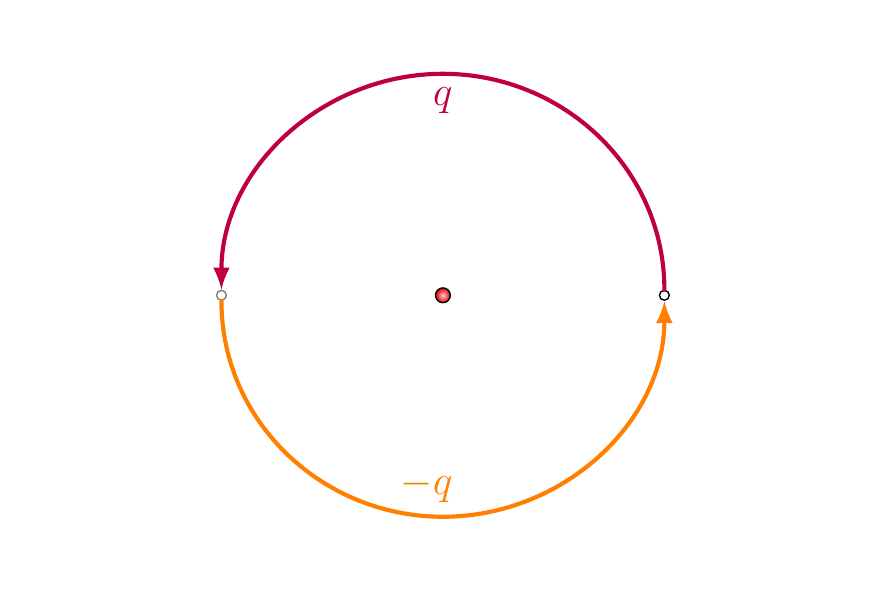}

\caption[Counter-clockwise circle broken into two half-circles]{A counter-clockwise circle is broken up into a half-circle,~$q$, and its inverse,~$-q$.}\label{fig:circletwain}
\end{figure}

We will now show that this choice does not generally allow the incorporation of symmetries of the space. We will do so for the important special case of the punctured plane,~$X=\R^2\setminus\{\Vec{0}\}$, an example we will return to in~\SecRef{sec:groupoidproof}, and will use in the next chapter. Let us define an inversion, ``minus'' operation, with~$-(a,b)\triangleq(-a,-b)$, where~$(a,b)$ are Cartesian coordinates. We break a circular loop counter-clockwise around the origin into two pieces,~$[q]$ and~$[-q]$, which are related by a simple inversion symmetry (not to be confused with the reversed path,~$q^{-1}$, defined in~\SecRef{sec:homotopy}):
\begin{equation}
(-q)(t)=-(q(t))\text{,}
\end{equation}
as seen in~\Figref{fig:circletwain}.

We then get:
\begin{equation}
\widetilde{\Weight}([q(-q)])=\widetilde{\Weight}([q])\widetilde{\Weight}([-q])\text{.}
\end{equation}
Now, $[q(-q)]$ generates the fundamental group at~$\Vec{x}_0$. We can then treat it as isomorphic to~$\Z$ with addition, $[q(-q)]$ then corresponding to~$1$. Let us write:~$\widetilde{\Weight}([q(-q)])=z$. This is the generator of the image of~$\Z$ under~$\Weight(\bullet)$. In order for this representation to respect inversion symmetry, we require:
\begin{align}
\widetilde{\Weight}([-q])&=\widetilde{\Weight}([q])=w\text{,}
\intertext{so:}
z&=w^2\text{.}
\intertext{However, since~$z$ generates the group, we must have~$w=z^{k}$, giving}
z&=z^{2k}\text{,}
\intertext{or}
1&=z^{2k-1}\text{.}
\end{align}
Therefore, this is only possible for special choices of~$z$, namely odd roots of unity. If we would instead like to incorporate symmetries of the space for~\emph{any} value of~$z$, we need to explore more degrees of freedom for the groupoid representation, other than re-ordering the group representations through a different choice of mesh.

The next section~(\SecRef{sec:groupoidproof}) will feature one of our novel results. We will start out by proving that the phases are a groupoid representation directly. We will then explore what the additional degrees of freedom are, beyond the choice of a fundamental group at a point. While they create physically indistinguishable solutions, they do allow the incorporation of more of the symmetries of the space at hand. The main simplifying tool in our treatment is that it starts from paths and their phases, rather than from partial propagators. It is bottom-up rather than top-down, and does not treat propagators as black boxes.
\section{A Direct Path to the Fundamental Groupoid}\label{sec:groupoidproof}
The aim of this section is to improve upon the work in~Refs.~\cite{LaidlawDeWitt1971,Laidlaw1971}, reviewed in the previous section, in a way which uses the fundamental~\emph{groupoid}, bearing in mind its most general representations. It will be seen that the result in~Refs.~\cite{LaidlawDeWitt1971,Laidlaw1971} is a special case of ours: we add a generalized gauge freedom. We will also provide an example of a groupoid representation which incorporates the rotational symmetry of the punctured plane,~$\R^2\setminus\{\Vec{0}\}$.

We start by exploring the origins of the topological degrees of freedom. Instead of starting from the top, that is, from the propagator and the quantum formalism, as do~\citet{Schulman1967} and its successors, we begin from the bottom, with Hamilton's Principle in classical physics: the least action principle, or, more accurately, the extremal action principle. 

Given an initial time~$t_1$ and point~$\Vec{x}_1$, and a final time~$t_2$ and point~$\Vec{x}_2$, we wish to find the path~$\Vec{x}_c(t)$ which is extremal (that is, a local minimum, maximum, possibly saddle) for the action functional:
\begin{equation}
\Action\{\Vec{x}(t)\}=\int\limits_{t_1}^{t_2}\Lagrangian(\Vec{x}(t),\dot{\Vec{x}}(t),t)\D{t}\text{.}
\end{equation}
We use $\{\bullet\}$ to denote a functional, rather than the usual~$[\bullet]$, since we have already reserved the latter for homotopy classes. More formally:
\begin{equation}
\delta\Action\{\Vec{x}_c(t)\}=0\text{,}
\end{equation}
with the variation keeping~$t_1,t_2,\Vec{x}_1,\Vec{x}_2$ fixed. Therefore, the same path would be extremal, even if a value depending only on these constants was added:
\begin{align}
\delta\Action\{\Vec{x}(t)\}&=\delta\left(\Action\{\Vec{x}(t)\}+\varsigma(\Vec{x}_2,t_2\,;\Vec{x}_1,t_1)\right)\text{.}
\intertext{We can go even further: extrema are only modified if the action functional is changed in such a way that its value for two paths which can be smoothly deformed into each other is altered; therefore, we can have this parameter also depend on the homotopy class of~$\Vec{x}(t)$, and write}
\delta\Action\{\Vec{x}(t)\}&=\delta\left(\Action\{\Vec{x}(t)\}+\varsigma_{[\Vec{x}(t)]}(\Vec{x}_2,t_2\,;\Vec{x}_1,t_1)\right)\text{,}
\end{align}
and this is independent of the specific Lagrangian, so long as it has the same singularity behavior (if applicable). In fact, we can write
\begin{equation}
\varsigma_{[\Vec{x}(t)]}(\Vec{x}_2,t_2\,;\Vec{x}_1,t_1)=\varsigma_{[\Vec{x}(t)]}(t_2\,;t_1)\text{,}
\end{equation}
as the end-points are implicit through~$\Source([\Vec{x}(t)])=\Vec{x}_1$ and~$\Target([\Vec{x}(t)])=\Vec{x}_2$, where the source and target functions~$\Source$ and~$\Target$ are as defined in~\SecRef{sec:homotopy}.

Finally, let us address time. Earlier literature widely agrees that the topological parameter should be time-independent,
 with~\cite{Dowker1972,DowkerCritchley1977,CasatiGuarneri1979,GerrySingh1979,Horvathy1980a,Horvathy1980b,Horvathy1980c,Isham1984,Wu1984,Dowker1985,SudarshanImboShahImbo1988,Morandi1992} all making similar assumptions. However, returning to the punctured plane, the topological degree of freedom is equivalent to a flux through a solenoid at the origin. That can be slowly changed, leading from one representation of the fundamental group at a point to another. \citet{GaveauNounouSchulman2011} go further to provide a solution for general changes of flux as a function of time. They find that the only strictly topological parameter is the enclosed flux at an arbitrary point in time. Since this result depends explicitly on the particular space being quantized, we expect a general solution to be more involved, and leave it to future work.
We therefore choose to focus on investigating a parameter which only depends on the homotopy class of the path:
\begin{equation}
\varsigma=\varsigma([\Vec{x}(t)])\text{;}\label{eq:topparamtimeind}
\end{equation}

Now we may calculate the appropriate Feynman amplitude for the path:
\begin{equation}
\exp\left(\frac{i}{\hbar}\left(\Action\{\Vec{x}(t)\}+\varsigma([\Vec{x}(t)])\right)\right)=\exp\left(\frac{i}{\hbar}\Action\{\Vec{x}(t)\}\right)\cdot\Weight([\Vec{x}])\text{,}\label{eq:ampwithweight}
\end{equation}
where, to remain consistent with the terms in Refs.~\cite{LaidlawDeWitt1971,Laidlaw1971},~$\Weight([\Vec{x}(t)])$ is called the~\emph{weight} corresponding to the homotopy class~$[\Vec{x}(t)]$, although we remove the ``scare tilde''. In fact, since this total amplitude must obey the Feynman product rule, and since we already know that the first factor does, as it is simply that corresponding to the original action, then these weights also obey the Feynman rules. 

We are no longer interested in the time-dependence of the paths, so we can revert to the lowercase~$[q]$ notation introduced in~\SecRef{sec:homotopy}. Combining concatenation, which gives~$[qp]=[q][p]$, with the product rule of the weights gives us:
\begin{equation}
\Weight([q][p])=\Weight([qp])=\Weight([q])\Weight([p])\text{;}
\end{equation}
meaning that~$\Weight(\bullet)$ is a representation of the fundamental groupoid. Working directly with the groupoid rather than the fundamental group allows us to more generally discuss the situation for arbitrary paths, instead of being forced to work strictly through loops, as in~Refs.~\cite{LaidlawDeWitt1971,Laidlaw1971}. As a subgroupoid of the fundamental groupoid, we have seen that fundamental groups in two points~$\Vec{x}_1$ and~$\Vec{x}_2$ are isomorphic, with isomorphisms induced by paths, and the relation between different isomorphisms being
\begin{equation}
[p^{-1} l p]=[p^{-1}q][q^{-1}l q][p^{-1}q]^{-1}\text{;}\tag{\ref{eq:inneriso}}
\end{equation}
this is an inner isomorphism, as we stated in~\SecRef{sec:homotopy}. If the group is abelian, or if we limit ourselves to a one-dimensional representation~$\Weight$, all inner isomorphisms collapse into the identity. In the former case:
\begin{align}
[p^{-1}q][q^{-1}l q][p^{-1}q]^{-1}&=[p^{-1}q][p^{-1}q]^{-1}[q^{-1}l q]=[q^{-1}l q]
\intertext{while in the latter case:}
\Weight([p^{-1}q][q^{-1}l q][p^{-1}q]^{-1})&=\Weight([p^{-1}q])\Weight([q^{-1}l q])\Weight([p^{-1}q]^{-1})\notag\\
&=\Weight([p^{-1}q]^{-1})\Weight([p^{-1}q])\Weight([q^{-1}l q])=\Weight([q^{-1}l q])\text{,}
\intertext{so all isomorphisms between~$\Vec{x}_1$ and~$\Vec{x}_2$ induced by paths are identical. Moreover:}
\Weight([q^{-1} l q])&=\Weight([q]^{-1} [l] [q])=\Weight([q]^{-1})\Weight([l])\Weight([q])\notag\\
&=\Weight([q])\Weight([q]^{-1})\Weight([l])=\Weight([q][q]^{-1}[l])\notag\\
&=\Weight([\Vec{x}_1][l])=\Weight([l])\text{.}
\end{align}

This means that once a fundamental group representation is chosen at one point, it has been set for all other points. However, this is not sufficient for describing the groupoid representation, since we have only set the representations for loops, while we have to assign values to~$\Weight([q])$ when~$\Source([q])\neq\Target([q])$. As we noted in~\Eqref{eq:ldwgroupoidrep}, the framework in~Ref.~\cite{LaidlawDeWitt1971} provides one way of assigning these weights. We intend to generalize that framework, using the fact that we already have the groupoid representation structure, in order to catalog all possible representations, allowing us to choose one that implements the symmetry of the space. To that end, we must introduce the notion of compatibility of representations. By saying that two representations~$\Weight$ and~$\Weight'$ of the fundamental groupoid are~\emph{compatible} we mean that, for any given~$\Vec{a}$ and~$\Vec{b}$, there is phase~$e^{i \delta(\Vec{a},\Vec{b})}$ such that:
\begin{equation}
\Weight'([q])=e^{i \delta(\Vec{a},\Vec{b})}\Weight([q]),\label{eq:compatscalreps}
\end{equation}
for all~$[q]$s such that~$\Source([q])=\Vec{a}$ and~$\Target([q])=\Vec{b}$; that is, $e^{i \delta(\Vec{a},\Vec{b})}$ only depends on the end-points.

When two representations are compatible, each one induces a propagator which results in the same physical predictions as the other, as an overall phase is removed by taking the absolute value:
\begin{equation}
\abs{\sum\limits_{[q]}\Weight'([q])\Kernel^{[q]}}=\abs{\sum\limits_{[q]}e^{i \delta}\Weight([q])\Kernel^{[q]}}=\abs{e^{i \delta}\sum\limits_{[q]}\Weight([q])\Kernel^{[q]}}=\abs{\sum\limits_{[q]}\Weight([q])\Kernel^{[q]}}
\end{equation}

Let us now explore exactly what kind of leeway we have with a fundamental groupoid representation once we have chosen a fundamental group representation,~$\Rep(\bullet)$. As we saw earlier, only paths between distinct points~$\Vec{a}$ and~$\Vec{b}$ can have any ambiguity to them. Let us choose a homotopy class~$[p_0]$ from~$\Vec{a}$ to~$\Vec{b}$. Then any path class~$[q]$ between these two points can be related to a loop around~$\Vec{a}$ through~$[q]\mapsto[qp_0^{-1}]$. The weight of this object is set by the fundamental group representation:
\begin{equation}
\Rep([qp_0^{-1}])=\Weight([qp_0^{-1}])=\Weight([q])\Weight([p_0]^{-1})\text{,}
\end{equation}
or:
\begin{equation}
\Weight([q])=\Rep([qp_0^{-1}])\Weight([p_0])\text{;}
\end{equation}
note that this is self-consistent, as~$\Rep([p_0p_0^{-1}])=\Rep(\Ident_{\Vec{a}})=1$.

Na{\"\i}vely it would seem that, for each pair of points~$\Vec{a}$ and $\Vec{b}$, we are free to choose an arbitrary class~$[p_{\Vec{a}\to\Vec{b}}]$, followed by a value~$\Weight([p_{\Vec{a}\to\Vec{b}}])$. 

This is not the case, however. First, this would lead to over-counting: as we have already chosen a fundamental group representation, a choice of~$\Weight([p_{\Vec{a}\to\Vec{b}}])$ is equivalent, for any other class~$[p'_{\Vec{a}\to\Vec{b}}]$ between these two points, to a choice of~$\Weight([p'_{\Vec{a}\to\Vec{b}}])$:
\begin{equation}
\Weight([p'_{\Vec{a}\to\Vec{b}}])=\Rep([p'_{\Vec{a}\to\Vec{b}}][p_{\Vec{a}\to\Vec{b}}]^{-1})\Weight([p_{\Vec{a}\to\Vec{b}}])\text{.}
\end{equation}
Furthermore, the groupoid structure imposes severe limitations on our freedom. If we made the following choices: 
\begin{equation}
[p_{\Vec{a}\to\Vec{b}}],\Weight([p_{\Vec{a}\to\Vec{b}}]);\quad[p_{\Vec{b}\to\Vec{c}}],\Weight([p_{\Vec{b}\to\Vec{c}}]);\quad[p_{\Vec{a}\to\Vec{c}}],\Weight([p_{\Vec{a}\to\Vec{c}}]);
\end{equation}
then they would be subject to the constraint:
\begin{equation}
\Weight([p_{\Vec{a}\to\Vec{b}}])\Weight([p_{\Vec{b}\to\Vec{c}}])=\Weight([p_{\Vec{a}\to\Vec{b}}][p_{\Vec{b}\to\Vec{c}}])=\Weight([p_{\Vec{a}\to\Vec{c}}])\Rep([p_{\Vec{a}\to\Vec{c}}]^{-1}[p_{\Vec{a}\to\Vec{b}}][p_{\Vec{b}\to\Vec{c}}])\text{,}
\end{equation}
so even with proper counting, only two of the three can be freely chosen. 

In short, we have a big set of degrees of freedom, but they are subject to over-counting and interdependencies. Fortunately, Ref.~\cite{LaidlawDeWitt1971} furnishes us with a way of organizing them: by using the base-point and mesh formalism we reviewed in~\SecRef{sec:lmdwexmotivation}, but in a more generalized fashion. 

If we single out any one point~$\Vec{x}_0$, and create a mesh from it to any other point,~$\Mesh(\Vec{x})$, then we can write, for any homotopy class~[q] between~$\Vec{a}$ to~$\Vec{b}$:
\begin{align}
\Weight([q])&=\Weight([\Mesh(\Vec{a})]^{-1}[\Mesh(\Vec{a})][q][\Mesh^{-1}(\Vec{b})][\Mesh(\Vec{b})])\notag\\
&=\Weight([\Mesh(\Vec{a})]^{-1})\Weight([\Mesh(\Vec{a})q\Mesh^{-1}(\Vec{b})])\Weight([\Mesh(\Vec{b})])\notag\\
&=\Weight([\Mesh(\Vec{a})]^{-1})\Rep([\Mesh(\Vec{a})q\Mesh^{-1}(\Vec{b})])\Weight([\Mesh(\Vec{b})])\text{;}\label{eq:oid2groupandmesh}
\intertext{we retrieve~\Eqref{eq:ldwgroupoidrep} when we choose~$\Weight([\Mesh(\Vec{x})])\equiv 1$. Generally, if we pick a different base-point,~$\Vec{x}'_0$,
and any mesh~$\Mesh'(\Vec{x})$, we have:}
\Weight([\Mesh'(\Vec{x})])&=\Weight([\Mesh(\Vec{x}'_0)]^{-1})\Rep([\Mesh(\Vec{x}'_0)\Mesh'(\Vec{x})\Mesh^{-1}(\Vec{x})])\Weight([\Mesh(\Vec{x})])\text{,}
\end{align}
so once this has been chosen for one base-point and mesh, we have set it up for all base-points and meshes. A special case of this is the one discussed in~Ref.~\cite{LaidlawDeWitt1971}, a change of mesh with the same base-point. Using our tools, we would also have to change the phases of the mesh so that~$\Weight([\Mesh(\Vec{x})])\equiv1$, which leads to a physically equivalent situation. 

Conversely, if we choose a mesh and phases for each path in the mesh (as well as the reciprocal for the reversed path:~$\Weight([\Mesh(\Vec{x})]^{-1})=\Weight^{-1}([\Mesh(\Vec{x})])$), and a representation of the fundamental group at~$\Vec{x}_0$,~\Eqref{eq:oid2groupandmesh} defines a groupoid representation; let~$q$ and~$p$ be such that~$\Source(q)=\Vec{a}$,~$\Source(p)=\Target(q)=\Vec{b}$,~$\Target(p)=\Vec{c}$; then:
\begin{align}
\Weight&([q])\Weight([p])\notag\\
&=\Weight([\Mesh(\Vec{a})]^{-1})\Rep([\Mesh(\Vec{a})q\Mesh^{-1}(\Vec{b})])\Weight([\Mesh(\Vec{b})])\Weight([\Mesh(\Vec{b})]^{-1})\Rep([\Mesh(\Vec{b})p\Mesh^{-1}(\Vec{c})])\Weight([\Mesh(\Vec{c})])\notag\\
&=\Weight([\Mesh(\Vec{a})]^{-1})\Rep([\Mesh(\Vec{a})q\Mesh^{-1}(\Vec{b})])\Weight([\Mesh(\Vec{b})])\Weight^{-1}([\Mesh(\Vec{b})])\Rep([\Mesh(\Vec{b})p\Mesh^{-1}(\Vec{c})])\Weight([\Mesh(\Vec{c})])\notag\\
&=\Weight([\Mesh(\Vec{a})]^{-1})\Rep([\Mesh(\Vec{a})q\Mesh^{-1}(\Vec{b})])\Rep([\Mesh(\Vec{b})p\Mesh^{-1}(\Vec{c})])\Weight([\Mesh(\Vec{c})])\notag\\
&=\Weight([\Mesh(\Vec{a})]^{-1})\Rep([\Mesh(\Vec{a})q\Mesh^{-1}(\Vec{b})][\Mesh(\Vec{b})p\Mesh^{-1}(\Vec{c})])\Weight([\Mesh(\Vec{c})])\notag\\
&=\Weight([\Mesh(\Vec{a})]^{-1})\Rep([\Mesh(\Vec{a})qp\Mesh^{-1}(\Vec{c})])\Weight([\Mesh(\Vec{c})])\notag\\
&=\Weight([qp])\text{.}
\end{align}

In fact, this is a generalization of the notion of the choice of gauge in quantum mechanics to multiply-connected spaces. Had this been a simply-connected space, this would be the same as saying we had some space-dependent phase ambiguity (as then all paths between two given points would be of the same homotopy class). More generally, there are additional degrees of freedom having to do with representations of the fundamental group at any point, which correspond to Aharonov-Bohm fluxes for the special case of spaces which can be represented as~$\R^2$ with holes.

Now, we have seen that every choice of phases for the mesh provides us with a valid groupoid representation. We have also seen that once these phases are set for one choice of base-point and mesh, the phases for any other choice are also set. We therefore proceed to prove that, given the fundamental group representation, a base-point, and a mesh, the choice of phases for the mesh has no  physical effect. Let us have two sets of mesh-weights,~$\Weight(\bullet)$ and~$\Weight'(\bullet)$, and calculate:
\begin{align}
\Weight([q])&=\Weight([\Mesh(\Vec{a})]^{-1})\Rep([\Mesh(\Vec{a})q\Mesh^{-1}(\Vec{b})])\Weight([\Mesh(\Vec{b})])\notag\\
&=\left(\Weight([\Mesh(\Vec{a})]^{-1})\Weight([\Mesh(\Vec{b})])\right)\Rep([\Mesh(\Vec{a})q\Mesh^{-1}(\Vec{b})])\label{eq:toscalcomp1}
\intertext{and:}
\Weight'([q])&=\Weight'([\Mesh'(\Vec{a})]^{-1})\Rep([\Mesh'(\Vec{a})q\Mesh'^{-1}(\Vec{b})])\Weight'([\Mesh'(\Vec{b})])\notag\\
&=\left(\Weight'([\Mesh'(\Vec{a})]^{-1})\Weight'([\Mesh'(\Vec{b})])\right)\Rep([\Mesh'(\Vec{a})q\Mesh'^{-1}(\Vec{b})])\label{eq:toscalcomp2}
\text{.}
\end{align}

However:
\begin{align}
\Rep([\Mesh'(\Vec{a})&q\Mesh'^{-1}(\Vec{b})])=\Rep([\Mesh'(\Vec{a})\Mesh^{-1}(\Vec{a})\Mesh(\Vec{a})q\Mesh^{-1}(\Vec{b})\Mesh(\Vec{b})\Mesh'^{-1}(\Vec{b})])\notag\\
&=\Rep([\Mesh'(\Vec{a})\Mesh^{-1}(\Vec{a})][\Mesh(\Vec{a})q\Mesh^{-1}(\Vec{b})][\Mesh(\Vec{b})\Mesh'^{-1}(\Vec{b})])\notag\\
&=\Rep([\Mesh'(\Vec{a})\Mesh^{-1}(\Vec{a})])\Rep([\Mesh(\Vec{a})q\Mesh^{-1}(\Vec{b})])\Rep([\Mesh(\Vec{b})\Mesh'^{-1}(\Vec{b})])\notag\\
&=\left(\Rep([\Mesh'(\Vec{a})\Mesh^{-1}(\Vec{a})])\Rep([\Mesh(\Vec{b})\Mesh'^{-1}(\Vec{b})])\right)\Rep([\Mesh(\Vec{a})q\Mesh^{-1}(\Vec{b})])\text{.}\label{eq:toscalcomp3}
\end{align}
Note that the coefficients in larger parentheses are independent of~$[q]$, so long as~$\Source([q])$ and~$\Target([q])$ remain unchanged. Therefore, there is a phase~$e^{i \delta}$, independent of~$[q]$, such that:
\begin{equation}
\Weight'([q])=e^{i \delta}\Weight([q])\text{,}
\end{equation}
meaning that, unless we alter the fundamental group representation, an arbitrary change in the weights for any mesh starting from any base-point will create physically indistinguishable groupoid representations. In particular, the choice of~$\Weight([q])\equiv1$ in~Ref.~\cite{LaidlawDeWitt1971} is, in fact, a legitimate special choice of gauge.

Finally, let us show that we can use this additional degree of freedom in order to directly incorporate the symmetry of a space, by returning to the case of the punctured plane,~$\R^2\setminus\{\Vec{0}\}$ from the end of the last section. We will make a choice which will allow for a~\emph{single} phase,~$e^{i \varphi}$, where~$\varphi=\phi/2$ and~$\Rep(1)=e^{i \phi}$, for~\emph{any} half-circle counter-clockwise path around the origin.

We will (ab)use complex polar coordinates to describe points in~$\R^2\setminus\{\Vec{0}\}$:
\begin{equation}
re^{i \theta}\triangleq(r\cos\!\theta,r\sin\!\theta)\text{,}
\end{equation}
where~$0\le\theta<2\pi$.

\begin{figure}[ht!]
\centering
\includegraphics{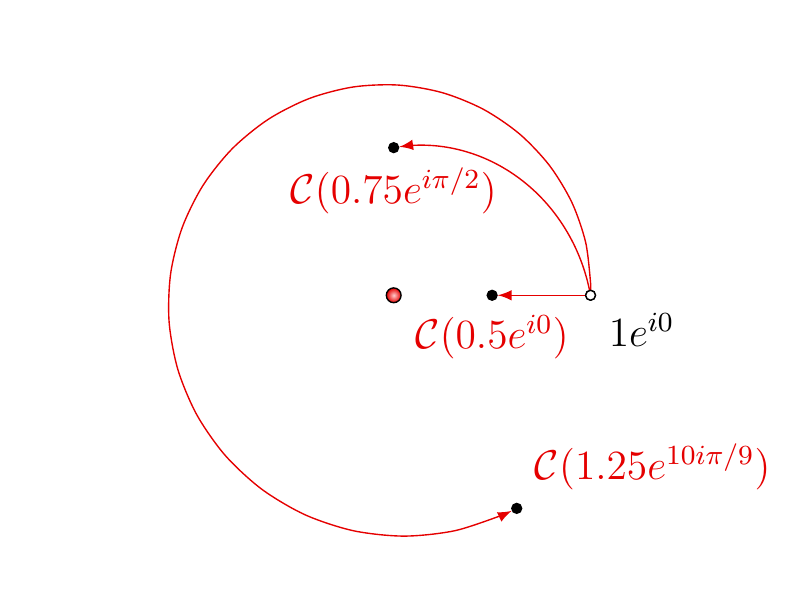}
\caption{Mesh for fundamental groupoid representation with rotational symmetry.}\label{fig:symmesh}
\end{figure}

Let~$\Vec{x}_0=(1,0)=1e^{i 0}$, and the mesh be as follows~(see~\Figref{fig:symmesh}):
\begin{align}
\Mesh(re^{i \theta})(t)&=(re^{i \theta})^t=r^t e^{i t\theta}\label{eq:abtopmesh}
\intertext{with the weights:}
\Weight(\Mesh(re^{i \theta}))&=e^{i \phi\theta/2\pi}\text{.}\label{eq:abtopmeshweights}
\end{align}

\begin{figure}[ht!]
\centering
\includegraphics{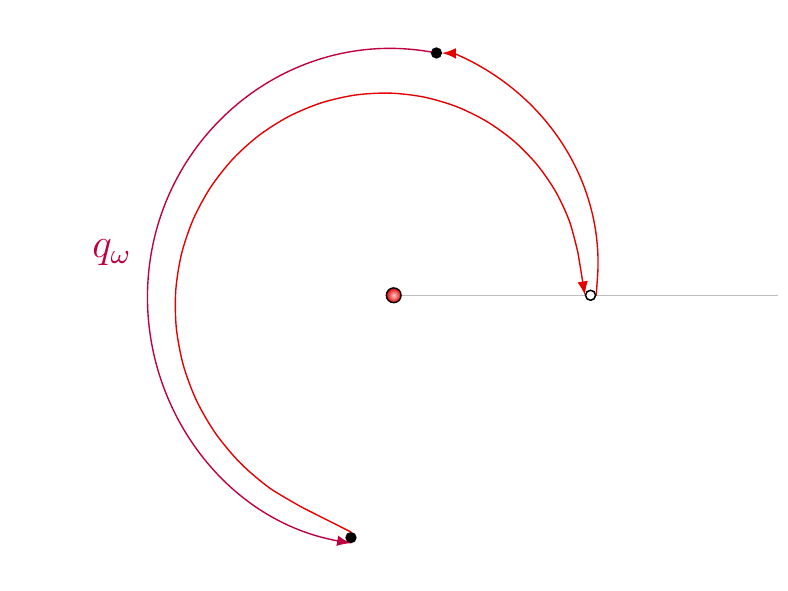}

\caption{Mesh takes a non-crossing half-circle into a contractible loop.}\label{fig:noncrossexch}
\end{figure}

\begin{figure}[ht!]
\centering
\includegraphics{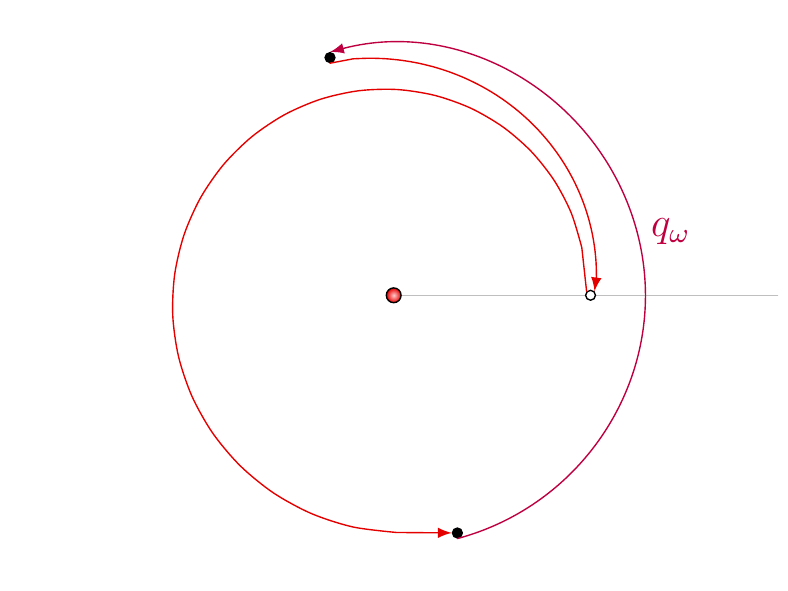}
\caption{Mesh takes a crossing half-circle into a loop with winding number~$1$.}\label{fig:crossexch}
\end{figure}

We will now show that, under this scheme, the amplitude of every counter-clockwise half-circle is equal to~$e^{i \phi/2}$. Let~$re^{i \omega}$ be the starting point, and let~$q_{\omega}$ be the counter-clockwise half-circle between it and~$(-re^{i \omega})$. Then:
\begin{align}
\Weight(q_{\omega})&=\Weight([\Mesh(re^{i \omega})]^{-1})\Rep([\Mesh(re^{i \omega})q_{\omega}\Mesh^{-1}(-re^{i \omega})])\Weight([\Mesh(-re^{i \omega})])\text{.}
\intertext{Now, there are two possibilities: either~$0\le\omega<\pi$, in which case~$0\le\omega+\pi<2\pi$, so we can write:}
-re^{i \omega}&=re^{i (\omega+\pi)}\text{,}
\intertext{and:}
\Weight(q_{\omega})&=e^{-i\phi\omega/2\pi}\Rep([\Mesh(re^{i \omega})q_{\omega}\Mesh^{-1}(re^{i (\omega+\pi)})])e^{i \phi(\omega+\pi)/2\pi}\notag\\
&=e^{i \phi\pi/2\pi}\Rep([\Mesh(re^{i \omega})q_{\omega}\Mesh^{-1}(re^{i (\omega+\pi)})])\notag\\
&=e^{i \phi/2}\Rep([\Mesh(re^{i \omega})q_{\omega}\Mesh^{-1}(re^{i (\omega+\pi)})])\text{;}
\intertext{the path~$\Mesh(re^{i \omega})q_{\omega}\Mesh^{-1}(re^{i (\omega+\pi)})$ is a path which starts at~$(1,0)$, goes to~$(r,0)$, circles to~$re^{i \omega}$, then retraces this path back to~$(1,0)$~(see~\Figref{fig:noncrossexch}); therefore, it can be contracted to a point, and:}
\Weight(q_{\omega})&=e^{i \phi/2}\cdot1=e^{i \phi/2}\text{.}
\intertext{The second possibility is that~$\pi\le\omega<2\pi$, so~$0\le\omega-\pi<\pi$, meaning we can write:}
-re^{i \omega}&=re^{i (\omega-\pi)}\text{,}
\intertext{so:}
\Weight(q_{\omega})&=e^{-i\phi\omega/2\pi}\Rep([\Mesh(re^{i \omega})q_{\omega}\Mesh^{-1}(re^{i (\omega-\pi)})])e^{i \phi(\omega-\pi)/2\pi}\notag\\
&=e^{-i\phi\pi/2\pi}\Rep([\Mesh(re^{i \omega})q_{\omega}\Mesh^{-1}(re^{i (\omega-\pi)})])\notag\\
&=e^{-i\phi/2}\Rep([\Mesh(re^{i \omega})q_{\omega}\Mesh^{-1}(re^{i (\omega-\pi)})])\text{;}
\intertext{the path~$\Mesh(re^{i \omega})q_{\omega}\Mesh^{-1}(re^{i (\omega-\pi)})$ is a path which starts at~$(1,0)$, goes to~$(r,0)$, circles clockwise to~$re^{i \omega}$, which causes it to \emph{cross the positive $x$-axis}, and so it is only the segment beyond which it retraces to~$(1,0)$~(see~\Figref{fig:crossexch}); therefore, it has winding number~$1$, and:}
\Weight(q_{\omega})&=e^{-i\phi/2}e^{i \phi}=e^{i \phi/2}\text{.}
\end{align}


Therefore, in both cases, which cover all counter-clockwise half-circles, the accrued phase is~$e^{i \phi/2}$, as expected. Furthermore, any fundamental groupoid representation which agrees with this one on the fundamental group is compatible with it. Finally, it embodies rotational invariance around the origin, so it is a natural choice. In the next chapter, we will use this representation to create distinguishable anyons.
\section{Conclusion}\label{sec:groupoidresultsconc}
In this chapter, we have found a direct derivation of the result in~Ref.~\cite{Schulman1968} for the Feynman path integral in multiply-connected spaces, and generalized upon the treatment in~Refs.~\cite{LaidlawDeWitt1971,Laidlaw1971} by presenting a novel way of generating the topological degrees of freedom as representations of the fundamental groupoid. This provided a well-organized way of assigning topological phases to paths rather than loops, which allowed us to directly incorporate the symmetry of a space into the representation.

In the next chapter, we will use this result to prove the strictly topological origin of anyons. The example of the punctured plane, running throughout this chapter, and culminating in our ability to provide a consistent topological phase for the exchange of distinguishable particles, will prove essential.
\chapter{Distinguishable and Identical Anyons}\label{ch:distidentanyons}
\section{Distinguishable Anyons}\label{sec:distanyons}
As we have mentioned in~\SecRef{sec:fqheanyons}, the experimental evidence for anyons in~\citet{CaminoZhouGoldman2005} involves interference between different types of particles, while that in~\citet{KimLawlerVishveshwaraFradkin2005} involves ones of the same type. Therefore, the aim of this chapter is to establish anyonic ``statistics'' between two~\emph{distinguishable} particles, and then to apply the results of~\citet{Goyal2015} to arrive at identical anyons. This is in contrast with approaches described in~\SecRef{sec:redconf}, where anyons were required to be~\emph{identical}. Since the exchange of distinguishable particles is not a closed loop, representations of the fundamental group, from which Ref.~\cite{LeinaasMyrheim1977} get their topological phases, are insufficient. We will therefore make use of the result of~\ChapRef{ch:groupoidresults}, where we showed that a topologically non-trivial space has an additional degree of freedom, which, for the case in question, corresponds to a choice of anyonic exchange amplitude.

The focus on two particles is in the interest of concreteness and simplicity of exposition. Since the operational results in~Ref.~\cite{Goyal2015} apply to an arbitrary number of particles, and since the topological situation for many particles involves the ``colored braids''~(see comment in~\citet{GoldinSharp1991}), which are generated by two-particle exchanges, we do not expect the generalization to~$N>2$ particles to introduce any significant complications.

The main assumption we must make is that the particles do not coincide. We will take from~Ref.~\cite{LeinaasMyrheim1977} the transition from the two-particle configuration space to the center of mass and relative spaces, thus helping us focus on the latter, as the former is topologically trivial. 

\begin{figure}[ht!]
\centering
\includegraphics{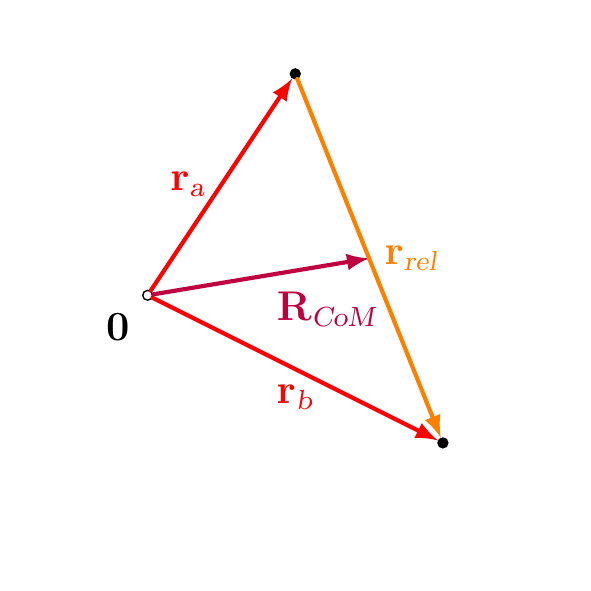}
\caption[Absolute to center of mass and relative coordinates]{Absolute coordinates transformed into center of mass and relative coordinates.}\label{fig:comandrelspce}
\end{figure}

Let us start from the original space:
\begin{equation}
\R^2\times\R^2\setminus\{(\Vec{r},\Vec{r})\mid\Vec{r}\in\R^2\}\text{.}
\end{equation}
If the original variables are of the form~$(\Vec{r}_{\textsl{a}},\Vec{r}_{\textsl{b}})$, then we may define the center of mass and relative coordinates, respectively, as (see~\Figref{fig:comandrelspce}):
\begin{align}
\Vec{R}_{\textsl{CoM}}&\triangleq\frac{\Vec{r}_{\textsl{b}}+\Vec{r}_{\textsl{a}}}{2}\label{eq:comvecdef}
\intertext{and:}
\Vec{r}_{\textsl{rel}}&\triangleq\Vec{r}_{\textsl{b}}-\Vec{r}_{\textsl{a}}\text{.}\label{eq:relvecdef}
\end{align}
In the variables~$(\Vec{R}_{\textsl{CoM}},\Vec{r}_{\textsl{rel}})$, the total space is now:
\begin{equation}
\R^2\times\left(\R^2\setminus\{\Vec{0}\}\right)\text{.}
\end{equation}

The center of mass space is thus simply-connected, so all features of topological interest will appear in the relative space, the punctured plane~$\R^2\setminus\{\Vec{0}\}$. We shall subsequently focus on physical systems where we can perform an appropriate separation of variables, for example, those where all interactions are between the two particles.

\begin{figure}[ht!]
\centering
\includegraphics{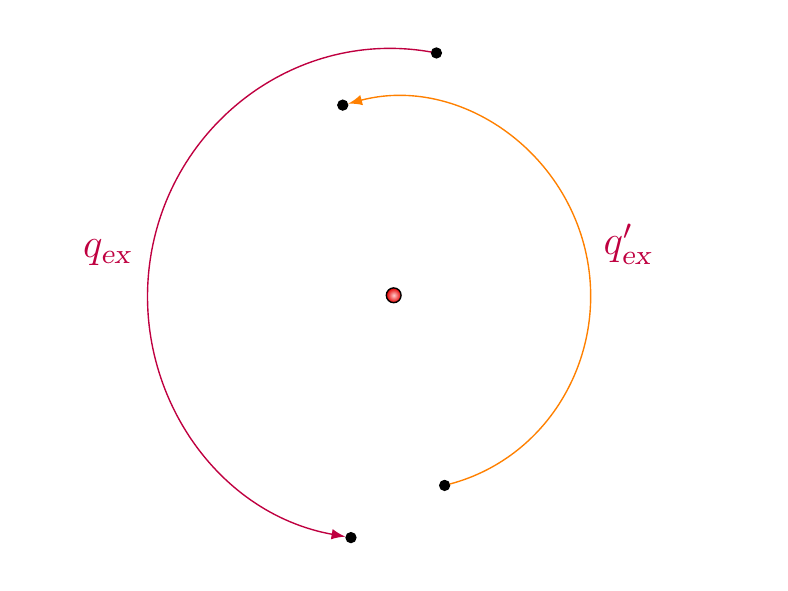}
\caption[Both half-circle counter-clockwise exchanges result in the same phase]{Both half-circle counter-clockwise exchanges result in the phase:~$\Weight([q_{\textsl{ex}}])=\Weight([q'_{\textsl{ex}}])=e^{i\varphi}$.}\label{fig:allexchanges}
\end{figure}

At the end of~\SecRef{sec:groupoidproof}, we characterized the topological degree of freedom of this very space through representations of the fundamental groupoid of the space. Therefore, for any path~$\Vec{x}(t)$, the amplitude associated with it is:
\begin{equation}
\Amp\{\Vec{x}(t)\}=\Weight([\Vec{x}(t)])\exp\left(\frac{i}{\hbar}\Action\{\Vec{x}(t)\}\right)\text{,}
\end{equation}
where~$[\Vec{x}(t)]$ is the homotopy class of the path, and~$\Weight([\Vec{x}(t)])$ is a representation of the fundamental groupoid. In this case, the representation of the fundamental group at an arbitrary point is parameterized by the amplitude corresponding to a single rotation around the origin,~$e^{i\phi}$, and we can choose a groupoid representation such that the amplitude corresponding to \emph{any} half-circular, counter-clockwise path~$q_{\textsl{ex}}$ is:
\begin{equation}
\Weight([q_{\textsl{ex}}])=e^{i\varphi}\text{,}
\end{equation}
where~$\varphi=\phi/2$~(see~\Figref{fig:allexchanges}). For our two-particle system, this corresponds to a simple, counter-clockwise exchange.

If we choose to perform an exchange in the opposite direction, then the path is~$q_{\textsl{ex}}^{-1}$, so:
\begin{equation}
\Weight([q_{\textsl{ex}}^{-1}])=\left(\Weight([q_{\textsl{ex}}])\right)^{-1}=e^{-i\varphi}
\end{equation}

Generally, if we have a path which half-circles the origin~$k$ times,~$q_{\Xi}$, we note that even if the exchange path does not pass through either the initial or final point other than during the start or the end, it is homotopic to a path which does, so that the product rule can be used:
\begin{equation}
[q_{\Xi}]=[q_{\textsl{ex}}]^k\text{,}
\end{equation} 
and we end up with:
\begin{equation}
\Weight([q_{\Xi}])=\Weight([q_{\textsl{ex}}]^k)=\Weight([q_x])^k=e^{ik\varphi}\text{.}
\end{equation}

Now, we have referred to this as an ``exchange'', but at this point we only have a representation for a homotopy class. An actual exchange would be a path~$\Vec{x}(t)$ such that~$[\Vec{x}(t)]=[q_{\Xi}]$, and the exchange amplitude would be:
\begin{align}
\Amp^{\text{D}}_{\Xi}&=\exp\left(\frac{i}{\hbar}\Action\{\Vec{x}(t)\}\right)\Weight([\Vec{x}(t)])\text{.}
\intertext{For simplicity let us treat the free-particle case:}
\Action\{\Vec{x}(t)\}&=\int\limits_{t_1}^{t_2}\frac{\mu}{2}\left(\dot{\Vec{x}}(t)\right)^2\D{t}\text{,}
\intertext{meaning that if we take the quasi-static limit,~$\Action\to 0$, and we get:}
\Amp^{\text{D}}_{\Xi}&\to e^{ik\varphi}\text{.}\label{eq:distexchendresult}
\end{align}

So anyonic behavior is entirely topological, having nothing to do with identical particles. In the rest of the chapter we will combine distinguishable anyons into identical anyons.
\section{Applying the Operational Results to Anyons}\label{sec:identanyons}
We now turn to show that identical particles can behave as anyons, in a way that is compatible with the operational results in~Ref.~\cite{Goyal2015}, which provide only two forms of combination of distinguishable-particle amplitudes into identical ones, and therefore appear to only admit bosons and fermions.

\begin{figure}[ht!]
\centering
\includegraphics{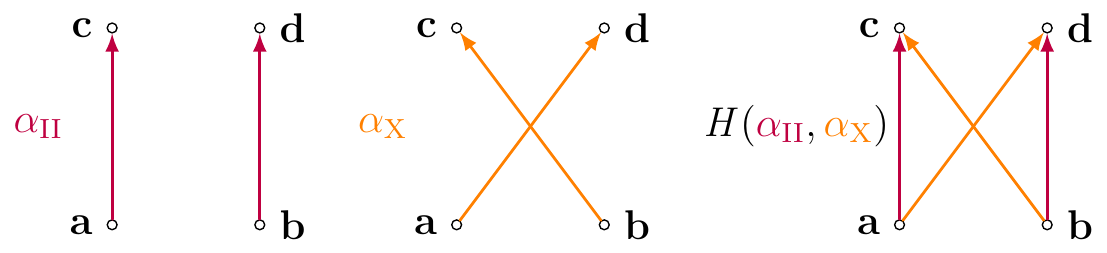}
\caption[Two distinguishable transitions for same identical transition]{Two transitions of the distinguishable-particle system are appropriate for the same identical-particle transition. The amplitudes are related through the function~$\Hfunc(\bullet,\bullet)$.}\label{fig:twodisttoidentamps}
\end{figure}

Let us recall the situation for two identical particles. Suppose that we had a way of finding the transition amplitudes for a system of two~\emph{distinguishable} particles, which are individually dynamically the same as one of the identical particles. For a single, given transition of the identical particles, we would have two transitions of the distinguishable system that would be indistinguishable if we were to ignore the particle labels: an original transition~$\mathrm{II}$, whose amplitude is~$\alpha_{\mathrm{II}}$, and a permuted transition~$\mathrm{X}$, for the situation where the two final particle labels are permuted with each other, whose amplitude is~$\alpha_{\mathrm{X}}$ ~(see~\Figref{fig:twodisttoidentamps}).
Then there are two possible expressions for the total amplitude for two subsequent measurements of a system of two identical particles:
\begin{equation}
\Amp_{\textsl{total}}=\Hfunc(\alpha_{\mathrm{II}},\alpha_{\mathrm{X}})=\alpha_{\mathrm{II}}\pm\alpha_{\mathrm{X}}\text{,}\label{eq:2idpartmainresrew}
\end{equation}
where~$+$ and~$-$ correspond to operational bosons and fermions, respectively. Note that this expression does not, at first, seem to allow for anyons. However, as we saw in the previous section, anyonic behavior arises from topological considerations for distinguishable particles. So both~$\alpha_{\mathrm{II}}$ and~$\alpha_{\mathrm{X}}$ separately incorporate this degree of freedom. Here we will show how these ideas are combined to form a pair of identical anyons.

\begin{figure}[ht!]
\centering
\includegraphics{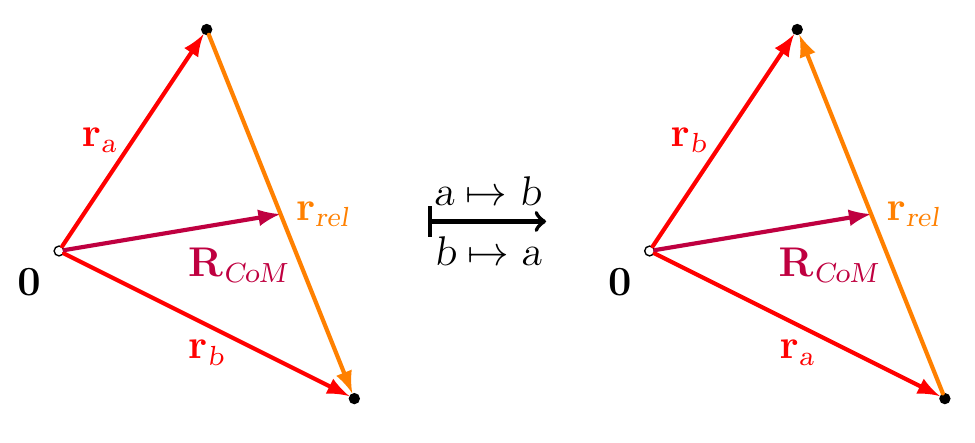}
\caption[Transposing reverses relative coordinate, center of mass unchanged]{Transposing the particles reverses~$\Vec{r}_{\textsl{rel}}$, while leaving~$\Vec{R}_{\textsl{CoM}}$ unchanged.}\label{fig:comrelvsabssymm}
\end{figure}

Let us write more explicit expressions for these amplitudes. We recall that we are discussing two particles in two dimensions, which cannot coincide. Now let us see how identity translates into the relative space. Given a pair of different locations in~$\R^2$:
\begin{align}
\Vec{r}_{\textsl{a}}&\neq\Vec{r}_{\textsl{b}}\text{;}
\intertext{if $(\Vec{r}_{\textsl{a}},\Vec{r}_{\textsl{b}})$ is, as in~\SecRef{sec:distanyons}, the outcome in which the first particle is in~$\Vec{r}_{\textsl{a}}$ and the second is in~$\Vec{r}_{\textsl{b}}$, then:}
(\Vec{r}_{\textsl{a}}&\SymmOp\Vec{r}_{\textsl{b}})
\end{align}
is an outcome in which one particle is found in each of these locations, without being able to distinguish the two. By the definition of the center of mass and relative coordinates in~\Eqref{eq:comvecdef} and~\Eqref{eq:relvecdef}, exchanging the particles:
\begin{align}
(\Vec{r}_{\textsl{a}},\Vec{r}_{\textsl{b}})&\mapsto(\Vec{r}_{\textsl{b}},\Vec{r}_{\textsl{a}})
\intertext{translates into a transformation reversing~$\Vec{r}_{\textsl{rel}}$, without changing~$\Vec{R}_{\textsl{CoM}}$:}
(\Vec{R}_{\textsl{CoM}},\Vec{r}_{\textsl{rel}})&\mapsto(\Vec{R}_{\textsl{CoM}},-\Vec{r}_{\textsl{rel}})\text{;}
\end{align}
see~\Figref{fig:comrelvsabssymm}. Therefore, when we focus on the relative space, distinguishable particle states may be designated by a single relative coordinate,~$(\Vec{r}_{\textsl{rel}})$, while the symmetrized space may be written as~$(\Vec{r}_{\textsl{rel}}\SymmOp-\Vec{r}_{\textsl{rel}})$.

\begin{figure}[ht!]
\centering
\includegraphics{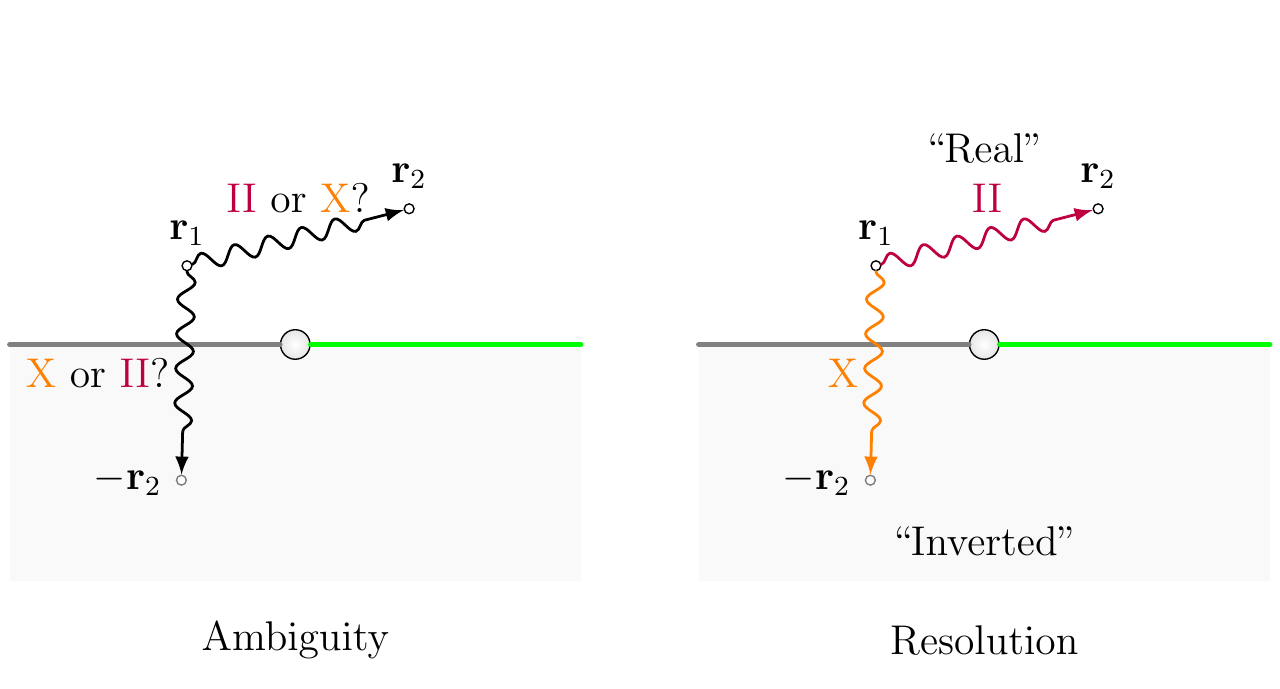}
\caption[Resolving ambiguity for identical particles in relative space]{Complete symmetrized space is~$\{y>0\}\cup\{y=0, x>0\}$. Unless an arbitrary choice is made, there is an ambiguity regarding which transition is original and which is permuted. This is resolved by deciding that the transition between~$\Vec{r}$'s in~``Real'' space is original, while that from~``Real'' to~``Inverted'' space is permuted.}\label{fig:2dposr}
\end{figure}

To simplify the notation, we can suppress the~$\textsl{rel}$ in the subscript, and simply use a subscript number for time-ordering purposes. Then a measurement outcome for the distinguishable-particle system is written as~$(\Vec{r}_n)$, while the corresponding outcome for the identified system is~$(\Vec{r}_n\SymmOp-\Vec{r}_n)$; however, the latter contains an ambiguity, as both~$(\Vec{r}_n\SymmOp-\Vec{r}_n)$ and~$(\Vec{r}_n\SymmOp\Vec{r}_n)$ refer to the same outcome of the identical particle system. As we can see from~\Eqref{eq:2idpartmainresrew}, in the case of fermions, this would lead to a multiply-defined value. We therefore must choose a subspace of the punctured plane of the ``Real'' ($+\Vec{r}$) values, so that they, along with the~``Inverted'' values ($-\Vec{r}$), cover the whole space. We choose:~$\{y>0\}\cup\{y=0, x>0\}$~(see~\Figref{fig:2dposr}). In this we are borrowing from the reduced configuration space approach in~Ref.~\cite{LeinaasMyrheim1977}; note, however, that we are dealing with particle identity \emph{after} quantization, as we assume that we have amplitudes to work with from the appropriate distinguishable-particle system.

If we then write, for two consecutive measurements indexed by~$1$ and~$2$:
\begin{align}
\alpha_{\mathrm{II}}&=\Amp\left((\Vec{r}_1)(\Vec{r}_2)\right)
\intertext{for the original transition,}
\alpha_{\mathrm{X}}&=\Amp\left((\Vec{r}_1)(-\Vec{r}_2)\right)
\intertext{for the permuted transition, and:}
\Amp_{\textsl{total}}&=\Amp\left((\Vec{r}_1\SymmOp-\Vec{r}_1)(\Vec{r}_2\SymmOp-\Vec{r}_2)\right)\text{;}
\end{align}
for the identical particle transition, we can then express~\Eqref{eq:2idpartmainresrew} as:
\begin{align}
\Amp\left((\Vec{r}_1\SymmOp-\Vec{r}_1)(\Vec{r}_2\SymmOp-\Vec{r}_2)\right)&=
\Hfunc\big(\Amp\left((\Vec{r}_1)(\Vec{r}_2)\right),\Amp\left((\Vec{r}_1)(-\Vec{r}_2)\right)\big)\notag\\
&=\Amp\left((\Vec{r}_1)(\Vec{r}_2)\right)\pm\Amp\left((\Vec{r}_1)(-\Vec{r}_2)\right)\text{.}
\end{align}

We define the original transition to be that which stays within the~``Real'' subspace, while the permuted transition crosses into the~``Inverted'' subspace~(see~\Figref{fig:2dposr} again).

\section{Exchanging Identical Anyons}\label{sec:identanyonexch}
\subsection{Circular Exchanges}\label{sec:identanyoncircexch}
\begin{figure}[ht!]
\centering
\includegraphics{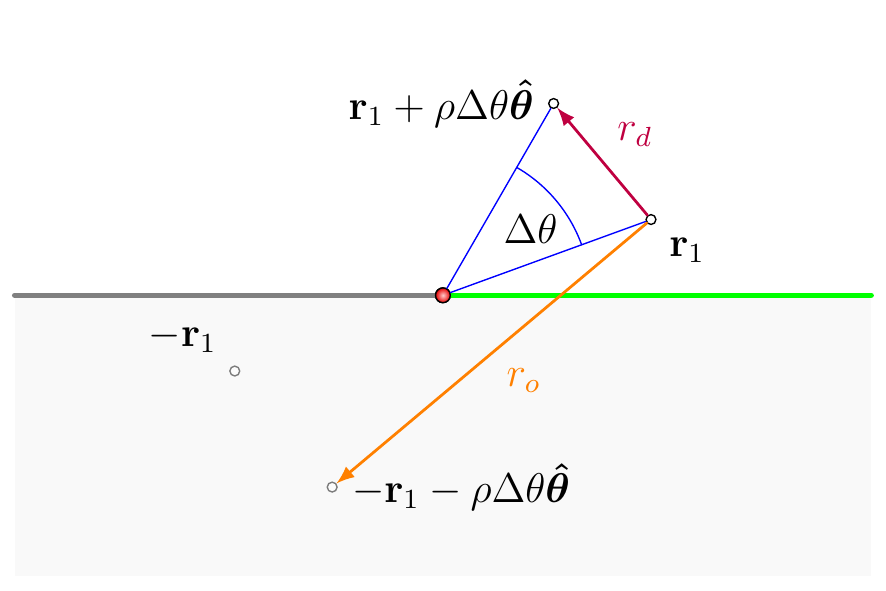}
\caption[Direct and opposite transitions for small angular change]{The direct and opposite transitions for a small angular change.
Note that $r_o$ does not pass through the hole at the origin.}\label{fig:2ddeltatheta}
\end{figure}

We turn to exchanging particles. In the~$\Delta t\to0$ limit, the only contribution to the amplitude comes from around the classical path~\cite{DeWitt1969}. So let us have the particles take very short steps, in a half-arc, around their center of mass~(see~\Figref{fig:2ddeltatheta}):
\begin{align}
\Vec{r}_1=(\rho\cos\!\theta,\rho\sin\!\theta)\mapsto(\rho\cos(\theta+\Delta\theta),\rho\sin(\theta+\Delta\theta))&=\Vec{r}_1+\rho\Delta\theta(-\sin\!\theta,\cos\!\theta)\notag\\
&=\Vec{r}_1+\rho\Delta\theta\ThetaHat\text{;}
\end{align}
we are dealing then with the~$\Delta t\to0$ limit of two distinguishable-particle amplitudes; direct:
\begin{align}
\Amp_d(\theta+\Delta\theta, t+\Delta t\,;\theta, t)&\triangleq\Amp\left((\Vec{r}_1)(\Vec{r}_1+\rho\Delta\theta\ThetaHat)\right)\text{,}
\intertext{over a classical line-segment we will call~$r_d$, and opposite:}
\Amp_o(\theta+\Delta\theta, t+\Delta t\,;\theta, t)&\triangleq\Amp\left((\Vec{r}_1)(-\Vec{r}_1-\rho\Delta\theta\ThetaHat)\right)\text{.}
\end{align}
over a classical line-segment we will call~$r_o$.

Classically, outside the origin, the two particles are free, so in both cases we are looking at particles moving uniformly, that is, in straight lines at constant speeds. In the case of the opposite transition, the straight line is within the relevant space because of the slight displacement from a diameter of the circle. For the direct transition, the velocity is:
\begin{align}
\Vec{v}_d&=\frac{\Vec{r}_1+\rho\Delta\theta\ThetaHat-\Vec{r}_1}{\Delta t}=\frac{\rho\Delta\theta}{\phantom{\rho}\Delta t}\ThetaHat\text{,}\\
\intertext{while the velocity for the opposite transition is very different:}
\Vec{v}_o&=\frac{-\Vec{r}_1-\rho\Delta\theta\ThetaHat-\Vec{r}_1}{\Delta t}=-\frac{2\Vec{r}_1}{\Delta t}-\frac{\rho\Delta\theta}{\phantom{\rho}\Delta t}\ThetaHat\text{;}
\end{align}
if we then take~$\Delta t,\Delta\theta\to0$ with~$\Delta\theta/\Delta t= \dot{\theta}$, a constant, then:
\begin{align}
\Vec{v}_d&=\rho\dot{\theta}\ThetaHat\text{,}
\intertext{while:}
\Vec{v}_o&=-\frac{2\Vec{r}_1}{\Delta t}\text{.}
\end{align}

We will find the misbehavior of the opposite transition's velocity very useful in the following. We now calculate the action functional for both cases (in this case~$m_1=m_2=m$), using the standard free-particle Lagrangian:
\begin{align}
\Action_d(\theta+\Delta\theta, t+\Delta t\,;\theta, t)&=\int\limits_{t}^{t+\Delta t}\frac{m}{2}\Vec{v}_d^2=
\int\limits_{t}^{t+\Delta t}\frac{m}{2}\rho^2\dot{\theta}^2=\frac{m}{2}\rho^2\dot{\theta}^2\Delta t\text{,}
\intertext{which is well-behaved, while:}
\Action_o(\theta+\Delta\theta, t+\Delta t\,;\theta, t)&=\int\limits_{t}^{t+\Delta t}\frac{2m\rho^2}{\Delta t^2}\D{t'}=\frac{2m\rho^2}{\Delta t}
\end{align}
is not. When passing to the amplitudes, recall that the distinguishable system incorporates the fundamental groupoid representation~$\Weight(\bullet)$ defined in~\SecRef{sec:distanyons}. Additionally, while the action functionals are time-independent, we will want to distinguish between the respective line segments, so we will index them:
\begin{align}
\Amp_{dn}&=\Weight([r_{dn}])\exp\left[i\frac{m\rho^2\dot{\theta}^2}{2\hbar}\Delta t\right]
\intertext{and}
\Amp_{on}&=\Weight([r_{on}])\exp\left[i\frac{2m\rho^2}{\hbar\Delta t}\right]\text{.}\label{eq:optrans}
\end{align}

\begin{figure}[ht!]
\centering
\includegraphics{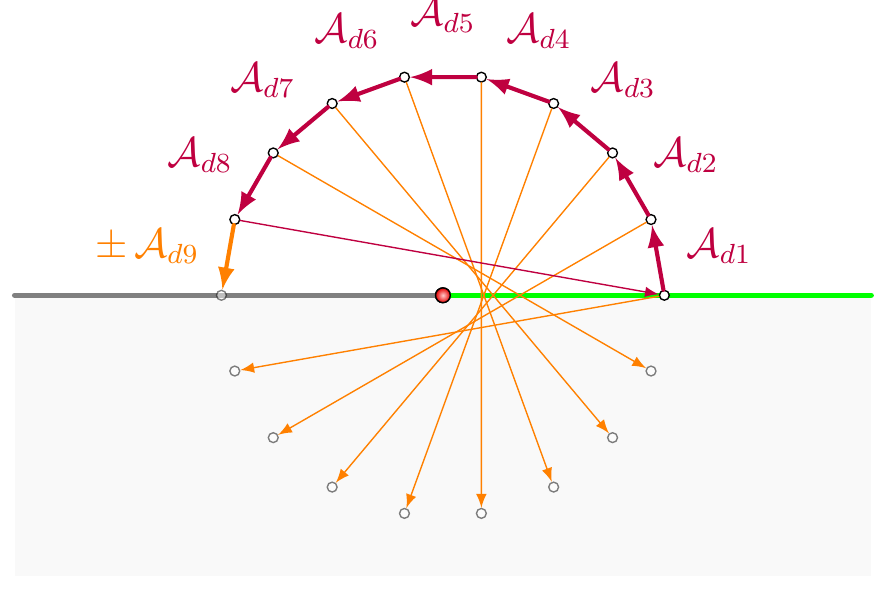}
\caption[Accumulation of direct amplitudes]{Accumulation of the direct amplitudes, with illustration of the opposites. Note that in the last transition, the direct transition, while retaining the same physical amplitude, crosses the boundary between ``real'' and ``inverted' space, while the opposite transition stays within ``real'' space, so that the former is ``permuted'' and the latter is ``original''. Therefore, the total exchange amplitude acquires a possible sign, depending on whether the particles are originally bosons ($+$) or fermions ($-$).}\label{fig:2dsteps}
\end{figure}

Now, let us concatenate~$N$ such transitions, with~$N\Delta t=T$,~$N\Delta \theta=\pi$, so that~$\dot{\theta}=\pi / T$, and we create an overall exchange amplitude:
\begin{equation}
\Amp^{\text{I}}_{\Xi}=(\Amp_{d1}\pm\Amp_{o1})(\Amp_{d2}\pm\Amp_{o2})\dotsm(\Amp_{d[N-1]}\pm\Amp_{o[N-1]})(\Amp_{oN}\pm\Amp_{dN})\text{;}
\end{equation}
the change in order in the last transition is not accidental: since it crosses out of the domain of definition of~$\Vec{r}\SymmOp\Vec{-r}$, the order of terms is then, and only then, reversed. This will always happen an odd number of times in an overall exchange, regardless of where the line of demarcation is chosen (see~\Figref{fig:2dsteps}).

Now, as seen in~\Eqref{eq:optrans},~$\Amp_{on}$ depends on~$\Delta t$ as~$\exp(\widetilde{C}i\Delta t^{-1})/\hbar$, or equivalently on~$N$ as~$\exp(CiN/\hbar)$, with~$C>0$. If we follow the procedure in~\citet{Feynman1948} and replace~$\hbar\mapsto(1-i\delta)\hbar$, with the limit~$\delta\to0^{+}$ to be taken after our calculations, then~$1/\hbar\mapsto(1+i\delta)\hbar$, and:
\begin{equation}
\exp(CiN/\hbar)\mapsto e^{CiN(1+i\delta)/\hbar}=e^{CiN/\hbar}\cdot e^{-CN\delta/\hbar}\text{.}
\end{equation}

Therefore, in each factor of~$\Amp^{\text{I}}_{\Xi}$,~$\Amp_{on}$ will become exponentially smaller than~$\Amp_{dn}$ as~$N\to\infty$, so that we can focus on the latter at the expense of the former. Another way of understanding this mathematical trick is by noting that in the~$\Amp_{on}$, the phase varies wildly as~$N\to\infty$, so that it will tend to cancel out, while the~$\Amp_{dn}$ remain manageable, and lead to a finite product among themselves.
 
Therefore, we are left with the amplitude of a circular path~$r_{\Xi}$ in the positive direction:
\begin{align}
\Amp^{\text{I}}_{\Xi}&=\pm\prod\limits_{n=1}^{N}\Amp_{dn}=\pm\prod\limits_{n=1}^{N}\Weight([r_{dn}])\exp\left[i\frac{m\rho^2\dot{\theta}^2}{2\hbar}\Delta t\right]\notag\\
&=\pm\Weight([r_{d1}\dotsm r_{dN}])\exp\left[ i\frac{m\rho^2\dot{\theta}^2}{2\hbar}N\Delta t\right]=\pm \Weight([r_{\Xi}]) e^{i\frac{m\pi^2\rho^2}{2\hbar T}}\text{,}
\end{align}
factoring nicely to an operational factor~$\pm$, a topological factor~$\Weight([r_{\Xi}])$, and a dynamical factor~$e^{i\frac{m\pi^2\rho^2}{2\hbar T}}$. We can finally take the quasi-static limit,~$T\to\infty$, removing the dynamical factor to find that:
\begin{equation}
\Amp^{\text{I}}_{\Xi}=\pm \Weight([r_{\Xi}])=\pm e^{i\phi}\text{,}
\end{equation}
that is, an exchange in the positive direction multiplies by the operational coefficient, as well as an Aharonov-Bohm phase. Since~$-1=e^{i\pi}$, we can incorporate the operational degree of freedom into a new phase, with angle~$\varphi$, and we have anyons:
\begin{equation}
\Amp^{\text{I}}_{\Xi}=e^{i\varphi}\text{.}
\end{equation}

\begin{figure}[ht!]
\centering
\includegraphics{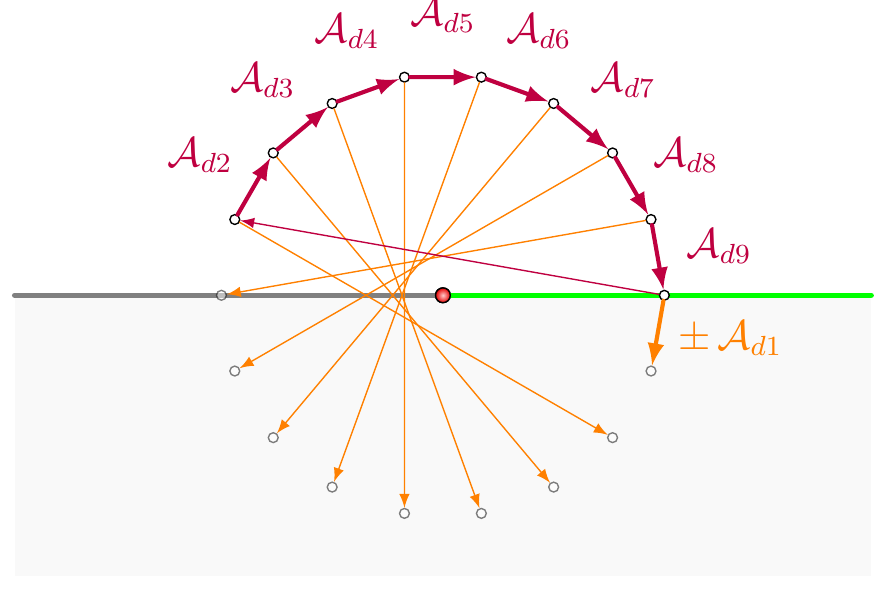}
\caption[Accumulation of direct amplitudes, counter-clockwise exchange]{Situation is as for~\Figref{fig:2dsteps}, but in the reverse direction. Here it is the~\emph{first} transition in which~``original'' and~``permuted'' must be reversed.}\label{fig:2dstepsbw}
\end{figure}

This was presented as a single exchange in the positive direction. The treatment in the negative direction only requires us to choose~$N\Delta\theta=-\pi$, and to now have the segments and the exchange itself be in the opposite direction. Then the exchange amplitude would be:
\begin{equation}
\Amp^{\text{I}}_{\Xi}=(\Amp_{o1}\pm\Amp_{d1})(\Amp_{d2}\pm\Amp_{o2})\dotsm(\Amp_{d[N-1]}\pm\Amp_{o[N-1]})(\Amp_{dN}\pm\Amp_{oN})\text{,}
\end{equation}
where the change in order is now in the~\emph{first} transition, which is the one crossing the line of demarcation, despite being shorter locally (see~\Figref{fig:2dstepsbw}).

The total amplitude would then be:
\begin{align}
\Amp^{\text{I}}_{\Xi\text{rev}}&=\pm \Weight([r_{\Xi\text{rev}}])e^{i\frac{m\pi^2\rho^2}{\hbar T}}=\pm e^{-i\phi}e^{i\frac{m\pi^2\rho^2}{\hbar T}}\text{,}
\intertext{or in the quasi-static limit (and noting that~$e^{i\pi}=e^{-i\pi}$):}
\Amp^{\text{I}}_{\Xi\text{rev}}&=e^{-i\varphi}\text{.}
\intertext{Generally, if one wishes to perform a~$k$-fold exchange, then if~$k\ge0$, one gets:}
\left(\Amp^{\text{I}}_{\Xi}\right)^k&=e^{ik\varphi}\text{,}
\intertext{while if~$k<0$, one would have:}
\left(\Amp^{\text{I}}_{\Xi\text{rev}}\right)^{\abs{k}}&=e^{-i\abs{k}\varphi}=e^{ik\varphi}\text{.}
\intertext{So, for the general~$k$-fold circular exchange:}
\Amp^{\text{I}}_{\Xi}&=e^{ik\varphi}\text{,}
\end{align}
as in~\Eqref{eq:distexchendresult}.
\subsection{More General Paths}\label{sec:identanyongenexch}
Now, let us generalize further. Suppose the exchange were performed along a more general path,~$\Vec{r}(t)$, such that:
\begin{equation}
\Vec{r}(0)=-\Vec{r}(T)=\Vec{x}_0\neq\Vec{0}\text{.}
\end{equation}
We will still impose a few regularity conditions on this path: it is smooth, with finite velocity, it stays at a minimal distance~$\rho_{\textsl{min}}>0$ from the excluded origin, and for simplicity, the angular velocity is never zero. We choose some~$\Delta t=T/N$, and look at short steps in time:
\begin{align}
\Vec{r}_n=(\rho_n\cos\!\theta_n,\rho_n\sin\!\theta_n)\mapsto\Vec{r}_{n+1}&=\Vec{r}_n+\Delta\rho_n\RhoHat_n+\rho_n\Delta\theta_n\ThetaHat_n\notag\\
&=\Vec{r}_n+\left(\dot{\rho}_n\RhoHat_n+\rho_n\dot{\theta}_n\ThetaHat_n\right)\Delta t\text{,}
\end{align}
where:
\begin{equation}
\{\textsl{variable}\}_n\triangleq\{\textsl{variable}\}(n\Delta t)\text{.}
\end{equation}

At the~$\Delta t\to0$ limit, the two relevant paths --- which do not cross the excluded origin, as we have specified that~$\dot{\theta}\neq 0$ --- have the following direct straight-line velocity:
\begin{align}
\Vec{v}_{dn}&=\dot{\rho}_n\RhoHat_n+\rho_n\dot{\theta}_n\ThetaHat_n\text{,}
\intertext{with the opposite velocity being:}
\Vec{v}_{on}&=-\frac{2\Vec{r}_n}{\Delta t}-\Vec{v}_d\text{;}
\intertext{as in~\SecRef{sec:identanyoncircexch}, the~$\Delta t^{-1}$ term dominates, and we may write:}
\Vec{v}_{on}&=-\frac{2\Vec{r}_n}{\Delta t}\text{.}
\end{align}
The action functionals for the two cases are:
\begin{align}
\Action_d(\rho_{n+1},\theta_{n+1}, t_{n+1}\,;\rho_n,\theta_n, t_n)&=\int\limits_{t_n}^{t_n+\Delta t}\frac{m}{2}\Vec{v}_{dn}^2\D{t'}=
\int\limits_{t_n}^{t_n+\Delta t}\frac{m}{2}\left[\dot{\rho}_n^2+\rho_n^2\dot{\theta}_n^2\right]\D{t'}\notag\\
&=\frac{m}{2}\dot{\Vec{r}}_n^2\Delta t\text{,}
\intertext{which is well-behaved, while:}
\Action_o(\rho_{n+1},\theta_{n+1}, t_{n+1}\,;\rho_n,\theta_n, t_n)&=\int\limits_{t_n}^{t_n+\Delta t}\frac{2m\rho_n^2}{\Delta t^2}\D{t'}=\frac{2m\rho_n^2}{\Delta t}\ge\frac{2m\rho_{\textsl{min}}^2}{\Delta t}
\intertext{is not. we can then write the amplitudes more succinctly as follows:}
\Amp_{dn}&=\Weight([r_{dn}])\exp\left[i\frac{m}{2\hbar}\dot{\Vec{r}}_n^2\Delta t\right]
\intertext{and}
\Amp_{on}&=\Weight([r_{on}])\exp\left[i\frac{4m\rho_n^2}{\hbar\Delta t}\right]\text{,}
\end{align}
where~$\Weight(\bullet)$ is defined as in~\SecRef{sec:identanyoncircexch}.

The next step is more subtle than for a simple, circular path. We have a product of mostly:
\begin{equation}
\Amp_{dn}\pm\Amp_{on}\text{,}
\end{equation}
where this is replaced by~$\Amp_{on}\pm\Amp_{dn}$ whenever we have a path segment crossing the boundary of the region of definition of the symmetrized space. If the exchange involves~$k$ half-rotations of the particles around each other, then there will be~$k$ such terms. Since, as in~\SecRef{sec:identanyoncircexch}, a~$\Amp_{on}$ factor suppresses any term which contains even one instance thereof, the total amplitude of exchange is:
\begin{align}
\Amp^{\text{I}}_{\Xi}&=(\pm1)^k\prod\limits_{n}\Amp_{dn}=(\pm1)^k\left\{\prod\limits_{n}\Weight([r_{dn}])\right\}\exp\left[\sum\limits_{n}i\frac{m\dot{\Vec{r}}_n^2}{2\hbar}\Delta t\right]\notag\\
&=(\pm1)^k  \Weight([r])e^{\Action_{\textsl{free}}[\Vec{r}(t)]}\text{;}
\end{align}
now, the~$k$ half-rotations have to be in the same direction, as, again, we assume~$\dot{\theta}\neq0$, and the sign of~$k$ can be taken to signify the direction of rotation, so that the total topological phase is:
\begin{equation}
\Weight([r])=e^{ik\phi}\text{.}
\end{equation}
Finally, we can take the quasi-static limit, in which the free particle's action functional is zero, and we define~$\varphi=\phi$ or~$\varphi=\phi+\pi$, depending on whether we started from a fermionic or bosonic total amplitude, so we find:
\begin{equation}
\Amp^{\text{I}}_{\Xi}=e^{ik\varphi}\text{,}
\end{equation}
which is again the same expression in~\Eqref{eq:distexchendresult}, for distinguishable particles.
\section{One Dimension Without Incidence Point}\label{sec:identanyon1dnoinc}
Before moving on to the discussion, we will note that the appropriate ``anyonic'' behavior is also found in one dimension, without the need for symmetrizing the space prior to quantization. For two particles in one dimension, if the incidence point is taken out, the particles cannot change places. Therefore,~$\alpha_{\mathrm{X}}=0$, and the total amplitude is:
\begin{equation}
\Amp=\alpha_{\mathrm{II}}\text{;}
\end{equation}
that is, the situation is the same as that for distinguishable particles without incidence. Nevertheless, whether or not the particles are identical, the removal of the incidence point creates a one-dimensional family of self-adjoint extensions for this problem, allowing for an interpolation between fermion- and boson-like behavior on each side, as described in~Ref.~\cite{LeinaasMyrheim1977}, as well as more rigorously, and with an incorporation of Feynman path integrals, in~\citet{FarhiGutmann1990}. The importance of non-connectivity was originally noted by~\citet{Girardeau1965}.
\section{Conclusion}\label{sec:distidentanyonsconc}
In this chapter we were able to use the groupoid-based approach to quantization of multiply-connected spaces developed in~\ChapRef{ch:groupoidresults} to express anyonic behavior in two distinguishable particles in the path integral formalism. We then showed how to use the identical particle amplitude combination results of~Ref.~\cite{Goyal2015} to turn distinguishable anyons into identical anyons. This should put to rest the idea that anyonic behavior is intrinsically tied to particle identicality; instead, it is a result of the topology of particles in two-dimensional space which cannot coincide.
\chapter{Historical and Philosophical Discussion of Particle Identity}\label{ch:histover}
Now that we have shown that anyonic behavior is a topological phenomenon rather than having to do with identical particles, let us look back at the historical development of this issue, compare it to our own results in~\ChapRef{ch:distidentanyons}, and see why anyons and identicality are so often intertwined in the literature.

All theoretical discussions and most experimental papers which use anyons see the origins of their topic in~\citet{LeinaasMyrheim1977}. In that paper, the original, classical configuration space of a set of identical particles was symmetrized: coordinates connected by a permutation were simply identified, and the resulting, non-Euclidean space, was quantized. They noted that this quantization would be difficult if the incidence points were retained, as the symmetrized space would then not be a manifold, so they took them out. They then found that in two dimensions, the path of exchange between particles, not just the fact that they are exchanged, matters. Not every two paths can be continuously deformed into each other, so that the space has become~\emph{multiply-connected}. This makes it the first paper to make use of the fact that the configuration space for two non-intersecting particles in two dimensions is multiply-connected, and the result was anyonic behavior.

The reduced configuration space project itself came about as a response to several issues raised with the way identical particles were tackled at the time: criticism in~\citet{Mirman1973}, which noted that particle labels are physically meaningless, and therefore should be abandoned in a more rigorous theory; the existence of the Gibbs paradox~\cite{Gibbs1960}, showing that it is necessary to reduce the size of the configuration space of a system of indistinguishable particles by the number of permutations even in classical systems; and, most importantly, the seeming arbitrariness of the Symmetrization Postulate~(see~\SecRef{sec:identpart}). Ref.~\cite{LeinaasMyrheim1977} decided to resolve all of these issues in one fell swoop, by imposing the identity of particles on a classical system, and then proceeding to quantize it with the tools dedicated to geometrically or topologically non-trivial spaces, with the hope that something like the symmetrization postulate would come up naturally. The  failure of this program in two dimensions led to the discovery of anyons.

However, as we have seen in~\SecRef{sec:distanyons}, anyons can be dealt with without referring to the reduced configuration space project, or indeed, to identical particles at all, and this has already been recognized in the past. Ref.~\cite{LeinaasMyrheim1977} was published in~\citeyear{LeinaasMyrheim1977}; as early as~\citeyear{GoldinMenikoffSharp1985}, we already find~\citet{GoldinMenikoffSharp1985} pointing out that unusual phases on exchange are possible even for distinguishable particles in two dimensions, since it is the connectivity of the space, rather than the identity of the particles, that is relevant.
\citet{Zee1995}, in notes from a conference in~1994, acknowledges that the ``statistics'' involved here can be generated from a phase interaction between the particles, rather than by requiring them to be a type of identical particles.

Despite these early admissions of the separability of anyons from particle identity, the vast majority of papers and reviews which were aware of either Ref.~\cite{LeinaasMyrheim1977} or~\citet{Wilczek1982b} (which cited the former while popularizing the term ``anyon''), introduced the issue in the context of the reduced configuration space, even though  in many cases they proceeded to use a particle-flux composite or Chern-Simons potential when performing calculations~\cite{Wilczek1990, Lerda1992,Khare1997, Stern2008}.

Ref.~\cite{Wilczek1982b} itself represents a clear hybrid phenomenon. On the one hand, the two-anyon case is treated with an Aharonov-Bohm potential, without using a reduced configuration space. On the other hand, the generalization to~$N>2$ particles is said to be difficult because the reduced configuration space would then become more complicated. More recently,~\citet{NayakSimonSternFreedmanDas2008}, while starting out with reduced configuration space ideas, acknowledge that there could be many different ``types'' of anyons, with arbitrary phases accrued by twisting them around each other, implicitly admitting non-identical anyons. They, with~\citet{Zee1995}, accepted that what is presented as ``statistics'' of anyons could be instead understood as a special kind of interaction.

Another curious phenomenon is the fact that none of the papers or books cited above seem to seriously address the \emph{geometrical} (rather than topological) consequences of reducing the configuration space in two dimensions. As noted in~Ref.~\cite{LeinaasMyrheim1977}, the reduced space is equivalent to that of a cone, which has non-zero curvature at the origin. Had this been important, we would expect to find many more citations and discussions of works dealing with this kind of system\footnote{This is usually known as the ``cosmic string''; see Refs.~\cite{Dowker1977,DeserJackiwHooft1984,Hooft1988,DeserJackiw1988, KayStuder1991} for in-depth discussion.}.

So the conceptual question is why, despite the fact that the reduced configuration space approach is ill-suited to applications, and would lead to fundamental additional problems if taken seriously, is it nevertheless invoked without causing any practical problems?

\begin{figure}[ht!]
\centering
\includegraphics{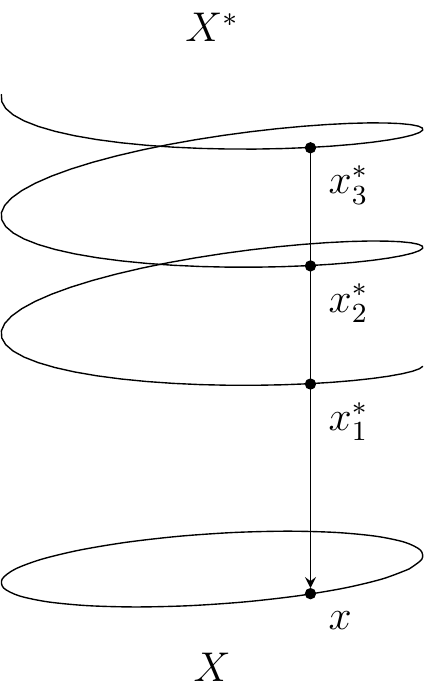}
\caption[Covering space]{$X^*$ is the covering space of~$X$. $x^*_i$ all project to the same~$x$.}\label{fig:coveringspace}
\end{figure}

The solution involves a tool essential for the conventional quantization of multiply-connected spaces: the use of covering spaces (see~\Figref{fig:coveringspace}; for a more extensive discussion, see~\citet{SingerThorpe1967}). These are spaces that are themselves simply-connected, contain several copies of the underlying multiply-connected space, properly patched together, and which project in a canonical manner down to the space in question. A wave-function or propagator is calculated for the covering space, and then further restricted so that the existence of a space larger than the physical space cannot be detected. While one may cite~\citet{LaidlawDeWitt1971} as bucking this trend, the more elaborate development of this work in~\citet{Laidlaw1971} concedes that the covering space approach is necessary in order to make any meaningful calculations, and from~\citet{Dowker1972} on, this is the method normally used\footnote{The other two resolutions in the literature are the use of multivalued functions and of fiber bundles. Multivalued functions themselves are best accommodated through the universal covering space~\cite{HorvathyMorandiSudarshan1989}, or its complex function analogue, the Riemann surface~\cite{Krantz2008}, while fiber bundles are a different form of covering space, and in any event, are applied to the non-symmetrized, electromagnetic problem anyway~\cite{Morandi1992}.}.

The relevance of this issue can be found in~\citet{FrenchKrause2006}, through an analysis of an argument between~\citet{Grunbaum1977} and~\citet{Stein1977a,Stein1977b}. The salient point is that the passage to a covering space is \emph{not innocent}. We see this most clearly when we look at~Ref.~\cite{LeinaasMyrheim1977}'s treatment of the three-dimensional quantum case. After going to the trouble of removing the origin and identifying exchanged point, the symmetrization postulate in three dimensions is retrieved by using the universal covering space, which, since the configuration space is now doubly-connected, is a doubling-up of the reduced space: in other words, the original space, unnecessary particle labels and all. In two dimensions the situation is more difficult, but there is nothing preventing a practitioner from ``unreducing'' the configuration space to that of distinguishable particles before applying the tools needed to deal with an Aharonov-Bohm effect. We ourselves approached this topologically in~\SecRef{sec:distanyons}. Previous treatments sometimes involved a universal covering space, but this special case is more easily handled through a composite flux or or Chern-Simons field, instead~\cite{Morandi1992,Lerda1992}. Most importantly for our argument, these tools do not concern themselves with whether the particles are intrinsically identical. Identity can be imposed in parallel or in concert with these other tools.

We can also see evidence for the work that the covering space implicitly performs by asking why issues of curvature hardly arise. Indeed, while the symmetrized space of two particles in two dimensions is, in fact, a cone, and thus has a non-trivial curvature at the removed origin, the particular form of the cone belongs to a family of special cases. In these special cases, the reduced range of the azimuthal angle is a whole fraction of~$2\pi$~\cite{Dowker1977,DeserJackiw1988}. These special cases are equivalent to solving the problem in a flat space, and then imposing an n-fold symmetry condition on the probability density: that is, requiring that~$\abs{\Psi(\rho,\theta+2\pi/n)}=\abs{\Psi(\rho,\theta)}$. For the relative space of two particles in two dimensions, this is seen to be equivalent to the classical way of deriving the symmetrization postulate.

The basic idea of approaching identical particles from~Refs.~\cite{NeoriGoyal2013,Goyal2015} completely sidesteps any criticism of the legitimacy of using particle labels, and does not result in a curved configuration space. Instead, it assumes that a certain problem has been solved for distinguishable particles, that is, with particle labels, and sees what consistent ways there are to combine these solutions into a solution to the identified system. So it is compatible both with the idea that particle labels are necessary, but then suppressed, and with the idea that there is a useful covering space using particle labels which can be harnessed for the purpose of also dealing with identical particles.

To conclude, we challenged the fundamental impetus for the reduced configuration space approach, namely that quantization after symmetrizing the space is more natural because it removes the need for using unphysical labels, by showing that successful quantization in the reduced space requires the addition of redundant structure, so that one unmeasurable family of parameters is exchanged with another. Reflecting on the results in~\ChapRef{ch:distidentanyons}, we found that this is also not necessary in order to provide the additional richness that is anyonic behavior.
\chapter{Conclusion}\label{ch:dissconc}
We began our research asking whether the Symmetrization Postulate was valid.
The results in~\citet{NeoriGoyal2013}, and their generalization in~\citet{Goyal2015}, only leave room for bosons and fermions.

On its surface, this seemed to run counter to the existence of anyons, which interpolate between bosons and fermions. Since there are theoretical models as well as physical evidence standing behind anyons, we were not satisfied to simply rule them out. Instead, we investigated the issue of identical particles in two dimensions. We found that the apparent paradox stems from the aforementioned assumption that anyons must be identical, when this is demonstrably not the case. We discovered a novel approach for quantization in general multiply-connected spaces using the fundamental groupoid, which is based upon the concatenation of classes of paths, rather than loops. This allowed us to provide consistent topological amplitudes for exchanges of distinguishable particles, which lead us to distinguishable anyons. Using the operational results from~Ref.~\cite{Goyal2015}, we showed that the resulting probability amplitudes can then be combined to form identical anyons.

Looking to future work, there are at least two important issues that have not yet been resolved, and await further developments.

The first is the question of paraparticles. While the operational formalism, with the assumptions we added to incorporate identical particles into it, rules them out, it would be interesting to see what would be needed to change in order to allow for them. One possibility is to weaken atomic measurements so we would allow additional parameters, which cannot be directly measured. That may result in vectors of numbers needed to express a measurement, rather than a single parameter, leading to the possibility of unitary operators instead of numerical amplitudes. Similar work will be needed to incorporate measurable internal degrees of freedom such as spin, so it may well be a good exercise even if it does not result in paraparticles. Additionally, generalizing the groupoid results to multiple-dimensional representations may have applications to non-abelian anyons, of the form used in topological quantum computing~\cite{NayakSimonSternFreedmanDas2008}.

Another issue is the removal of incidence points. This is a recurring problem in the literature. The operational results are general, meaning that they should apply even to particles with small, discrete configuration spaces. In those situations, particle incidence become a significant part of the space, and a serious challenge for bosons, which should be able to occupy them in an unlimited manner. We hope future work will resolve these issues.

Finally, a potentially minor point regarding our anyon proofs: strictly speaking, we only discussed systems with two anyons, and have not generalized to any number of particles. However, since the operational results do generalize, and since the braid group, which forms the foundation for anyonic behavior, is generated by individual exchanges, we do not expect there to be serious challenges when more than two particles are involved.
%
%
%
\begingroup
\renewcommand{\thechapter}{A}
\chapter*{Appendix: Groupoid Source and Target Proof}\label{ch:grpdappdx}
\addcontentsline{toc}{chapter}{Appendix: Groupoid Source and Target Proof}
Let us show that any groupoid will have~$\Source$ and~$\Target$ functions as described in~\ChapRef{ch:groupoidresults}.

Minimally, a groupoid is a set with a sometimes-defined operation, satisfying:
\begin{enumerate}[label=(\Roman*)]
\item If~$(ab)c$ is defined then, in particular,~$ab$ is defined; futhermore,~$a(bc)$ is defined, so that~$bc$ is also defined. Finally,~$(ab)c=a(bc)$.
\item For each~$a$ there is always a~$a^{-1}$ such that~$aa^{-1}$ and~$a^{-1}a$ are defined, and:\label{it:inverse}
\begin{enumerate}[label=(\roman*)]
\item $(a^{-1}a)b=b$ when~$ab$ is defined\label{it:leftinverse}
\item $(aa^{-1})b=b$ when~$a^{-1}b$ is defined\label{it:leftdoubleinverse}
\item $b(aa^{-1})=b$ when~$ba$ is defined\label{it:rightinverse}
\item $b(a^{-1}a)=b$ when~$ba^{-1}$ is defined\label{it:rightdoubleinverse}
\end{enumerate}
\end{enumerate}
If we replace~\ref{it:inverse} by a weaker condition: 
\begin{enumerate}[label=(\Roman*$'$)]
\setcounter{enumi}{1}
\item For each~$a$ there are~$a^{-1}_{\text{L}}$ and~$a^{-1}_{\text{R}}$ such that~$a^{-1}_{\text{L}} a$ and~$aa^{-1}_{\text{R}}$ are defined, and:
\begin{enumerate}[label=(\roman*$'$)]
\item $(a^{-1}_{\text{L}} a)b=b$ when~$ab$ is defined
\item$b(aa^{-1}_{\text{R}})=b$ when~$ba$ is defined
\end{enumerate}
\end{enumerate}
we can retrieve it as follows. We note that:
\begin{equation}
a^{-1}_{\text{L}}=a^{-1}_{\text{L}} (aa^{-1}_{\text{R}})=(a^{-1}_{\text{L}} a)a^{-1}_{\text{R}}=a^{-1}_{\text{R}}\text{,}
\end{equation}
so~$a^{-1}_{\text{L}}=a^{-1}_{\text{R}}$ and we can write both as~$a^{-1}$, giving us~\ref{it:leftinverse} and~\ref{it:rightinverse}. Now, there also exists a~$(a^{-1})^{-1}$, and we find that:
\begin{equation}
a=a(a^{-1}(a^{-1})^{-1})=(a a^{-1})(a^{-1})^{-1}=(a^{-1})^{-1}\text{,}
\end{equation}
and we get~\ref{it:leftdoubleinverse} and~\ref{it:rightdoubleinverse}.~$\blacksquare$

Now, let us define:
\begin{align}
\Target(a)&=\{b: ab\text{ is defined}\}\\
\Source(b)&=\{c: c^{-1}b\text{ is defined}\}=\Target(b^{-1})\text{.}
\end{align}
We can prove that~$\Target(a)=\Source(b)$ iff~$ab$ is defined. Note that~$a^{-1}=\Target(a)$, since~$aa^{-1}$ is defined, and~$b\in\Source(b)$, since~$b^{-1}b$ is defined. Then:
\begin{itemize}
\item[$\Rightarrow$:] If~$\Target(a)=\Source(b)$, then~$b\in\Target(a)$, so~$ab$ is defined.
\item[$\Leftarrow$:] If~$ab$ is defined, then:
\begin{itemize}
\item[$\subseteq$:] If~$c\in\Target(a)$ then~$ac$ is defined, so~$a(cc^{-1})=a$ is defined, and:
\begin{equation}
ab=(a(cc^{-1}))b=a((cc^{-1})b)=a(c(c^{-1}b))\text{,}
\end{equation}
meaning~$c^{-1}b$ is defined, so~$c\in\Source(b)$.
\end{itemize}
Therefore,~$\Target(a)=\Source(b)$.
\item[$\supseteq$:] If~$c\in\Source(b)$ then~$c^{-1}b$ is defined, so~$(cc^{-1})b=b$ is defined, and:
\begin{equation}
ab=a((cc^{-1})b)=(a(cc^{-1}))b=((ac)c^{-1})b\text{,}
\end{equation}
meaning~$ac$ is defined, so~$c\in\Target(a)$.
\end{itemize}
And we are done.~$\blacksquare$
\endgroup
\bibliographystyle{unsrtnat}
\bibliography{NeoriKlilHDissertation}

\begin{thebibliography}{109}
\providecommand{\natexlab}[1]{#1}
\providecommand{\url}[1]{\texttt{#1}}
\expandafter\ifx\csname urlstyle\endcsname\relax
  \providecommand{\doi}[1]{doi: #1}\else
  \providecommand{\doi}{doi: \begingroup \urlstyle{rm}\Url}\fi

\bibitem[Goyal(2015)]{Goyal2015}
P.~Goyal.
\newblock {Informational approach to the quantum symmetrization postulate}.
\newblock \emph{New J. Phys.}, 17\penalty0 (1):\penalty0 013043, 2015.
\newblock \doi{10.1088/1367-2630/17/1/013043}.
\newblock URL \url{http://dx.doi.org/10.1088/1367-2630/17/1/013043}.
\newblock Preprint: \url{http://arxiv.org/abs/1309.0478}.

\bibitem[Neori and Goyal(2013)]{NeoriGoyal2013}
K.~H. Neori and P.~Goyal.
\newblock {On the origin of the quantum rules for identical particles}.
\newblock \emph{AIP Conf. Proc.}, 1553\penalty0 (1):\penalty0 220--227, 2013.
\newblock \doi{10.1063/1.4820003}.
\newblock URL \url{http://dx.doi.org/10.1063/1.4820003}.
\newblock Preprint: \url{http://arxiv.org/abs/1211.1994}.

\bibitem[Souriau(1967)]{Souriau1967}
J.-M. Souriau.
\newblock {Quantification g{\'e}om{\'e}trique. {A}pplications}.
\newblock \emph{Ann. Inst. H. P. Sec. A}, 6\penalty0 (4):\penalty0 311--341,
  1967.
\newblock URL \url{http://eudml.org/doc/75556}.

\bibitem[Laidlaw and DeWitt-Morette(1971)]{LaidlawDeWitt1971}
M.~G.~G. Laidlaw and C.~DeWitt-Morette.
\newblock {Feynman functional integrals for systems of indistinguishable
  particles}.
\newblock \emph{Phys. Rev. D}, 3\penalty0 (6):\penalty0 1375--1378, 1971.
\newblock \doi{10.1103/PhysRevD.3.1375}.
\newblock URL \url{http://dx.doi.org/10.1103/PhysRevD.3.1375}.

\bibitem[Leinaas and Myrheim(1977)]{LeinaasMyrheim1977}
J.~M. Leinaas and J.~Myrheim.
\newblock {On the theory of identical particles}.
\newblock \emph{Nuovo Cimento B}, 37\penalty0 (1):\penalty0 1--23, 1977.
\newblock \doi{10.1007/BF02727953}.
\newblock URL \url{http://dx.doi.org/10.1007/BF02727953}.

\bibitem[Neori and Goyal(2015{\natexlab{a}})]{NeoriGoyal2015a}
K.~H. Neori and P.~Goyal.
\newblock {Anyons in the operational formalism}.
\newblock \emph{AIP Conf. Proc.}, 1641:\penalty0 189--196, 2015{\natexlab{a}}.
\newblock \doi{10.1063/1.4905978}.
\newblock URL \url{http://dx.doi.org/10.1063/1.4905978}.
\newblock Preprint: \url{http://arxiv.org/abs/1410.8080}.

\bibitem[Neori and Goyal(2015{\natexlab{b}})]{NeoriGoyal2015b}
K.~H. Neori and P.~Goyal.
\newblock {Fundamental groupoids in quantum mechanics: a new approach to
  quantization in multiply-connected spaces}.
\newblock Submitted to J. Phys. A. Preprint:
  \url{http://arxiv.org/abs/1501.00231}, 2015{\natexlab{b}}.

\bibitem[Neori and Goyal(2015{\natexlab{c}})]{NeoriGoyal2015c}
K.~H. Neori and P.~Goyal.
\newblock {Distinguishable and identical anyons}.
\newblock In preparation, 2015{\natexlab{c}}.

\bibitem[Dirac(1926)]{Dirac1926}
P.~A.~M. Dirac.
\newblock {On the Theory of Quantum Mechanics}.
\newblock \emph{Proc. Roy. Soc. Lon. Ser. A}, 112\penalty0 (762):\penalty0
  661--677, 1926.
\newblock \doi{10.1098/rspa.1926.0133}.
\newblock URL \url{http://dx.doi.org/10.1098/rspa.1926.0133}.

\bibitem[Heisenberg(1926)]{Heisenberg1926}
W.~Heisenberg.
\newblock {Mehrk{\"o}rperproblem und Resonanz in der Quantenmechanik}.
\newblock \emph{Z. Phys.}, 38\penalty0 (6--7):\penalty0 411--426, 1926.
\newblock ISSN 0044-3328.
\newblock \doi{10.1007/BF01397160}.
\newblock URL \url{http://dx.doi.org/10.1007/BF01397160}.

\bibitem[Wigner(1927{\natexlab{a}})]{Wigner1927a}
E.~P. Wigner.
\newblock {{\"U}ber nicht kombinierende Terme in der neueren Quantentheorie}.
\newblock \emph{Z. Phys.}, 40\penalty0 (7):\penalty0 492--500,
  1927{\natexlab{a}}.
\newblock \doi{10.1007/BF01440826}.
\newblock URL \url{http://dx.doi.org/10.1007/BF01440826}.

\bibitem[Wigner(1927{\natexlab{b}})]{Wigner1927b}
E.~P. Wigner.
\newblock {{\"U}ber nicht kombinierende Terme in der neueren Quantentheorie. II
  Teil}.
\newblock \emph{Z. Phys.}, 40\penalty0 (11):\penalty0 883--892,
  1927{\natexlab{b}}.
\newblock \doi{10.1007/BF01390906}.
\newblock URL \url{https://dx.doi.org/10.1007/BF01390906}.

\bibitem[Pauli(1980)]{Pauli1980}
W.~Pauli.
\newblock \emph{{General Principles of Quantum Mechanics}}.
\newblock Springer-Verlag, 1980.

\bibitem[Dirac(1930)]{Dirac1930}
P.~A.~M. Dirac.
\newblock \emph{{Principles of Quantum Mechanics}}.
\newblock Oxford University Press, 1 edition, 1930.

\bibitem[Gentile(1940)]{Gentile1940}
G.~Gentile, Jr.
\newblock {Osservazioni sopra le statistiche intermedie}.
\newblock \emph{Nuovo Cimento}, 17\penalty0 (10):\penalty0 493--497, 1940.
\newblock \doi{10.1007/BF02960187}.
\newblock URL \url{http://dx.doi.org/10.1007/BF02960187}.

\bibitem[Green(1953)]{Green1953}
H.~S. Green.
\newblock {A generalized method of field quantization}.
\newblock \emph{Phys. Rev.}, 90\penalty0 (2):\penalty0 270--273, 1953.
\newblock \doi{10.1103/PhysRev.90.270}.
\newblock URL \url{http://dx.doi.org/10.1103/PhysRev.90.270}.

\bibitem[Jauch(1960)]{Jauch1960}
J.~M. Jauch.
\newblock {Systems of observables in quantum mechanics}.
\newblock \emph{Helv. Phys. Acta}, 33\penalty0 (8):\penalty0 711--726, 1960.
\newblock \doi{10.5169/seals-113092}.
\newblock URL \url{http://dx.doi.org/10.5169/seals-113092}.

\bibitem[Jauch and Misra(1961)]{JauchMisra1961}
J.~M. Jauch and B.~Misra.
\newblock {Supersymmetries and essential observables}.
\newblock \emph{Helv. Phys. Acta}, 34\penalty0 (6--7):\penalty0 699--709, 1961.
\newblock \doi{10.5169/seals-113193}.
\newblock URL \url{http://dx.doi.org/10.5169/seals-113193}.

\bibitem[Galindo et~al.(1962)Galindo, Morales, and
  Nu{\~n}ez-Lagos]{Galindoetal1962}
A.~Galindo, A.~Morales, and R.~Nu{\~n}ez-Lagos.
\newblock {Superselection principle and pure states of {$n$}-identical
  particles}.
\newblock \emph{J. Math. Phys.}, 3\penalty0 (2):\penalty0 324--328, 1962.
\newblock \doi{10.1063/1.1703808}.
\newblock URL \url{http://dx.doi.org/10.1063/1.1703808}.

\bibitem[Pandres(1962)]{Pandres1962}
D.~Pandres, Jr.
\newblock {Derivation of admissibility conditions for wave functions from
  general quantum-mechanical principles}.
\newblock \emph{J. Math. Phys.}, 3\penalty0 (2):\penalty0 305--308, 1962.
\newblock \doi{10.1063/1.1703804}.
\newblock URL \url{http://dx.doi.org/10.1063/1.1703804}.

\bibitem[Messiah and Greenberg(1964)]{MessiahGreenberg1964}
A.~M.~L. Messiah and O.~W. Greenberg.
\newblock {Symmetrization postulate and its experimental foundation}.
\newblock \emph{Phys. Rev.}, 136\penalty0 (1B):\penalty0 248--267, 1964.
\newblock \doi{10.1103/PhysRev.136.B248}.
\newblock URL \url{http://dx.doi.org/10.1103/PhysRev.136.B248}.

\bibitem[Greenberg(1964)]{Greenberg1964}
O.~W. Greenberg.
\newblock {Spin and unitary-spin independence in a paraquark model of baryons
  and mesons}.
\newblock \emph{Phys. Rev. Lett.}, 13\penalty0 (20):\penalty0 598--602, 1964.
\newblock \doi{10.1103/PhysRevLett.13.598}.
\newblock URL \url{http://dx.doi.org/10.1103/PhysRevLett.13.598}.

\bibitem[Greenberg(2015)]{Greenberg2015}
O.~W. Greenberg.
\newblock {The origin of quark color}.
\newblock \emph{Phys. Today}, 68\penalty0 (1):\penalty0 33--37, 2015.
\newblock \doi{10.1063/PT.3.2655}.
\newblock URL \url{http://dx.doi.org/10.1063/PT.3.2655}.

\bibitem[Greenberg and Zwanziger(1966)]{GreenbergZwanziger1966}
O.~W. Greenberg and D.~Zwanziger.
\newblock {Saturation in triplet models of hadrons}.
\newblock \emph{Phys. Rev.}, 150\penalty0 (4):\penalty0 1177--1180, 1966.
\newblock \doi{10.1103/PhysRev.150.1177}.
\newblock URL \url{http://dx.doi.org/10.1103/PhysRev.150.1177}.

\bibitem[Greenberg(1991)]{Greenberg1991}
O.~W. Greenberg.
\newblock {Particles with small violations of Fermi or Bose statistics}.
\newblock \emph{Phys. Rev. D}, 43\penalty0 (12):\penalty0 4111--4120, 1991.
\newblock \doi{10.1103/PhysRevD.43.4111}.
\newblock URL \url{http://dx.doi.org/10.1103/PhysRevD.43.4111}.

\bibitem[Tikochinsky(1988{\natexlab{a}})]{Tikochinsky1988}
Y.~Tikochinsky.
\newblock {From single-particle states to {$n$}-particle states}.
\newblock \emph{Phys. Rev. A}, 37\penalty0 (9):\penalty0 3553--3556,
  1988{\natexlab{a}}.
\newblock \doi{10.1103/PhysRevA.37.3553}.
\newblock URL \url{http://dx.doi.org/10.1103/PhysRevA.37.3553}.

\bibitem[Feynman(1948)]{Feynman1948}
R.~P. Feynman.
\newblock {Space-time approach to non-relativistic quantum mechanics}.
\newblock \emph{Rev. Mod. Phys.}, 20\penalty0 (2):\penalty0 367--387, 1948.
\newblock \doi{10.1103/RevModPhys.20.367}.
\newblock URL \url{http://dx.doi.org/10.1103/RevModPhys.20.367}.

\bibitem[Goyal et~al.(2010)Goyal, Knuth, and Skilling]{GoyalKnuthSkilling2010}
P.~Goyal, K.~H. Knuth, and J.~Skilling.
\newblock {Origins of complex quantum amplitudes and {Feynman}'s rules}.
\newblock \emph{Phys. Rev. A}, 81\penalty0 (2):\penalty0 022109, 2010.
\newblock \doi{10.1103/PhysRevA.81.022109}.
\newblock URL \url{http://dx.doi.org/10.1103/PhysRevA.81.022109}.
\newblock Preprint: \url{http://arxiv.org/abs/0907.0909}.

\bibitem[Mirman(1973)]{Mirman1973}
R.~Mirman.
\newblock {Experimental meaning of the concept of identical particles}.
\newblock \emph{Nuovo Cimento B}, 18\penalty0 (1):\penalty0 110--122, 1973.
\newblock \doi{10.1007/BF02832643}.
\newblock URL \url{http://dx.doi.org/10.1007/BF02832643}.

\bibitem[Wilczek(1982)]{Wilczek1982b}
F.~Wilczek.
\newblock {Quantum mechanics of fractional-spin particles}.
\newblock \emph{Phys. Rev. Lett.}, 49\penalty0 (14):\penalty0 957--959, 1982.
\newblock \doi{10.1103/PhysRevLett.49.957}.
\newblock URL \url{http://dx.doi.org/10.1103/PhysRevLett.49.957}.

\bibitem[Girardeau(1965)]{Girardeau1965}
M.~D. Girardeau.
\newblock {Permutation symmetry of many-particle wave functions}.
\newblock \emph{Phys. Rev.}, 139\penalty0 (2B):\penalty0 B500--B508, 1965.
\newblock \doi{10.1103/PhysRev.139.B500}.
\newblock URL \url{http://dx.doi.org/10.1103/PhysRev.139.B500}.

\bibitem[Tsui et~al.(1982)Tsui, Stormer, and Gossard]{TsuiStormerGossard1982}
D.~C. Tsui, H.~L. Stormer, and A.~C. Gossard.
\newblock {Two-dimensional magnetotransport in the extreme quantum limit}.
\newblock \emph{Phys. Rev. Lett.}, 48\penalty0 (22):\penalty0 1559--1562, 1982.
\newblock \doi{10.1103/PhysRevLett.48.1559}.
\newblock URL \url{http://dx.doi.org/10.1103/PhysRevLett.48.1559}.

\bibitem[Laughlin(1983{\natexlab{a}})]{Laughlin1983a}
R.~B. Laughlin.
\newblock {Anomalous quantum Hall effect: an incompressible quantum fluid with
  fractionally charged excitations}.
\newblock \emph{Phys. Rev. Lett.}, 50\penalty0 (18):\penalty0 1395--1398,
  1983{\natexlab{a}}.
\newblock \doi{10.1103/PhysRevLett.50.1395}.
\newblock URL \url{http://dx.doi.org/10.1103/PhysRevLett.50.1395}.

\bibitem[Arovas et~al.(1984)Arovas, Schrieffer, and
  Wilczek]{ArovasSchriefferWilczek1984}
D.~Arovas, J.~R. Schrieffer, and F.~Wilczek.
\newblock {Fractional statistics and the quantum Hall effect}.
\newblock \emph{Phys. Rev. Lett.}, 53\penalty0 (7):\penalty0 722--723, 1984.
\newblock \doi{10.1103/PhysRevLett.53.722}.
\newblock URL \url{http://dx.doi.org/10.1103/PhysRevLett.53.722}.

\bibitem[Stormer(1999)]{Stormer1999}
H.~L. Stormer.
\newblock {Nobel Lecture: The fractional quantum Hall effect}.
\newblock \emph{Rev. Mod. Phys.}, 71\penalty0 (4):\penalty0 875--889, 1999.
\newblock \doi{10.1103/RevModPhys.71.875}.
\newblock URL \url{http://dx.doi.org/10.1103/RevModPhys.71.875}.

\bibitem[Laughlin(1999)]{Laughlin1999}
R.~B. Laughlin.
\newblock {Nobel Lecture: Fractional quantization}.
\newblock \emph{Rev. Mod. Phys.}, 71\penalty0 (4):\penalty0 863--874, 1999.
\newblock \doi{10.1103/RevModPhys.71.863}.
\newblock URL \url{http://dx.doi.org/10.1103/RevModPhys.71.863}.

\bibitem[Camino et~al.(2005)Camino, Zhou, and Goldman]{CaminoZhouGoldman2005}
F.~E. Camino, W.~Zhou, and V.~J. Goldman.
\newblock {Realization of a Laughlin quasiparticle interferometer: observation
  of fractional statistics}.
\newblock \emph{Phys. Rev. B}, 72\penalty0 (7):\penalty0 075342, 2005.
\newblock \doi{10.1103/PhysRevB.72.075342}.
\newblock URL \url{http://dx.doi.org/10.1103/PhysRevB.72.075342}.

\bibitem[Kim et~al.(2005)Kim, Lawler, Vishveshwara, and
  Fradkin]{KimLawlerVishveshwaraFradkin2005}
E.-A. Kim, M.~Lawler, S.~Vishveshwara, and E.~Fradkin.
\newblock {Signatures of fractional statistics in noise experiments in quantum
  hall fluids}.
\newblock \emph{Phys. Rev. Lett.}, 95\penalty0 (17):\penalty0 176402, 2005.
\newblock \doi{10.1103/PhysRevLett.95.176402}.
\newblock URL \url{http://dx.doi.org/10.1103/PhysRevLett.95.176402}.

\bibitem[Goldin et~al.(1985)Goldin, Menikoff, and
  Sharp]{GoldinMenikoffSharp1985}
G.~A. Goldin, R.~Menikoff, and D.~H. Sharp.
\newblock {Comments on ``{General theory for quantum statistics in two
  dimensions}''}.
\newblock \emph{Phys. Rev. Lett.}, 54\penalty0 (11):\penalty0 603--603, 1985.
\newblock \doi{10.1103/PhysRevLett.54.603}.
\newblock URL \url{http://dx.doi.org/10.1103/PhysRevLett.54.603}.

\bibitem[Lerda(1992)]{Lerda1992}
A.~Lerda.
\newblock \emph{{Anyons: Quantum Mechanics of Particles with Fractional
  Statistics}}.
\newblock Springer-Verlag, 1992.

\bibitem[Morandi(1992)]{Morandi1992}
G.~Morandi.
\newblock \emph{{The Role of Topology in Classical and Quantum Physics}}.
\newblock Springer-Verlag, 1992.

\bibitem[Zee(1995)]{Zee1995}
A.~Zee.
\newblock {Quantum Hall fluids}.
\newblock In H.~B. Geyer, editor, \emph{{Field Theory, Topology and Condensed
  Matter Physics}}, volume 456 of \emph{Lecture Notes in Physics}, pages
  99--153. Springer Berlin Heidelberg, 1995.
\newblock \doi{10.1007/BFb0113369}.
\newblock URL \url{http://dx.doi.org/10.1007/BFb0113369}.

\bibitem[Sudarshan et~al.(1988)Sudarshan, Imbo, and
  Shah-Imbo]{SudarshanImboShahImbo1988}
E.~C.~G. Sudarshan, T.~D. Imbo, and C.~Shah-Imbo.
\newblock {Topological and algebraic aspects of quantization: symmetries and
  statistics}.
\newblock \emph{Ann. Inst. H. P.}, 49\penalty0 (3):\penalty0 387--396, 1988.
\newblock URL \url{http://eudml.org/doc/76427}.

\bibitem[Schulman(1993)]{Schulman1993}
L.~S. Schulman.
\newblock {Homotopy and statistics: of spin and diagonals}.
\newblock \emph{J. Phys. A}, 26\penalty0 (20):\penalty0 L1093--L1097, 1993.
\newblock \doi{10.1088/0305-4470/26/20/006}.
\newblock URL \url{http://dx.doi.org/10.1088/0305-4470/26/20/006}.

\bibitem[Wu(1984)]{Wu1984}
Y.-S. Wu.
\newblock {General theory for quantum statistics in two-dimensions}.
\newblock \emph{Phys. Rev. Lett.}, 52\penalty0 (24):\penalty0 2103--2106, 1984.
\newblock \doi{10.1103/PhysRevLett.52.2103}.
\newblock URL \url{http://dx.doi.org/10.1103/PhysRevLett.52.2103}.

\bibitem[Stern(2008)]{Stern2008}
A.~Stern.
\newblock {Anyons and the quantum {Hall} effect --- {A} pedagogical review}.
\newblock \emph{Ann. Phys. (N. Y.)}, 323\penalty0 (1):\penalty0 204--249, 2008.
\newblock \doi{10.1016/j.aop.2007.10.008}.
\newblock URL \url{http://dx.doi.org/10.1016/j.aop.2007.10.008}.
\newblock Preprint: \url{http://arxiv.org/abs/0711.4697}.

\bibitem[Born(1926)]{Born1926}
M.~Born.
\newblock {Zur Quantenmechanik der Sto{\ss}vorg{\"a}nge}.
\newblock \emph{Z. Phys.}, 37\penalty0 (12):\penalty0 863--867, 1926.
\newblock ISSN 0044-3328.
\newblock \doi{10.1007/BF01397477}.
\newblock URL \url{http://dx.doi.org/10.1007/BF01397477}.

\bibitem[Bohr(1937)]{Bohr1937}
N.~Bohr.
\newblock {Causality and Complementarity}.
\newblock \emph{Philos. Sci.}, 4\penalty0 (3):\penalty0 289--298, 1937.
\newblock ISSN 00318248.
\newblock \doi{10.1086/286465}.
\newblock URL \url{http://dx.doi.org/10.1086/286465}.

\bibitem[Einstein et~al.(1935)Einstein, Podolsky, and Rosen]{EPR1935}
A.~Einstein, B.~Podolsky, and N.~Rosen.
\newblock {Can quantum-mechanical description of physical reality be considered
  complete?}
\newblock \emph{Phys. Rev.}, 47\penalty0 (10):\penalty0 777--780, 1935.
\newblock \doi{10.1103/PhysRev.47.777}.
\newblock URL \url{http://dx.doi.org/10.1103/PhysRev.47.777}.

\bibitem[Pauli(1925)]{Pauli1925}
W.~Pauli.
\newblock {{\"U}ber den Zusammenhang des Abschlusses der Elektronengruppen im
  Atom mit der Komplexstruktur der Spektren}.
\newblock \emph{Z. Phys.}, 31\penalty0 (1):\penalty0 765--783, 1925.
\newblock \doi{10.1007/BF02980631}.
\newblock URL \url{http://dx.doi.org/10.1007/BF02980631}.

\bibitem[Bose(1924)]{Bose1924}
S.~N. Bose.
\newblock {Plancks Gesetz und Lichtquantenhypothese}.
\newblock \emph{Z. Phys.}, 26\penalty0 (1):\penalty0 178--181, 1924.
\newblock \doi{10.1007/BF01327326}.
\newblock URL \url{http://dx.doi.org/10.1007/BF01327326}.

\bibitem[Einstein(1924)]{Einstein1924}
A.~Einstein.
\newblock {Quantentheorie des einatomigen idealen {Gases}}.
\newblock \emph{Sitzungsberichte Preussiche Akademie der Wissenschaften,
  Physikalisch--Mathematische Klasse}, pages 261--67, 1924.
\newblock URL
  \url{http://web.physik.rwth-aachen.de/~meden/boseeinstein/einstein1924.pdf}.

\bibitem[Bohm(1952{\natexlab{a}})]{Bohm1952a}
D.~Bohm.
\newblock {A suggested interpretation of the quantum theory in terms of
  ``hidden'' variables. I}.
\newblock \emph{Phys. Rev.}, 85\penalty0 (2):\penalty0 166--179,
  1952{\natexlab{a}}.
\newblock \doi{10.1103/PhysRev.85.166}.
\newblock URL \url{http://dx.doi.org/10.1103/PhysRev.85.166}.

\bibitem[Bohm(1952{\natexlab{b}})]{Bohm1952b}
D.~Bohm.
\newblock {A suggested interpretation of the quantum theory in terms of
  ``hidden'' variables. II}.
\newblock \emph{Phys. Rev.}, 85\penalty0 (2):\penalty0 180--193,
  1952{\natexlab{b}}.
\newblock \doi{10.1103/PhysRev.85.180}.
\newblock URL \url{http://dx.doi.org/10.1103/PhysRev.85.180}.

\bibitem[Nelson(1966)]{Nelson1966}
E.~Nelson.
\newblock {Derivation of the Schr{\"o}dinger equation from Newtonian
  mechanics}.
\newblock \emph{Phys. Rev.}, 150\penalty0 (4):\penalty0 1079--1085, 1966.
\newblock \doi{10.1103/PhysRev.150.1079}.
\newblock URL \url{http://dx.doi.org/10.1103/PhysRev.150.1079}.

\bibitem[Bell(1964)]{Bell1964}
J.~S. Bell.
\newblock {On the Einstein-Podolsky-Rosen paradox}.
\newblock \emph{Physics (N.Y.)}, 1\penalty0 (3):\penalty0 195--200, 1964.
\newblock URL
  \url{http://philoscience.unibe.ch/documents/TexteHS10/bell1964epr.pdf}.

\bibitem[Kochen and Specker(1968)]{KochenSpecker1968}
S.~B. Kochen and E.~Specker.
\newblock {The problem of hidden variables in quantum mechanics}.
\newblock \emph{Indiana Univ. Math. J.}, 17\penalty0 (1):\penalty0 59--87,
  1968.
\newblock ISSN 0022-2518.
\newblock \doi{10.1007/978-94-010-1795-4_17}.
\newblock URL
  \url{http://www.iumj.indiana.edu/IUMJ/fulltext.php?year=1968&volume=17&artid=17004}.

\bibitem[Hardy(2001)]{Hardy2001}
L.~Hardy.
\newblock {Quantum theory from five reasonable axioms}.
\newblock Unpublished. Preprint: \url{http://arxiv.org/abs/quant-ph/0101012},
  2001.

\bibitem[Caticha(2011)]{Caticha2011}
A.~Caticha.
\newblock {Entropic dynamics, time and quantum theory}.
\newblock \emph{J. Phys. A}, 44\penalty0 (22):\penalty0 225303, 2011.
\newblock \doi{10.1088/1751-8113/44/22/225303}.
\newblock URL \url{http://dx.doi.org/10.1088/1751-8113/44/22/225303}.

\bibitem[Tikochinsky(1988{\natexlab{b}})]{Tikochinsky1988a}
Y.~Tikochinsky.
\newblock {On the generalized multiplication and addition of complex numbers}.
\newblock \emph{J. Math. Phys.}, 29\penalty0 (2):\penalty0 398--399,
  1988{\natexlab{b}}.
\newblock \doi{10.1063/1.528026}.
\newblock URL \url{http://dx.doi.org/10.1063/1.528026}.

\bibitem[Tikochinsky(1988{\natexlab{c}})]{Tikochinsky1988b}
Y.~Tikochinsky.
\newblock {Feynman rules for probability amplitudes}.
\newblock \emph{Int. J. Theor. Phys.}, 27\penalty0 (5):\penalty0 543--549,
  1988{\natexlab{c}}.
\newblock ISSN 0020-7748.
\newblock \doi{10.1007/BF00668836}.
\newblock URL \url{http://dx.doi.org/10.1007/BF00668836}.

\bibitem[Caticha(1998)]{Caticha1998}
A.~Caticha.
\newblock {Consistency, amplitudes, and probabilities in quantum theory}.
\newblock \emph{Phys. Rev. A}, 57\penalty0 (3):\penalty0 1572--1582, 1998.
\newblock \doi{10.1103/PhysRevA.57.1572}.
\newblock URL \url{http://dx.doi.org/10.1103/PhysRevA.57.1572}.

\bibitem[Cox(1946)]{Cox1946}
R.~T. Cox.
\newblock {Probability, frequency and reasonable expectation}.
\newblock \emph{Am. J. Phys.}, 14\penalty0 (1):\penalty0 1--13, 1946.
\newblock \doi{10.1119/1.1990764}.
\newblock URL \url{http://dx.doi.org/10.1119/1.1990764}.

\bibitem[Tikochinsky and Gull(2000)]{TikochinskyGull2000}
Y.~Tikochinsky and S.~F. Gull.
\newblock {Consistency, amplitudes and probabilities in quantum theory}.
\newblock \emph{J. Phys. A}, 33\penalty0 (31):\penalty0 5615--5618, 2000.
\newblock \doi{10.1088/0305-4470/33/31/314}.
\newblock URL \url{http://dx.doi.org/10.1088/0305-4470/33/31/314}.

\bibitem[Gibbs(1960)]{Gibbs1960}
J.~W. Gibbs.
\newblock \emph{{Elementary Principles in Statistical Mechanics}}, chapter~XV.
\newblock Dover Publications, Inc, New York, New York, 1960.

\bibitem[Bacciagaluppi and Valentini(2009)]{BacciagaluppiValentini2009}
G.~Bacciagaluppi and A.~Valentini.
\newblock \emph{{Quantum Theory at the Crossroads: Reconsidering the 1927
  Solvay Conference}}.
\newblock Cambridge University Press, 2009.

\bibitem[Schulman(1971)]{Schulman1971}
L.~S. Schulman.
\newblock {Approximate topologies}.
\newblock \emph{J. Math. Phys.}, 12\penalty0 (2):\penalty0 304--308, 1971.
\newblock \doi{10.1063/1.1665592}.
\newblock URL \url{http://dx.doi.org/10.1063/1.1665592}.

\bibitem[Imbo et~al.(ca. 1989)Imbo, Shah-Imbo, Mahajan, and
  Sudarshan]{ImboShah-ImboMahajanSudarshanca.1990}
T.~D. Imbo, C.~Shah-Imbo, R.~S. Mahajan, and E.~C.~G. Sudarshan.
\newblock {A user's guide to exotic statistics}.
\newblock Preliminary draft. Preprint:
  \url{http://wildcard.ph.utexas.edu/~sudarshan/pub/Unpu_061.pdf}, ca. 1989.

\bibitem[Hall(1879)]{Hall1879}
E.~H. Hall.
\newblock {On a new action of the magnet on electric currents}.
\newblock \emph{Am. J. Math.}, 2\penalty0 (3):\penalty0 287--292, 1879.
\newblock ISSN 00029327.
\newblock \doi{10.2307/2369245}.
\newblock URL \url{http://dx.doi.org/10.2307/2369245}.

\bibitem[von Klitzing et~al.(1980)von Klitzing, Dorda, and
  Pepper]{KlitzingDordaPepper1980}
K.~von Klitzing, G.~Dorda, and M.~Pepper.
\newblock {New method for high-accuracy determination of the fine-structure
  constant based on quantized hall resistance}.
\newblock \emph{Phys. Rev. Lett.}, 45\penalty0 (6):\penalty0 494--497, 1980.
\newblock \doi{10.1103/PhysRevLett.45.494}.
\newblock URL \url{http://dx.doi.org/10.1103/PhysRevLett.45.494}.

\bibitem[Laughlin(1983{\natexlab{b}})]{Laughlin1983b}
R.~B. Laughlin.
\newblock {Quantized motion of three two-dimensional electrons in a strong
  magnetic field}.
\newblock \emph{Phys. Rev. B}, 27\penalty0 (6):\penalty0 3383--3389,
  1983{\natexlab{b}}.
\newblock \doi{10.1103/PhysRevB.27.3383}.
\newblock URL \url{http://dx.doi.org/10.1103/PhysRevB.27.3383}.

\bibitem[Halperin(1984{\natexlab{a}})]{Halperin1984}
B.~I. Halperin.
\newblock {Statistics of quasiparticles and the hierarchy of fractional
  quantized Hall states}.
\newblock \emph{Phys. Rev. Lett.}, 52\penalty0 (18):\penalty0 1583--1586,
  1984{\natexlab{a}}.
\newblock \doi{10.1103/PhysRevLett.52.1583}.
\newblock URL \url{http://dx.doi.org/10.1103/PhysRevLett.52.1583}.

\bibitem[Halperin(1984{\natexlab{b}})]{Halperin1984e}
B.~I. Halperin.
\newblock {Statistics of Quasiparticles and the Hierarchy of Fractional
  Quantized Hall States}.
\newblock \emph{Phys. Rev. Lett.}, 52\penalty0 (26):\penalty0 2390,
  1984{\natexlab{b}}.
\newblock \doi{10.1103/PhysRevLett.52.2390.4}.
\newblock URL \url{http://dx.doi.org/10.1103/PhysRevLett.52.2390.4}.
\newblock Erratum.

\bibitem[Aharonov and Bohm(1959)]{AharonovBohm1959}
Y.~Aharonov and D.~Bohm.
\newblock {Significance of electromagnetic potentials in the quantum theory}.
\newblock \emph{Phys. Rev.}, 115\penalty0 (3):\penalty0 485--491, 1959.
\newblock \doi{10.1103/PhysRev.115.485}.
\newblock URL \url{http://dx.doi.org/10.1103/PhysRev.115.485}.

\bibitem[Ehrenberg and Siday(1949)]{EhrenbergSiday1949}
W.~Ehrenberg and R.~E. Siday.
\newblock {The refractive index in electron optics and the principles of
  dynamics}.
\newblock \emph{Proc. Phys. Soci. Sec. B}, 62\penalty0 (1):\penalty0 8--21,
  1949.
\newblock \doi{10.1088/0370-1301/62/1/303}.
\newblock URL \url{http://dx.doi.org/10.1088/0370-1301/62/1/303}.

\bibitem[Spanier(1981)]{Spanier1981}
E.~H. Spanier.
\newblock \emph{{Algebraic Topology}}.
\newblock Springer-Verlag, 1981.

\bibitem[Singer and Thorpe(1967)]{SingerThorpe1967}
I.~M. Singer and J.~A. Thorpe.
\newblock \emph{{Lecture Notes on Elementary Topology and Geometry}}.
\newblock Scott, Foresman and Company, Glenview, Illinois, 1967.

\bibitem[Higgins(1971)]{Higgins1971}
P.~J. Higgins.
\newblock \emph{{Notes on Categories and Groupoids}}.
\newblock Van Nostrand Reinhold Company, 1971.
\newblock URL \url{http://www.tac.mta.ca/tac/reprints/articles/7/tr7abs.html}.

\bibitem[Schulman(1968)]{Schulman1968}
L.~S. Schulman.
\newblock {A path integral for spin}.
\newblock \emph{Phys. Rev.}, 176\penalty0 (5):\penalty0 1558--1569, 1968.
\newblock \doi{10.1103/PhysRev.176.1558}.
\newblock URL \url{http://dx.doi.org/10.1103/PhysRev.176.1558}.

\bibitem[Schulman(1967)]{Schulman1967}
L.~S. Schulman.
\newblock \emph{{A Path Integral for Spin}}.
\newblock PhD thesis, Princeton University, 1967.

\bibitem[Laidlaw(1971)]{Laidlaw1971}
M.~G.~G. Laidlaw.
\newblock \emph{{Quantum Mechanics in Multiply Connected Spaces}}.
\newblock PhD thesis, University of North Carolina at Chapel Hill, 1971.

\bibitem[Dowker(1972)]{Dowker1972}
J.~S. Dowker.
\newblock {Quantum mechanics and field theory on multiply connected and on
  homogeneous spaces}.
\newblock \emph{J. Phys. A}, 5\penalty0 (7):\penalty0 936--943, 1972.
\newblock \doi{10.1088/0305-4470/5/7/004}.
\newblock URL \url{http://dx.doi.org/10.1088/0305-4470/5/7/004}.

\bibitem[Dowker and Critchley(1977)]{DowkerCritchley1977}
J.~S. Dowker and R.~Critchley.
\newblock {Vacuum stress tensor in an Einstein universe. Finite temperature
  effects}.
\newblock \emph{Phys. Rev. D}, 15\penalty0 (6):\penalty0 1484--1493, 1977.
\newblock \doi{10.1103/PhysRevD.15.1484}.
\newblock URL \url{http://dx.doi.org/10.1103/PhysRevD.15.1484}.

\bibitem[Casati and Guarneri(1979)]{CasatiGuarneri1979}
G.~Casati and I.~Guarneri.
\newblock {The Aharonov-Bohm effect from the `hydrodynamical' viewpoint}.
\newblock \emph{Phys. Rev. Lett.}, 42\penalty0 (24):\penalty0 1579--1581, 1979.
\newblock \doi{10.1103/PhysRevLett.42.1579}.
\newblock URL \url{http://dx.doi.org/10.1103/PhysRevLett.42.1579}.

\bibitem[Gerry and Singh(1979)]{GerrySingh1979}
C.~C. Gerry and V.~A. Singh.
\newblock {Feynman path integral approach to the Aharonov-Bohm effect}.
\newblock \emph{Phys. Rev. D}, 20\penalty0 (10):\penalty0 2550--2554, 1979.
\newblock \doi{10.1103/PhysRevD.20.2550}.
\newblock URL \url{http://dx.doi.org/10.1103/PhysRevD.20.2550}.

\bibitem[Horv{\'a}thy(1980{\natexlab{a}})]{Horvathy1980a}
P.~A. Horv{\'a}thy.
\newblock {Quantisation in multiply connected spaces }.
\newblock \emph{Phys. Lett. A}, 76\penalty0 (1):\penalty0 11--14,
  1980{\natexlab{a}}.
\newblock ISSN 0375-9601.
\newblock \doi{10.1016/0375-9601(80)90133-4}.
\newblock URL \url{http://dx.doi.org/10.1016/0375-9601(80)90133-4}.

\bibitem[Horv{\'a}thy(1980{\natexlab{b}})]{Horvathy1980b}
P.~A. Horv{\'a}thy.
\newblock {Prequantization from path integral viewpoint}.
\newblock In H.-D. Doebner, S.~I. Andersson, and H.~R. Petry, editors,
  \emph{Differential Geometric Methods in Mathematical Physics}, volume 905 of
  \emph{Lecture Notes in Mathematics}, pages 197--206. Springer-Verlag,
  1980{\natexlab{b}}.
\newblock \doi{10.1007/BFb0092438}.
\newblock URL \url{http://dx.doi.org/10.1007/BFb0092438}.

\bibitem[Horv{\'a}thy(1980{\natexlab{c}})]{Horvathy1980c}
P.~A. Horv{\'a}thy.
\newblock {Classical action, the {Wu-Yang} phase factor and prequantization}.
\newblock In P.~L. Garc{\'\i}a, A.~P{\'e}rez-Rend{\'o}n, and J.-M. Souriau,
  editors, \emph{Differential Geometrical Methods in Mathematical Physics},
  volume 836 of \emph{Lecture Notes in Mathematics}, pages 67--90,
  1980{\natexlab{c}}.
\newblock \doi{10.1007/BFb0089727}.
\newblock URL \url{http://dx.doi.org/10.1007/BFb0089727}.

\bibitem[Isham(1984)]{Isham1984}
C.~J. Isham.
\newblock {Topological and global aspects of quantum theory}.
\newblock In B.~S. DeWitt and R.~Stora, editors, \emph{Relativity, Groups and
  Topology II}, volume~40 of \emph{Les Houches}, pages 1059--1290.
  North-Holland, 1984.

\bibitem[Dowker(1985)]{Dowker1985}
J.~S. Dowker.
\newblock {Remarks on non-standard statistics}.
\newblock \emph{J. Phys. A}, 18\penalty0 (18):\penalty0 3521--3530, 1985.
\newblock \doi{10.1088/0305-4470/18/18/015}.
\newblock URL \url{http://dx.doi.org/10.1088/0305-4470/18/18/015}.

\bibitem[Gaveau et~al.(2011)Gaveau, Nounou, and
  Schulman]{GaveauNounouSchulman2011}
B.~Gaveau, A.~M. Nounou, and L.~S. Schulman.
\newblock {Homotopy and path integrals in the time-dependent Aharonov-Bohm
  effect}.
\newblock \emph{Found. Phys.}, 41\penalty0 (9):\penalty0 1462--1474, 2011.
\newblock ISSN 0015-9018.
\newblock \doi{10.1007/s10701-011-9559-y}.
\newblock URL \url{http://dx.doi.org/10.1007/s10701-011-9559-y}.

\bibitem[Goldin and Sharp(1991)]{GoldinSharp1991}
G.~A. Goldin and D.~H. Sharp.
\newblock {The diffeomorphism group approach to anyons}.
\newblock \emph{Int. J. Mod. Phys. B}, 5\penalty0 (16--17):\penalty0
  2625--2640, 1991.
\newblock \doi{10.1142/S0217979291001048}.
\newblock URL \url{http://dx.doi.org/10.1142/S0217979291001048}.

\bibitem[DeWitt-Morette(1969)]{DeWitt1969}
C.~DeWitt-Morette.
\newblock {L'int{\'e}grale fonctionnelle de {Feynman}. {Une} introduction}.
\newblock \emph{Ann. Inst. H. P. Sec. A}, 11\penalty0 (2):\penalty0 153--206,
  1969.
\newblock URL \url{http://eudml.org/doc/75637}.

\bibitem[Farhi and Gutmann(1990)]{FarhiGutmann1990}
E.~Farhi and S.~Gutmann.
\newblock {The functional integral on the half line}.
\newblock \emph{Int. J. Mod. Phys. A}, 5\penalty0 (15):\penalty0 3029--3052,
  1990.
\newblock \doi{10.1142/S0217751X90001422}.
\newblock URL \url{http://dx.doi.org/10.1142/S0217751X90001422}.

\bibitem[Wilczek(1990)]{Wilczek1990}
F.~Wilczek.
\newblock \emph{{Fractional Statistics and Anyon Superconductivity}}.
\newblock World Scientific, 1990.

\bibitem[Khare(1997)]{Khare1997}
A.~Khare.
\newblock \emph{{Fractional Statistics and Quantum Theory}}.
\newblock World Scientific, 1997.

\bibitem[{Nayak} et~al.(2008){Nayak}, {Simon}, {Stern}, {Freedman}, and {Das
  Sarma}]{NayakSimonSternFreedmanDas2008}
C.~{Nayak}, S.~H. {Simon}, A.~{Stern}, M.~{Freedman}, and S.~{Das Sarma}.
\newblock {Non-Abelian anyons and topological quantum computation}.
\newblock \emph{Rev. Mod. Phys.}, 80\penalty0 (3):\penalty0 1083--1159, 2008.
\newblock \doi{10.1103/RevModPhys.80.1083}.
\newblock URL \url{http://dx.doi.org/10.1103/RevModPhys.80.1083}.

\bibitem[Dowker(1977)]{Dowker1977}
J.~S. Dowker.
\newblock {Quantum field theory on a cone}.
\newblock \emph{J. Phys. A}, 10\penalty0 (1):\penalty0 115--124, 1977.
\newblock \doi{10.1088/0305-4470/10/1/023}.
\newblock URL \url{http://dx.doi.org/10.1088/0305-4470/10/1/023}.

\bibitem[Deser et~al.(1984)Deser, Jackiw, and 't~Hooft]{DeserJackiwHooft1984}
S.~Deser, R.~Jackiw, and G.~'t~Hooft.
\newblock {Three-dimensional {Einstein} gravity: {Dynamics} of flat space}.
\newblock \emph{Ann. Phys. (N. Y.)}, 152\penalty0 (1):\penalty0 220--235, 1984.
\newblock ISSN 0003-4916.
\newblock \doi{10.1016/0003-4916(84)90085-X}.
\newblock URL \url{http://dx.doi.org/10.1016/0003-4916(84)90085-X}.

\bibitem['t~Hooft(1988)]{Hooft1988}
G.~'t~Hooft.
\newblock {Non-perturbative 2 particle scattering amplitudes in {$2+1$}
  dimensional quantum gravity}.
\newblock \emph{Commun. Math. Phys.}, 117\penalty0 (4):\penalty0 685--700,
  1988.
\newblock \doi{10.1007/BF01218392}.
\newblock URL \url{http://projecteuclid.org/euclid.cmp/1104161824}.

\bibitem[Deser and Jackiw(1988)]{DeserJackiw1988}
S.~Deser and R.~Jackiw.
\newblock {Classical and quantum scattering on a cone}.
\newblock \emph{Commun. Math. Phys.}, 118\penalty0 (3):\penalty0 495--509,
  1988.
\newblock \doi{10.1007/BF01466729}.
\newblock URL \url{https://projecteuclid.org/euclid.cmp/1104162096}.

\bibitem[Kay and Studer(1991)]{KayStuder1991}
B.~S. Kay and U.~M. Studer.
\newblock {Boundary conditions for quantum mechanics on cones and fields around
  cosmic strings}.
\newblock \emph{Commun. Math. Phys.}, 139\penalty0 (1):\penalty0 103--139,
  1991.
\newblock \doi{10.1007/BF02102731}.
\newblock URL \url{https://projecteuclid.org/euclid.cmp/1104203138}.

\bibitem[Horv{\'a}thy et~al.(1989)Horv{\'a}thy, Morandi, and
  Sudarshan]{HorvathyMorandiSudarshan1989}
P.~A. Horv{\'a}thy, G.~Morandi, and E.~C.~G. Sudarshan.
\newblock {Inequivalent quantizations in multiply connected spaces}.
\newblock \emph{Nuovo Cimento D}, 11\penalty0 (1--2):\penalty0 201--228, 1989.
\newblock \doi{10.1007/BF02450240}.
\newblock URL \url{http://wildcard.ph.utexas.edu/~sudarshan/pub/1989_005.pdf}.

\bibitem[Krantz(2008)]{Krantz2008}
S.~G. Krantz.
\newblock \emph{{A Guide to Complex Variables}}, volume~1 of \emph{MAA Guides}.
\newblock The Mathematical Association of America, 2008.

\bibitem[French and Krause(2006)]{FrenchKrause2006}
S.~French and D.~Krause.
\newblock \emph{{Identity in Physics: A Historical, Philosophical, and Formal
  Analysis}}.
\newblock Oxford University Press, 2006.

\bibitem[Gr{\"u}nbaum(1977)]{Grunbaum1977}
A.~Gr{\"u}nbaum.
\newblock {Absolute and relational theories of space and space-time}.
\newblock In  \citet{EarmanGlymourStachel1977}, pages 303--373.
\newblock URL \url{http://mcps.umn.edu/assets/pdf/8.11_Grunbaum.pdf}.

\bibitem[Stein(1977{\natexlab{a}})]{Stein1977a}
H.~Stein.
\newblock {Some philosophical prehistory of general relativity}.
\newblock In  \citet{EarmanGlymourStachel1977}, pages 3--49.
\newblock URL \url{http://mcps.umn.edu/assets/pdf/8.1_stein.pdf}.

\bibitem[Stein(1977{\natexlab{b}})]{Stein1977b}
H.~Stein.
\newblock {On space-time and ontology: extracts from a letter to Adolf
  Gr{\"u}nbaum}.
\newblock In  \citet{EarmanGlymourStachel1977}, pages 374--402.
\newblock URL \url{http://mcps.umn.edu/assets/pdf/8.12_Stein.pdf}.

\bibitem[Earman et~al.(1977)Earman, Glymour, and
  Stachel]{EarmanGlymourStachel1977}
J.~Earman, C.~Glymour, and J.~Stachel, editors.
\newblock \emph{{Foundations of Space-Time Theories}}, volume~8 of
  \emph{Minnesota Studies in the Philosophy of Science}, 1977. University of
  Minnesota Press.
\newblock URL \url{http://mcps.umn.edu/philosophy/completeVol8.html}.

\end{thebibliography}
\end{document}